\documentclass{npj}

\usepackage{verbatim}
\usepackage{xcolor}
\usepackage{lineno} 

\title{Observation of ultrafast interfacial Meitner-Auger energy transfer in a van der Waals heterostructure}

\author{Shuo Dong$^{1*}$, 
Samuel Beaulieu$^{1,2}$,
Malte Selig$^3$,
Philipp Rosenzweig$^4$,
Dominik Christiansen$^3$,
Tommaso Pincelli$^1$,
Maciej Dendzik$^{1,5}$,
Jonas D. Ziegler$^6$,
Julian Maklar$^1$, 
R. Patrick Xian$^{1,7}$,
Alexander Neef$^1$,
Avaise Mohammed$^4$, 
Armin Schulz$^4$, 
Mona Stadler$^8$, 
Michael Jetter$^8$, 
Peter Michler$^8$, 
Takashi Taniguchi$^9$,
Kenji Watanabe$^{10}$,
Hidenori Takagi$^{4,11,12}$,
Ulrich Starke$^4$,
Alexey Chernikov$^{6,13}$,
Martin Wolf$^1$, 
Hiro Nakamura$^{4,14}$,
Andreas Knorr$^3$,
Laurenz Rettig$^{1*}$ 
\& Ralph Ernstorfer$^{1,15*}$}

\begin{document}

\maketitle

\begin{affiliations}
 \item Fritz-Haber-Institut der Max-Planck-Gesellschaft, Faradayweg 4-6, 14195 Berlin, Germany
 \item Université de Bordeaux - CNRS - CEA, CELIA, UMR5107, F33405, Talence, France
  \item Nichtlineare Optik und Quantenelektronik, Institut für Theoretische Physik, Technische Universität Berlin, 10623 Berlin, Germany
 \item Max Planck Institute for Solid State Research, 70569 Stuttgart, Germany
  \item Department of Applied Physics, KTH Royal Institute of Technology, Hannes Alfvéns väg 12, 114 19 Stockholm, Sweden
 \item Department of Physics, University of Regensburg, Regensburg D-93053, Germany
 \item Department of Engineering, University of Cambridge, Trumpington Street, Cambridge CB2 1PZ, United Kingdom
 \item Institute of Semiconductor Optics and Functional Interfaces, Research Center SCoPE and IQST, University of Stuttgart, 70569 Stuttgart, Germany
 \item International Center for Materials Nanoarchitectonics, National Institute for Materials Science,  1-1 Namiki, Tsukuba 305-0044, Japan
  \item Research Center for Functional Materials, National Institute for Materials Science, 1-1 Namiki, Tsukuba 305-0044, Japan
 \item Department of Physics, University of Tokyo, 113-0033 Tokyo, Japan
 \item Institute for Functional Matter and Quantum Technologies, University of Stuttgart, 70569 Stuttgart, Germany
 \item Institute for Applied Physics, Dresden University of Technology, Dresden, 01187, Germany
 \item Department of Physics, University of Arkansas, Fayetteville, Arkansas 72701, USA
  \item Institut für Optik und Atomare Physik, Technische Universität Berlin, 10623 Berlin, Germany
 
\end{affiliations}
\begin{center}
\normalsize{$^\ast$To whom correspondence should be addressed;}\\
\normalsize{E-mail: dong@fhi-berlin.mpg.de, rettig@fhi-berlin.mpg.de, ernstorfer@fhi-berlin.mpg.de.} 
\end{center}

\newpage
\setlength{\parskip}{1em}
\begin{abstract}
Atomically thin layered van der Waals heterostructures feature exotic and emergent optoelectronic properties. With growing interest in these novel quantum materials, the microscopic understanding of fundamental interfacial coupling mechanisms is of capital importance. Here, using multidimensional photoemission spectroscopy, we provide a layer- and momentum-resolved view on ultrafast interlayer electron and energy transfer in a monolayer-WSe$_2$/graphene heterostructure. Depending on the nature of the optically prepared state, we find the different dominating transfer mechanisms: while electron injection from graphene to WSe$_2$ is observed after photoexcitation of quasi-free hot carriers in the graphene layer, we establish an interfacial Meitner-Auger energy transfer process following the excitation of excitons in WSe$_2$. By analysing the time-energy-momentum distributions of excited-state carriers with a rate-equation model, we distinguish these two types of interfacial dynamics and identify the ultrafast conversion of excitons in WSe$_2$ to valence band transitions in graphene. Microscopic calculations find interfacial dipole-monopole coupling underlying the Meitner-Auger energy transfer to dominate over conventional Förster- and Dexter-type interactions, in agreement with the experimental observations. The energy transfer mechanism revealed here might enable new hot-carrier-based device concepts with van der Waals heterostructures.
\end{abstract}

The unique physical properties of atomically thin two-dimensional (2D) materials\cite{novoselov20162d,akinwande2019graphene} and constantly improving fabrication methods\cite{liu2020disassembling,emtsev2009towards} have lead to a great interest in novel quantum materials based on van der Waals (vdW) heterostructures\cite{jin2018ultrafast}. 
By stacking 2D materials, vdW heterostructures inherit the properties from individual constituents, and exotic physical phenomena may emerge due to the interfacial interaction\cite{lee2014atomically,jin2018ultrafast,mcgilly2020visualization}. 
An emblematic example is the emergence of superconductivity in twisted bilayer graphene when stacked at the so-called ‘magic angle’\cite{cao2018unconventional}. As another example, interlayer excitons, which are spatially separated yet Coulomb-bound electron-hole pairs in semiconducting transition metal dichalcogenide (TMDC) heterostructures allow exceptional control of optoelectronic properties\cite{kunstmann2018momentum,rivera2015observation,binder2019upconverted}.
Out of the vdW heterostructure library, a basic optoelectronic building block is a monolayer (ML) semiconducting TMDC in contact with graphene\cite{avsar2014spin}. 
This hybrid structure represents a model system as it combines the strong light-matter coupling of TMDCs and the high mobility of massless Dirac carriers of graphene\cite{garcia2018spin}. 
The gapless electronic structure of graphene allows for harvesting low-energy photons, extending the spectral range covered by conventional photodetectors to the near-infrared wavelength, which is highly beneficial for photovoltaic applications\cite{massicotte2016photo}.

Optoelectronic functionality in vdW heterostructures arises from careful design and control of optical transitions and interfacial transfer processes.
Particularly, interfacial charge (ICT) and energy transfer (IET) are key processes which have triggered extensive experimental and theoretical efforts\cite{selig2019theory,froehlicher2018charge,aeschlimann2020direct,hill2017exciton,ferrante2021picosecond,luo2021twist}
Using time-resolved optical spectroscopies, a strong reduction of the exciton lifetime\cite{he2014electron} and optically active charge-transfer excitations of TMDC/graphene heterostructures have been observed\cite{krause2021ultrafast,zhou2021deciphering}, suggesting strong interlayer coupling and the underlying mechanisms have been discussed\cite{yuan2018photocarrier,fu2021long,krause2020microscopic}.
Moreover, the efficiency of IET processes like Förster-type coupling (based on electronic dipole-dipole interaction) has recently been investigated theoretically, pointing out the importance of energy-momentum conservation between participating quasiparticles\cite{selig2019theory}. 
Therefore, a momentum resolved probe is required to directly monitor the dynamics and reveal the mechanism of interfacial transfer process in vdW heterostructures, including those involving momentum-forbidden dark states.

\begin{figure}
\centering
\includegraphics[width=1\linewidth,keepaspectratio=true]{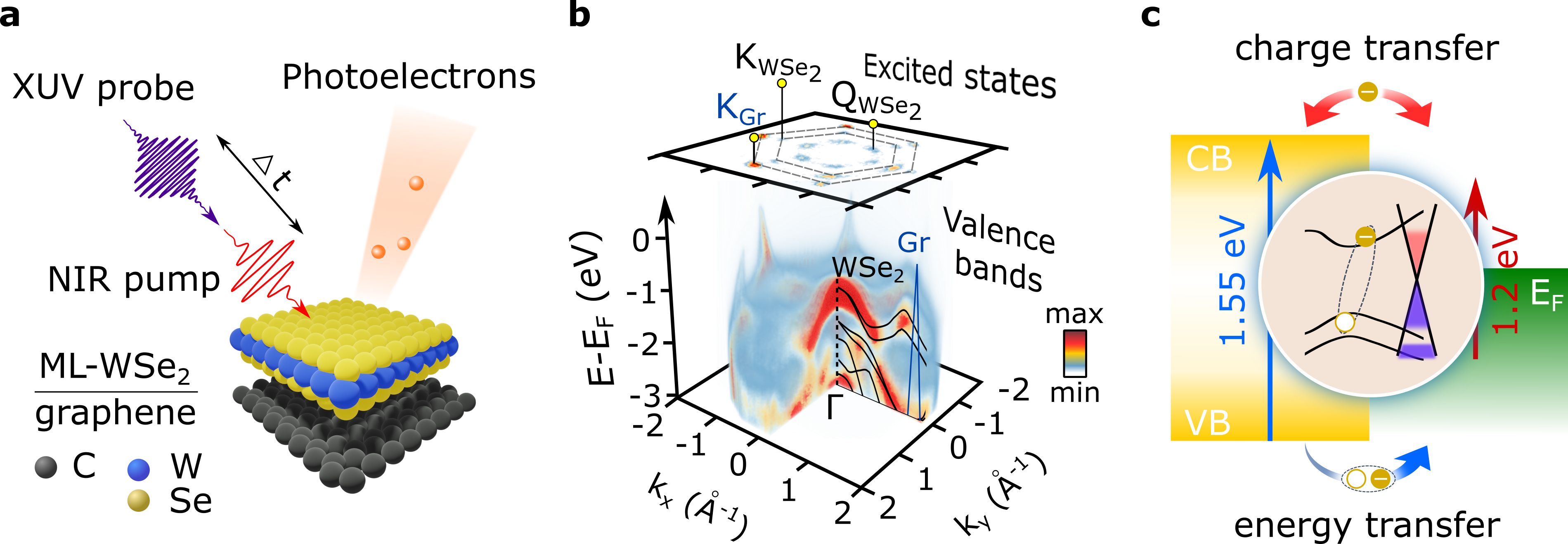}
\caption{\textbf{Time- and angle-resolved photoemission measurement of interlayer charge and energy transfer in a ML-WSe$_2$/graphene heterostructure.} \textbf{a}, Following the near-infrared pump, electrons are photoionized by the delayed XUV probe pulses and collected by a three-dimensional (3D) ($E_{kin}$, $k_x$, $k_y$) detector as a function of pump-probe delay $\Delta t$. \textbf{b}, The 3D snapshot of the 4D data, $I$($E_{kin}$, $k_x$, $k_y$, $\Delta t=0 ~\mathrm{fs}$) presents the valence band structures from the $\Gamma$ point to the Brillouin zone boundary of WSe$_2$, as well as the linearly dispersing graphene bands. The excited state population can be clearly mapped at the $\mathrm{K_{WSe_2}}$ and $\mathrm{Q_{WSe_2}}$ valleys, and the $\pi^*$ band of graphene ($\mathrm{K_{Gr}}$). \textbf{c}, By changing the pump wavelength, we can selectively prepare different initial excited states: quasi-free carriers in graphene with below-bandgap excitation (red arrow) or excitons in WSe$_2$ using excitation on the excitonic resonance (blue arrow).} 
\label{fig1}
\end{figure}

Here, we use time- and angle-resolved photoemission spectroscopy (trARPES) to investigate ultrafast interlayer carrier interactions in an epitaxially grown ML-WSe$_2$/graphene heterostructure. Our trARPES setup combines a high-repetition-rate (500~kHz) femtosecond extreme ultraviolet (XUV) source~\cite{puppin2019time} coupled to a time-of-flight momentum microscope\cite{maklar2020quantitative} (see Methods). It allows the measurement of the four-dimensional (4D) photoemission intensity $I$($E_{kin}$, $k_x$, $k_y$, $\Delta t$), where $E_{kin}$ is the outgoing photoelectron kinetic energy, $k_x$,$k_y$ are the in-plane momenta and $\Delta t$ is the pump-probe delay, as shown in Fig.\ref{fig1}\textbf{a,b}. The probe photon energy of 21.7~eV allows accessing the entire Brillouin zone of the heterostructure and the variable pump wavelength allows us to photoexcite the heterostructure in a state-resolved manner. In the following, we present a time-, energy-, and momentum-resolved study on the excited-state dynamics in the heterostructure with two different pump photon energies: below the optical bandgap of WSe$_2$ (1.2 eV) and in resonance with its first excitonic transition (1.55 eV).

\newpage

\section*{Interlayer quasi-free carrier transfer}

\begin{figure}
\centering
\includegraphics[width=16cm,keepaspectratio=true]{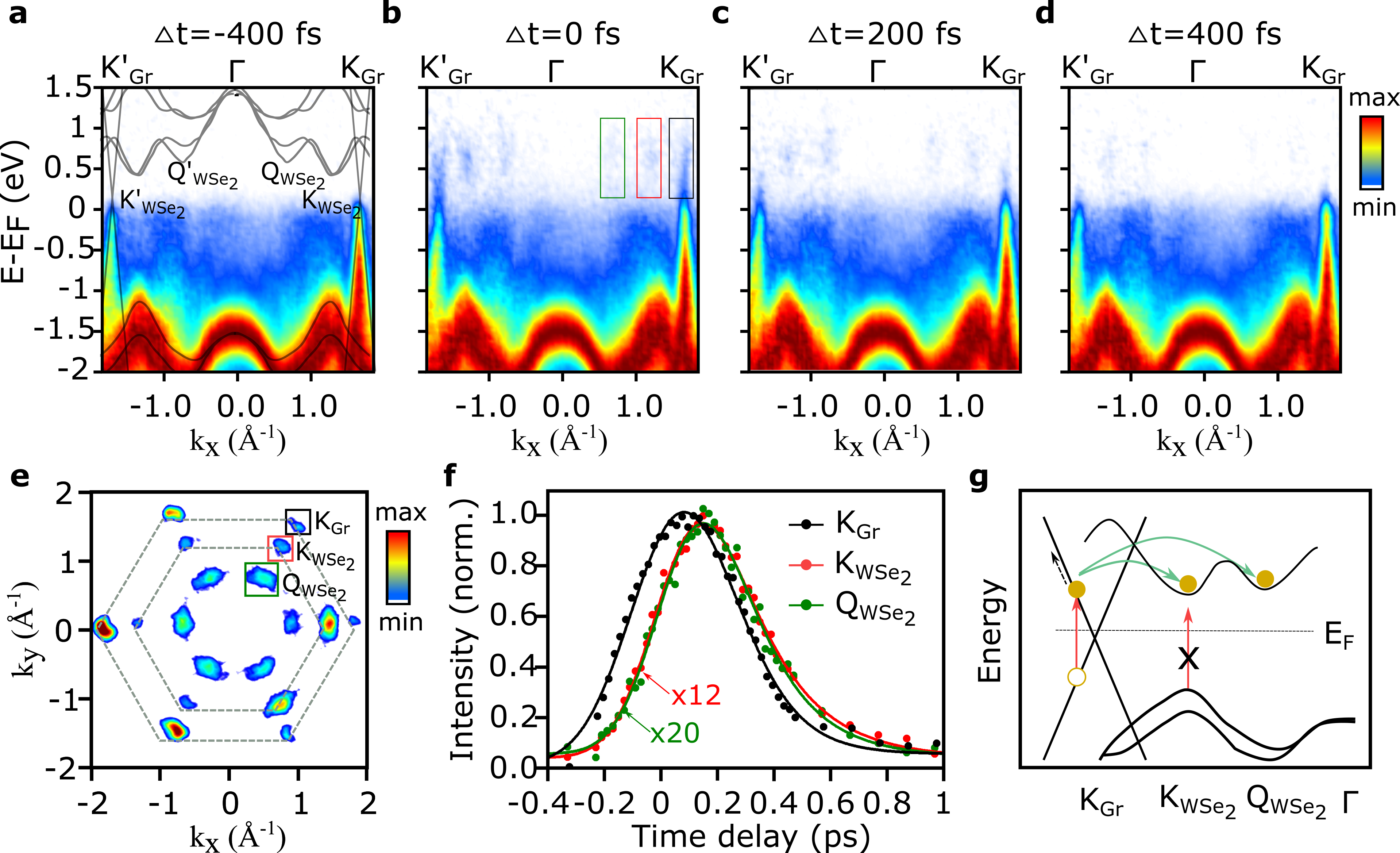}
\caption{\textbf{Layer- and valley-resolved ultrafast dynamics upon below-bandgap pumping.} \textbf{a-d}, Energy-momentum cuts of the photoemission signal along the $\mathrm{K'_{Gr}}$- $\mathrm{K'_{WSe_2}}$-$\mathrm{\Gamma}$-$\mathrm{K_{WSe_2}}$-$\mathrm{K_{Gr}}$ high symmetry direction, at selected pump-probe time delays. \textbf{a}, The 2D spectrum at negative time delay reveals the equilibrium band structure of ML-WSe$_2$ as well as the linearly dispersing $\pi$ band of graphene. The grey lines represent the DFT-calculated band structures (details in methods). Snapshots of the energy-momentum cuts at time delays of \textbf{b} $\Delta t=0~\mathrm{fs}$, \textbf{c} $\Delta t=200~\mathrm{fs}$ and \textbf{d} $\Delta t=400~\mathrm{fs}$, respectively. \textbf{e}, Momentum map of the excited states (energy integrated for $E>E_\mathrm{F}$ and time integrated for the first 400 fs), showing the $\mathrm{K_{Gr}}$ points of graphene (black box) as well as the $\mathrm{K_{WSe_2}}$ and $\mathrm{Q_{WSe_2}}$ valleys (red and green boxes, respectively). The dashed grey lines represent the hexagonal Brillouin zones of both layers. \textbf{f}, Normalized population dynamics within the three ROIs defined in \textbf{e}: $\mathrm{K_{WSe_2}}$ (red) and $\mathrm{Q_{WSe_2}}$ (green) are populated with a $\sim$50~fs delay with respect to $\mathrm{K_{Gr}}$ (black). \textbf{g}, Schematic of the early time carrier dynamics upon below-bandgap excitation: photogenerated hot carriers within the graphene layer are transferred to the conduction bands of WSe$_2$ via hot electron injection after the thermalization.}
\label{fig2}
\end{figure}

First, we photoexcite the heterostructure with the pump photon energy centred at $\hbar\omega_{pump}$=1.2~eV (pump pulse duration 200~fs FWHM), well below the optical bandgap of WSe$_2$\cite{li2014measurement}. The NIR-pump/XUV-probe experiments were performed with a pump fluence of $F=5.3 ~\mathrm{mJ/cm^2}$ and at room temperature. Fig.\ref{fig2}\textbf{a-d} show energy-resolved photoemission signals along the $\mathrm{K'}-\mathrm{K}$ cut of the Brillouin zone, at selected time delays. The momentum distributions above $E_F$ within the first 400~fs reveal that the excited states are localized in three different types of valleys: the Dirac cones of graphene at its $\mathrm{K}$ points ($\mathrm{K_{Gr}}$) and the $\mathrm{K}$ and $\mathrm{Q}$ valleys of WSe$_2$ ($\mathrm{K_{WSe_2}, \mathrm{Q_{WSe_2}}}$), as shown in Fig.\ref{fig2}\textbf{e}. The $\mathrm{Q_{WSe_2}}$ valley localizes between the $\mathrm{K_{WSe_2}}$ valley and the $\mathrm{\Gamma}$ point. By performing energy-momentum integration in selected regions of interest (ROIs), we extracted excited-state dynamics within these three valleys (Fig.\ref{fig2}\textbf{f}). 
Upon arrival of the pump pulses, the excited-state population rapidly builds up at $\mathrm{K_{Gr}}$ (black curve) and decays with a time scale of $\sim$ 200~fs. Strikingly, the conduction band minima (CBMs) at $\mathrm{K_{WSe_2}}$ (red curve) and $\mathrm{Q_{WSe_2}}$ valleys (green curve) are also being populated, however, with a delay of $\Delta t=51\pm9$ fs (see SI) compared to the rise of hot-carrier population in graphene. 
Since below-bandgap pump photon energy does not allow the direct photoexcitation of WSe$_2$, the delayed electron populations in the conduction bands arise through charge transfer from graphene to WSe$_2$. Two/multiple photon excitation can safely be ruled out (details see SI). The excited-state population of the $\mathrm{Q_{WSe_2}}$ valleys (Fig.\ref{fig2}\textbf{f}) could be raised via ICT from the graphene layer and the intervalley scattering from the $\mathrm{K_{WSe_2}}$ valleys\cite{wallauer2021momentum,madeo2020directly}.

These observations support the following picture of the underlying processes with a below-bandgap excitation: light is absorbed by graphene and populates unoccupied states at $E_{Gr}^{el}=E_D + \hbar\omega_{pump}/2$, leaving holes at $E_{Gr}^{h}=E_D - \hbar\omega_{pump}/2$ (Dirac energy $E_D>0$ for a p-doped system or $E_D<0$ for an n-doped system). The energy position of the Dirac point in our heterostructure is estimated to be $\sim -0.1~\mathrm{eV}$ below the Fermi level, obtained from the conical crossing\cite{bostwick2007quasiparticle,nakamura2020spin} (see SI). The photoexcited carriers quickly reach a quasi-thermalized states in $\sim$ 10 fs\cite{baudisch2018ultrafast} and could further increase their energy via intraband electron-electron scattering and interband Auger recombination in few tens of femtoseconds\cite{breusing2009ultrafast,chen2019highly}. Once electrons gained a sufficient amount of energy to overcome the energy barrier, they scatter to WSe$_2$ via a phonon-assisted tunneling process, filling the single-particle CBMs at $\mathrm{K_{WSe_2}}$ and $\mathrm{Q_{WSe_2}}$. This ICT mechanism is called \textit{interlayer hot carrier injection}, and is schematically illustrated in Fig.\ref{fig2}\textbf{g}. The excited electrons in WSe$_2$ may subsequently scatter back to graphene and relax down towards the Fermi energy ($E_F$). Based on the observed carrier dynamics, we performed microscopic calculations of the phonon-assisted interlayer tunneling process, allowing us to estimate the electronic wavefunction overlap between the involved conduction bands of WSe$_2$ and graphene to be around 4\% (see SI for details).

\newpage

\section*{Interlayer energy transfer: from excitons in WSe$_2$ to intraband transitions in graphene}

\begin{figure}
\centering
\includegraphics[width=12cm,keepaspectratio=true]{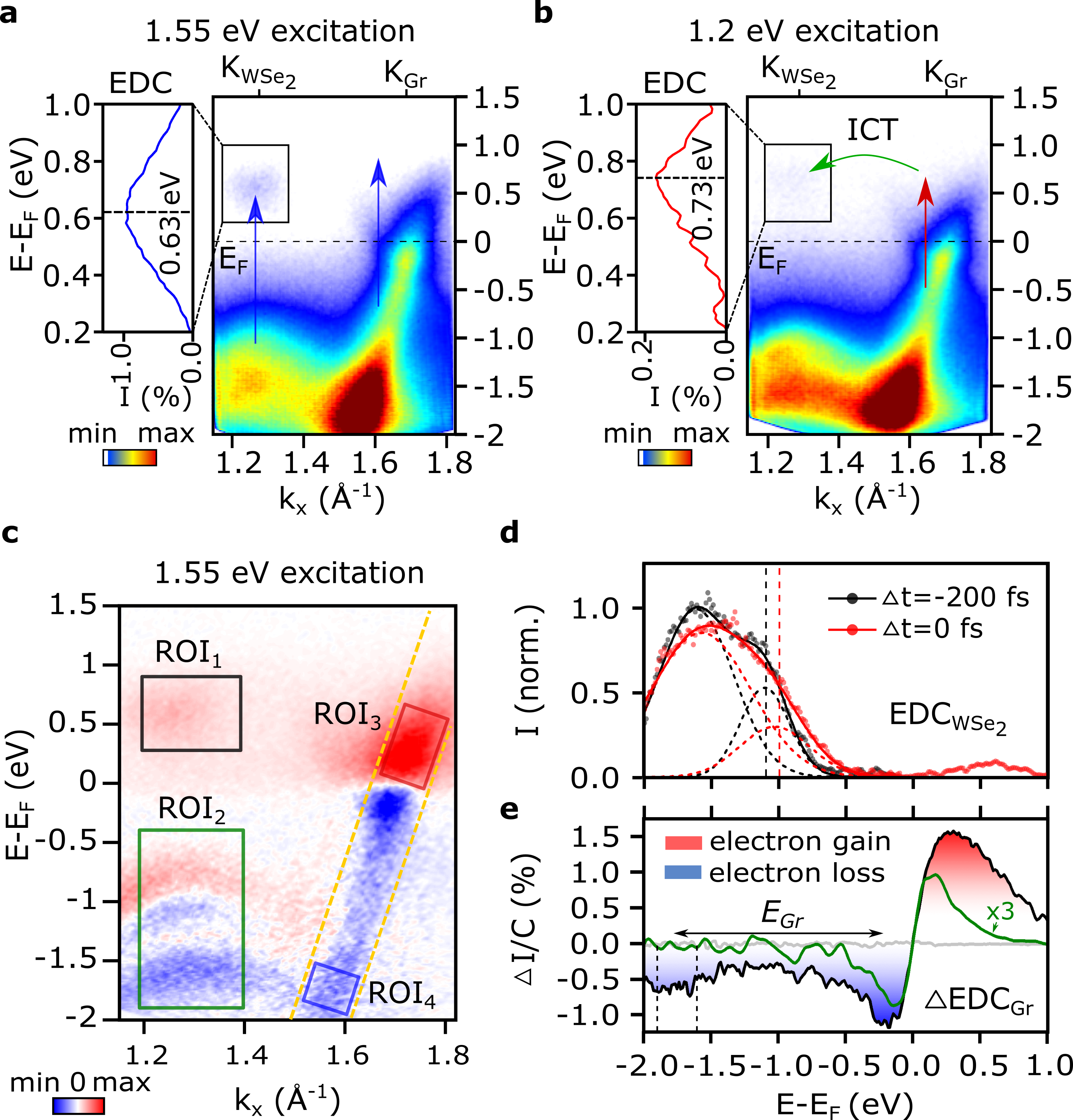}
\caption{\textbf{Photoemission signatures of exciton formation and interfacial interactions.} \textbf{a}, With near-resonant A-exciton pump (1.55 eV), carriers within both the WSe$_2$ and the graphene layer are photoexcited (time integration of 100 fs). The energy of the excited-states carriers at $\mathrm{K_{WSe_2}}$ is 0.63~eV, shown in the EDC (left panel figure). \textbf{b}, With below-bandgap excitation (1.2 eV), the local CBM of $\mathrm{K_{WSe_2}}$ is filled with ICT-induced electrons and centered at 0.73 eV. \textbf{c}, Differential energy-momentum cut with 1.55 eV pump at time zero, obtained by subtracting the negative time delay spectrum. \textbf{d}, The normalized EDC of $\mathrm{K_{WSe_2}}$ (momentum integration of 0.2 $\mathrm{\AA}^{-1}$) at $\Delta t=-200~\mathrm{fs}$ (black) and $\Delta t=0~\mathrm{fs}$ (red). The VBs are fitted with two Gaussian functions (dashed curves) and the positions of VB1 are indicated by the dash lines. \textbf{e}, The momentum-integrated spectrum of graphene Dirac bands (between the dashed yellow lines in \textbf{c}) shows the electron gain (positive, red area) and loss (negative, blue area) following photoexcitation. The intensity is normalized by the total electron count C obtained from negative time delay spectrum. Apart from the carriers accumulation near the $E_F$, the hole population forms another prominent peak around $E-E_F=-1.8~\mathrm{eV}$, indicated between the dash lines. The EDC of graphene with 1.2 eV pump (green) is also shown as a comparison}.
\label{fig3}
\end{figure}

Next, we select a pump photon energy of $\hbar\omega_{pump}$=1.55~eV (pump pulse duration: 35~fs FWHM, pump fluence: $F=1.7 ~\mathrm{mJ/cm^2}$), near-resonant to the A-excitonic transition of WSe$_2$. In this case, the pump photon energy allows both the WSe$_2$ and the graphene layer to be simultaneously photoexcited. One striking observation is that the energy distribution of excited carriers at the $\mathrm{K_{WSe_2}}$ valleys is centred at 0.63 eV (Fig.\ref{fig3}\textbf{a}), $\sim$100 meV lower than with below-bandgap excitation (Fig.~\ref{fig3}\textbf{b}), as apparent from the energy distribution curves (EDCs) (first 100 fs). As discussed above, with 1.2~eV excitation, the $\mathrm{K_{WSe_2}}$ valleys are filled with quasi-free electrons that have tunneled from the graphene layer. Therefore, this $\sim$ 100~meV energy difference is a direct photoemission signature of exciton formation, when near-resonantly pumping using 1.55 eV photons\cite{dong2021direct}: the bound electron-hole (el-h) pair reduces the quasi-free particle bandgap by the exciton binding energy. 
In addition to this excitonic feature, we also observe a transient shift of WSe$_2$ valence bands. In Fig.~\ref{fig3}\textbf{d}, EDCs at $\mathrm{K_{WSe_2}}$ are shown at $\Delta t=0~\mathrm{fs}$ (red) and $\Delta t=-200~\mathrm{fs}$ (black), in which the top two valence bands, VB1 and VB2, are fitted using Gaussian lineshape functions (see SI). The peak position of VB1 shifts towards the conduction band within the first 100 fs, transiently shrinking the electronic bandgap. This is due to the arrival of ICT-induced charge carriers from the graphene layer. With near-resonantly pumping the A-exciton, the occurrence of ICT and injection of quasi-free carriers from graphene to WSe$_2$ is expected, similar to the case of below-bandgap excitation. This could lead to dynamical screening effect and the observed bandgap renormalization, as reported in highly-excited or doped ML TMDC materials\cite{chernikov2015population,liu2019direct,gao2016dynamical,dendzik2020observation,steinhoff2014influence}. As the magnitude of such a transient bandgap renormalization has been shown to scale with the excited charge carrier density\cite{gao2016dynamical, liang2015carrier}, we utilize the VB shift in the following as a measure of the ICT transferred carriers dynamics from graphene layer.

\begin{figure}
\centering
\includegraphics[width=16cm,keepaspectratio=true]{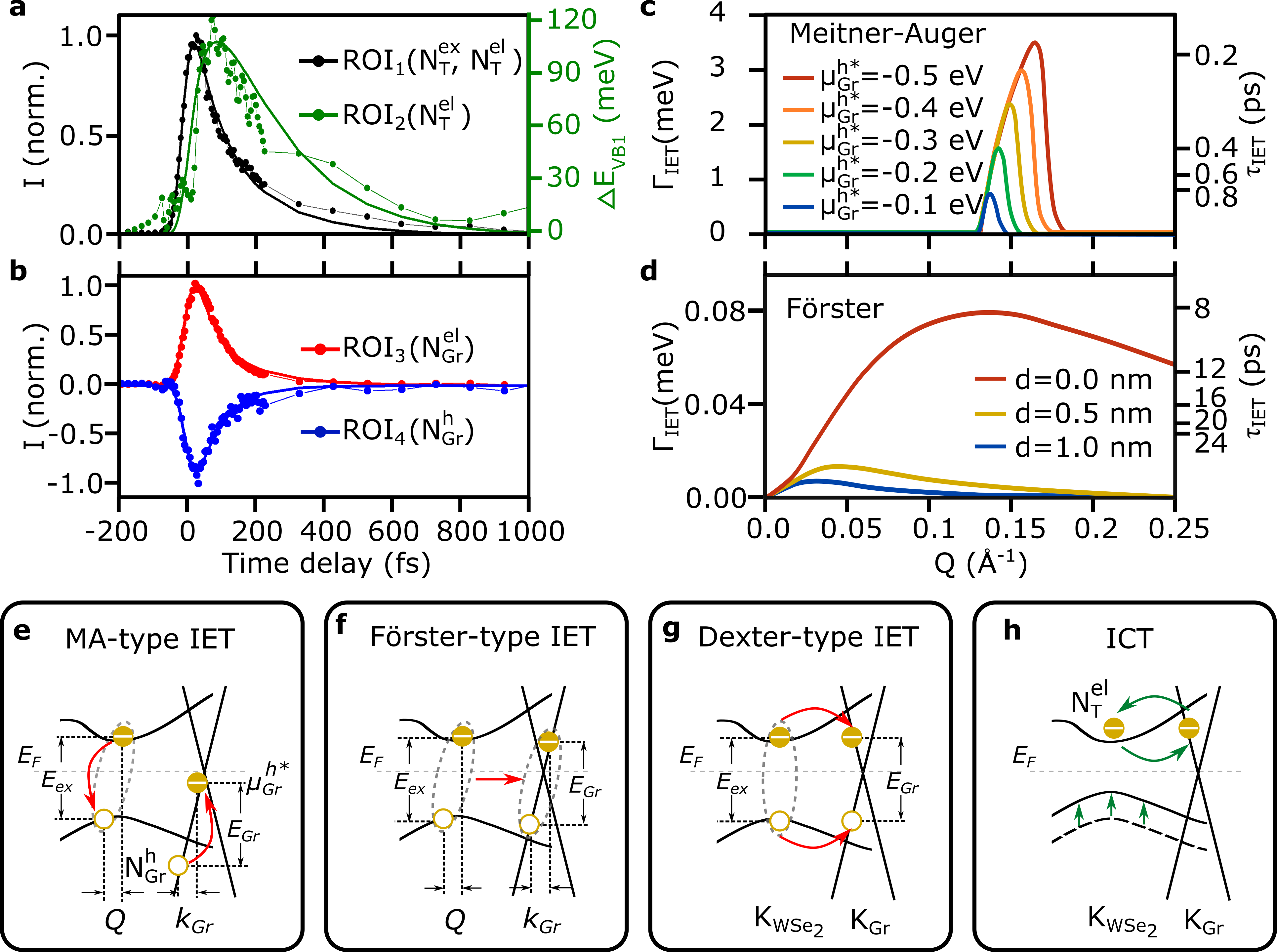}
\caption{\textbf{Interlayer charge and energy transfer upon near-resonant A-exciton excitation.} \textbf{a}, By integrating the ROI$_1$ in Fig.\ref{fig3}\textbf{c}, the time trace of the normalized photoemission intensity of excited-state carriers at the CBM of WSe$_2$ (black)  contains the dynamics of excitons ($N_T^{ex}$) and ICT-induced quasi-free carriers ($N_T^{el}$). The energy shift of VB1 (green) mainly reflects the dynamic of $N_T^{el}$, which are extracted from time-dependent EDCs in ROI$_2$. \textbf{b}, The time traces of hot electrons (red) and hot holes in the deep VB (blue) in graphene are extracted from the ROI$_3$ and ROI$_4$ in Fig.\ref{fig3}\textbf{c}, respectively. The time traces in \textbf{a-b} are fitted globally based on a rate equation model (see text).} \textbf{c}, Calculated Meitner-Auger mediated IET transfer rate as a function of COM momentum $\mathbf{Q}$ with different photo-induced hole vacancy at $E=\mu_{Gr}^{h*}$. \textbf{d}, Calculated Förster coupling rate as a function of $\mathbf{Q}$ with varied interlayer distance of $d$. Sketch of the underlying carrier dynamics: \textbf{e}, Meitner-Auger IET with creation of intraband electron-hole pairs in graphene by absorbing the exciton energy. \textbf{f}, Förster-type energy transfer with generation of interband electron-hole pairs in graphene. \textbf{g}, Dexter-type energy transfer with electrons and holes injection to graphene simultaneously. \textbf{h}, ICT-induced hot electron injection into WSe$_2$ and transient energy shift of its valence band.
\label{fig4}
\end{figure}

In addition to the excited-state dynamics in WSe$_2$, important insight can be drawn from the energy-momentum distribution of hot carriers in graphene.
As shown in the early-time 2D differential spectrum $\Delta I(E,k,\Delta t=0~\mathrm{fs})$  (Fig.~\ref{fig3}\textbf{c}), obtained by subtracting the spectrum at negative time, hot carriers distribute in a broad energy range. The momentum-integrated spectrum along the linearly dispersing band in Fig.~\ref{fig3}\textbf{e} clearly features the energy distribution of net electron gain (positive; red area) and loss (negative; blue area) following near-resonant photoexcitation. 
Remarkably, besides the modification of the distribution function near the Fermi level, we notice a strong negative peak at $E-E_F=-1.8~\mathrm{eV}$. As noted earlier, for direct photoexcitation in graphene the photoexcited carriers are expected to be spread $\pm 0.77 ~\mathrm{eV}$ ($\hbar\omega_{pump}/2$) around the Dirac point and quickly relax back to the Fermi level. Thus, this simple excitation mechanism cannot explain this peculiar feature in the valence band spectrum. The electron-electron scattering and Auger recombination could lead to a transient broadening of the momentum-space carrier distribution, but without any preferential energy localization\cite{tomadin2013nonequilibrium,chen2019highly,gierz2014non}.
Interlayer hot \emph{hole} transfer can also be ruled out, as the top valence band of WSe$_2$ lies at $E-E_F=-1.0~\mathrm{eV}$. It would require a multi-phonon absorption to populate the hole-states localized deeply in the valence band, taking the typical phonon energy of $\sim$0.17 eV in graphene\cite{na2019direct}, a process of very low probability. 
However, the energy difference of deep-lying valence holes ($E-E_F=-1.8~\mathrm{eV}$) and states near $E_F$ ($E-E_F=-0.2~\mathrm{eV}$) in graphene well matches the energy of the A-exciton in WSe$_2$  ($E_{ex}\sim1.6~\mathrm{eV}$). Combined with the fast depletion of exciton population shown in Fig.\ref{fig4}\textbf{a} (black curve) extracted from the excited state of WSe$_2$ (ROI$_1$ in Fig.\ref{fig3}\textbf{c}), this brings about the following scenario for the excitation of these carriers: annihilation of excitons in WSe$_2$ drives the intraband excitation of deep-lying valence electrons in graphene into empty hole states below the Dirac point. In more detail, this exciton energy transfer process, which we term Meitner-Auger energy transfer\cite{meitner1922connection,auger1923rayons}, considers recombination of excitons in WSe$_2$ with center-of-mass (COM) momentum $\mathbf{Q}$ and exciton energy $E_{ex}$. The photoexcitation prepares the required hot hole vacancy below $E_F$ in graphene, thus enabling the intraband excitation. The momentum of the valence electron-hole pair $k_{Gr}$ is determined by the Fermi velocity of the graphene bands and the transition energy $E_{Gr}$. This required momentum is provided by the optically pumped excitons which gain finite COM momenta during the population formation process via phonon-mediated dephasing and intravalley thermalization\cite{selig2016excitonic,christiansen2019theory,selig2019ultrafast,pollmann2015resonant} (see the discussion in SI). 
The highly efficient IET of the excitons and intraband electron-hole pairs is thus possible under the conservation of energy and momentum, \emph{i.e.}, $E_{ex}=E_{Gr}$ and $\mathbf{Q}=k_{Gr}$. 
In a similar trARPES study of a ML WS$_2$/graphene heterostructure, dominating interfacial charge transfer has been observed\cite{aeschlimann2020direct}. Compared with our study, the different charge transfer rates could be raised from the different band structure alignment near the interface and the density of defect-sites\cite{krause2020microscopic}. While the additional exciton energy transfer was not excluded, its relative efficiency might be reduced due to the larger COM momentum required at the larger A-exciton energy of WS$_2$ and the energy level alignment of these specific samples.    

In order to gain information on the time scales of the energy and charge transfer processes, next we analyze the dynamics of excited-state populations extracted from the ROIs shown in Fig.~\ref{fig3}\textbf{c}, including the excited-state carriers in WSe$_2$ (ROI$_1$), VB1 shifting (ROI$_2$), hot electrons in graphene (ROI$_3$) and IET-driven deep valence band holes (ROI$_4$). The time trace of hot carriers in the CBM of WSe$_2$ (black curve in Fig.\ref{fig4}\textbf{a}) contains two types of quasi-particles dynamics: the photo-generated excitons $N_T^{ex}$ and the ICT-induced quasi-free electrons $N_T^{el}$. 
The decay of excitons excite the valence band electrons in graphene via IET with a transfer time of $\tau_{IET}$ (Fig.~\ref{fig4}\textbf{f}). On the other hand, the arrival of ICT-induced electrons transiently shifts the VBs of WSe$_2$ (green curve in Fig.\ref{fig4}\textbf{a}) which therefore represents the dynamics of $N_T^{el}$ as discussed before.We assume VB1 and VB2 shift in the same way (fitting details see SI). The VB1 shifting shows a time delay of $\sim 65$ fs compared to the CB signal, evidencing the occurrence of interlayer hot electron injection after photoexcitation. The population of $N_T^{el}$ subsequently relaxes back to $\mathrm{K_{Gr}}$, refilling the excited-states of graphene (Fig.~\ref{fig4}\textbf{h}). 
From the graphene side, the photoexcited hot electrons $N_{Gr}^{el}$ (red curve in Fig.\ref{fig4}\textbf{b}) could either scatter to conduction bands of WSe$_2$ or relax by interband decay channels in graphene. Therefore, the relaxation of $N_{Gr}^{el}$ could be characterized with the charge transfer time of $\tau_{ICT}$ and a decay time of $\tau_{Gr}^{el}$.
The deep valence band holes $N_{Gr}^{h}$ (blue curve in Fig.\ref{fig4}\textbf{b}) are populated by exciton energy transfer on a time scale of $\tau_{IET}$, which would relax back to the Fermi level with a lifetime of $\tau_{Gr}^h$.

The complete dynamics across the interface can be described with a set of coupled rate equations based on a multi-level scheme (details see SI).
By numerically solving the rate equation model, we disentangle the dynamics of IET and ICT. Our global fit describes the data well and yields the transfer times of $\tau_{IET}=67\pm 7~\mathrm{fs}$ and $\tau_{ICT}=118\pm 18~\mathrm{fs}$.
The lifetimes of electrons and IET-populated hot holes in graphene are simultaneously extracted as $\tau_{Gr}^{el}=84\pm 7~\mathrm{fs}$ and $\tau_{Gr}^h=7\pm 4~\mathrm{fs}$.     
Combining all our observations and analysis of the energy-momentum dynamics in WSe$_2$ and graphene, we summarize the interfacial phenomena governing the non-equilibrium behaviour of our heterostructure: first, the optical pump generates excitons in WSe$_2$ and quasi-free carriers in graphene (Fig.~\ref{fig4}\textbf{e}). Following photoexcitation, the exciton annihilation excites deep valence electrons in graphene via an IET process (Fig.~\ref{fig4}\textbf{f-g}). 
Simultaneously, hot electrons in graphene are injected to the conduction bands of WSe$_2$ via ICT which transiently shift the valence bands of WSe$_2$ (Fig.~\ref{fig4}\textbf{h}).

To elucidate the interfacial coupling mechanism at play in our experiment, in particular the observed ultrafast energy transfer rate, we perform microscopic calculations of three types of IET mechanisms: Meitner-Auger, Förster and Dexter energy transfer. The interlayer MA process is described by the dipole-monopole energy transfer from excitons to valence band excitation, schematically shown in Fig.~\ref{fig4}\textbf{e}. The photoexcited hot holes in graphene quickly relax and distribute below $E_F$ near a transient chemical potential $\mu_{Gr}^{h*}$. This allows an MA-type transition from the deep valence band to the hot hole vacancy by absorbing the exciton energy. The microscopically calculated transfer rate is plotted as a function of $\mathbf{Q}$ in Fig.\ref{fig4}\textbf{c} with different transient chemical potentials for the hole distributions $\mu_{Gr}^{h*}$. When the hole vacancy is located around $\mu_{Gr}^{h*}=-0.3~\mathrm{eV}$, the maximum transfer rate reaches $\Gamma_{IET}=2.4~\mathrm{meV}$, corresponding to a $\tau_{IET}=270~\mathrm{fs}$ transfer time. The MA-type IET process could describe the observed energy-momentum distribution of intraband transition of valence electrons in a reasonable quantitative agreement with the extracted transfer rate.

Another IET mechanism is Förster energy transfer (Fig.\ref{fig4}\textbf{f}). The energy of the exciton excites an \textit{interband} transition from valence bands to above Dirac point via the dipole-dipole coupling.\cite{forster1948intermolecular}  In contrast to the MA-type IET process, the interband excitation via Förster-type energy transfer populates the conduction bands of graphene above the Fermi level, independent of the photon-induced hot carriers distribution. The coupling strength is explicitly evaluated (for derivation, see SI) and determined by the momentum $\mathbf{Q}$ and interlayer distance $d$. The strong exciton oscillator strength and intrinsic in-plane exciton dipole moment in many 2D materials favor the Förster-type IET\cite{kozawa2016evidence}. However, the calculated transfer rate is only $0.08~\mathrm{meV}$ (a transfer time of $\sim 8.1$ ps), even assuming a tightly stacked heterostructure with interlayer distance of $d=0~\mathrm{nm}$ (Fig.\ref{fig4}\textbf{d}). Our calculations reveal that the IET process preferably excites an \textit{intraband} rather than an \textit{interband} transition. The experimentally observed energy-momentum distribution of excited-state hot holes supports this conclusion.  
In addition, we also performed calculations of Dexter-type IET (Fig.\ref{fig4}\textbf{g}), in which scenario the electron and hole components of excitons in WSe$_2$ scatter to the graphene layer simultaneously. However, due to the small wavefunction overlap and the finite momentum distance between $\mathrm{K_{WSe_2}}$ and $\mathrm{K_{Gr}}$, we found a very weak Dexter-type interlayer coupling strength, more than three orders of magnitude smaller compared to the other two mechanisms (see SI).
We can thus identity the MA-type conversion of excitons in WSe$_2$ to intraband excitations in graphene as the dominant IET mechanism. 

In this work, we provide a detailed microscopic picture of interfacial charge and energy transfer processes in photoexcited ML-WSe$_2$/graphene heterostructures.
Optical excitation of electrons in graphene leads to inter-layer charge transfer of quasi-free electrons from the graphene layer to the K and Q valleys of the semiconductor's conduction bands on a time scale of $\sim$50~fs.
In contrast, excitons in WSe$_2$ decay through an interfacial Meitner-Auger energy transfer process with a time constant of $\sim$70~fs. This previously unidentified process is governed by inter-layer dipole-monopole interactions leading to annihilation of an exciton in WSe$_2$ and \textit{non-vertical intraband} excitations in graphene. 
The momentum of the electron-hole pair in graphene originates from the finite center of mass momentum of the hot excitons in WSe$_2$. The interfacial Meitner-Auger mechanism is found to dominate the energy transfer process over established mechanisms like Förster- and Dexter-type transfer.
This mechanism results in transient hole distributions as low as 2~eV below the Dirac points. These observations enrich the physical toolbox for designing van der Waals heterostructures and might be utilized in hot-carrier photovoltaic device concepts to harness the ultrafast and efficient carrier transfer processes at interfaces\cite{paul2021hot}.

\begin{methods}

\subsection{Time- and angle-resolved photoemission spectroscopy}

We used a 500kHz tabletop femtosecond optical parametric chirped pulse amplification (OPCPA) laser system operated at a center wavelength of 800 nm and delivering average power up to 15 W. The high harmonic generation is produced in a vacuum chamber by tight focusing (10 $\mathrm{\mu m}$) the second harmonic (400 nm) of the OPCPA fundamental on a thin and dense argon gas jet. We select the photons around 21.7 eV (110 meV FWHM bandwidth) as the probe arm for trARPES experiment\cite{puppin2019time}. Concerning the pump arm, we used two different beams for this study. One pump beam is directly obtained from the OPCPA (800 nm, FWHM=35 fs) and another one is the residual power of the compressed fiber amplifier (1030 nm, FWHM=200 fs). The pump and probe beams are coupled into an ultra-high-vacuum (UHV) chamber and spatially overlapped at the sample position which is controlled by a six-axis manipulator (Carving, SPECS GmbH). The main UHV chamber is equipped with an unique combination of a hemispherical electron energy analyzer (PHOIBOS150, SPECS GmbH) and time-of-flight (ToF) momentum microscope (METIS1000, SPECS GmbH)\cite{maklar2020quantitative}. On the one hand, the hemispherical analyzer, which can work in a multi-electrons per laser shot regime, provides high statistic energy/momentum cuts along a given momentum direction, as shown in Fig.\ref{fig3}. On the other hand, the momentum microscope allows for efficient, parallel, momentum-resolved detection of the full photoemission horizon from the surface as shown in Fig.\ref{fig1}\textbf{b} and Fig.\ref{fig2}\textbf{a-e}. All the experiments are performed at room temperature.

\subsection{ML-WSe$_2$/ML-Graphene vdW heterostructure fabrication}
Monolayer graphene on SiC (Si-terminated surface) was grown using the well-established recipe of sublimation growth at elevated temperatures in an argon atmosphere\cite{emtsev2009towards}. Note that, on SiC\cite{riedl2010structural}, the graphene monolayer resides on top of a $(6\sqrt{3} \times 6\sqrt{3})$R30\si{\degree} reconstructed carbon buffer layer that is covalently bound to the SiC substrate. WSe$_2$ films were grown on the thus prepared MLG/SiC substrates via hybrid-pulsed-laser deposition (hPLD) in ultra-high vacuum\cite{nakamura2020spin}. Pure tungsten (99.99\%) was ablated using a pulsed KrF excimer laser (248 nm) with a repetition rate of 10 Hz, while pure selenium (99.999\%) was evaporated from a Knudsen cell at a flux rate of around 1.5 \si{\angstrom/s} as monitored by a quartz crystal microbalance. The deposition was carried out at 450\si{\degree}C for 6 h, followed by two-step annealing at 640\si{\degree}C and 400\si{\degree}C for 1 h each.

\subsection{DFT calculation of band structure}
We performed density functional theory (DFT) calculation of suspended ML WSe$_2$ and graphene with the projector augmented wave code GPAW\cite{mortensen2005real} using GLLB-SC xc-functional, separately. The GLLB-SC is an orbital-dependent exact exchange-based functional including the spin-orbital coupling\cite{gritsenko1995self}.The relaxed lattice constant of WSe$_2$ is a=3.25 $\mathrm{\AA}$. We sample the Brillouin Zone with a ($15\times15\times1$) k-point mesh, and set the cutoff energy for the plane-wave expansion at 600 eV. The bandgap is adjusted to fit our data. The calculated band structures of both materials are superimposed on each other and shown in Fig.\ref{fig2}\textbf{a}.

\end{methods}

\section*{Data availability}
All data underlying this study are available from the Zenodo repository. Source data are provided with this paper.

\newpage
\section*{References}
\bibliographystyle{plain}

\begin{addendum}
 \item[Acknowledgements] This work was funded by the Max Planck Society, the European Research Council (ERC) under the European Union’s Horizon 2020 research and innovation program (Grant No. ERC-2015-CoG-682843), the German Research Foundation (DFG) within the Emmy Noether program (Grant No. RE 3977/1), through Projektnummer 18208777-SFB 951 "Hybrid Inorganic/Organic Systems for Opto-Electronics (HIOS)" (CRC 951 project B12, M.S., D.C., A.K.), and the SFB/TRR 227 "Ultrafast 264 Spin Dynamics" (projects A09 and B07). S.B. acknowledges financial support from the NSERC-Banting Postdoctoral Fellowships Program. M.D. acknowledges financial support from the Göran Gustafssons Foundation. K.W. and T.T. acknowledge support from the Elemental Strategy Initiative conducted by the MEXT, Japan (Grant Number JPMXP0112101001) and JSPS KAKENHI (Grant Numbers 19H05790 and JP20H00354).

 \item[Author contributions] S.D., S.B., T.P., M.D., J.M., A.N. and L.R.~performed the trARPES measurement. S.D.~analyzed the data and wrote the first draft of the manuscript. R.E., L.R. and M.W.~were responsible for developing all the experimental infrastructures. M.S. and D.C.~performed the microscopic calculation with the guidance of A.K.. R.P.X.~developed the 4D data processing code. P.R. and H.N. provided the epitaxially-grown heterostructure, with support from U.S. and H.T.. A.M., A.S., and M.S. conducted Raman and photoluminescence measurements, with guidance from M.J and P.M.. J.D.Z and A.C.~prepared the exfoliated ML sample with the hBN substrate provided by K.W. and T.T.. All authors contributed to the final version of the manuscript.
 \item[Competing Interests] The authors declare that they have no
competing financial interests.
 \item[Correspondence] Correspondence and requests for materials
should be addressed to dong@fhi-berlin.mpg.de, rettig@fhi-berlin.mpg.de and ernstorfer@fhi-berlin.mpg.de.

\end{addendum}



\end{document}


\maketitle

\begin{affiliations}
 \item Fritz-Haber-Institut der Max-Planck-Gesellschaft, Faradayweg 4-6, 14195 Berlin, Germany
 \item Universit\'e de Bordeaux - CNRS - CEA, CELIA, UMR5107, F33405, Talence, France
  \item Nichtlineare Optik und Quantenelektronik, Institut f\"ur Theoretische Physik, Technische Universit\"at Berlin, 10623 Berlin, Germany
 \item Max Planck Institute for Solid State Research, 70569 Stuttgart, Germany
  \item Department of Applied Physics, KTH Royal Institute of Technology, Hannes Alfv\'ens v\"ag 12, 114 19 Stockholm, Sweden
 \item Department of Physics, University of Regensburg, Regensburg D-93053, Germany
 \item Department of Engineering, University of Cambridge, Trumpington Street, Cambridge CB2 1PZ, United Kingdom
 \item Institute of Semiconductor Optics and Functional Interfaces, Research Center SCoPE and IQST, University of Stuttgart, 70569 Stuttgart, Germany
 \item International Center for Materials Nanoarchitectonics, National Institute for Materials Science,  1-1 Namiki, Tsukuba 305-0044, Japan
  \item Research Center for Functional Materials, National Institute for Materials Science, 1-1 Namiki, Tsukuba 305-0044, Japan
 \item Department of Physics, University of Tokyo, 113-0033 Tokyo, Japan
 \item Institute for Functional Matter and Quantum Technologies, University of Stuttgart, 70569 Stuttgart, Germany
 \item Institute for Applied Physics, Dresden University of Technology, Dresden, 01187, Germany
 \item Department of Physics, University of Arkansas, Fayetteville, Arkansas 72701, USA
 \item Institut f\"ur Optik und Atomare Physik, Technische Universit\"at Berlin, 10623 Berlin, Germany
\end{affiliations}

\newpage
This file includes:

Supplementary Notes 

Supplementary Methods 

Supplementary Equation (1) to (88) 

Supplementary Fig.1 to Supplementary Fig.14

Supplementary Table 1 

References 

\newpage
\noindent
\begin{Large}  
Supplementary Notes  
\end{Large} 

\noindent
\underline{\textbf{Characterization of the ML-WSe$_2$/graphene heterostructure}}

\noindent
Raman measurements were carried out at room temperature using a 532 nm laser with a power of 1 mW and a spot size of 5 to 10 $\mathrm{\mu m}$. As shown in Supplementary Fig.~\ref{figs1}\textbf{a}, ML-WSe$_2$ is confirmed by an intense peak at 250 cm$^{-1}$ which comes from essentially degenerate A$_{1g}$ and E$_{2g}$ lattice vibration modes\cite{zhao2013lattice,terrones2014new}. Photoluminescence (PL) measurements are performed using another system with a 532 nm excitation laser, power of 1 mW and spot size of 1~$\mu$m at room temperature. In Supplementary Fig.~\ref{figs1}\textbf{b}, ML-WSe$_2$ on graphene presents two weak PL peaks (778 nm and 914 nm) only slightly above the background. The peak at 778 nm is close in energy to the A-exciton transition energy\cite{tonndorf2013photoluminescence,he2014tightly}. The origin of the peak at higher wavelength is unknown, and may come from the existence of in-gap defect states. The weakness of the PL signals is consistent with the quenching of PL known to occur for ML-TMDCs adjacent to graphene\cite{froehlicher2018charge}.

The sample is protected by the Se capping layer before sending to our lab. After introducing the sample into our ultrahigh vacuum (UHV) photoemission end-station, we have annealed the sample for 15 minutes at 400$^{\circ}$C  through direct current heating to remove the Se capping. After annealing, we recorded a low energy electron diffraction (LEED) pattern with the incident beam energy of 95 eV, to verify the surface cleanliness and ordering (Supplementary Fig.~\ref{figs1}\textbf{c}). The six outer sharp LEED spots come from the bottom ML graphene layer (yellow box) and the inner six arc-shaped diffraction spots originate from the top ML-WSe$_2$ layer (red box). The occurrence of a well-oriented hexagonal pattern of WSe$_2$ spots aligned to the graphene pattern attests the epitaxial nature of our heterostructure and single-domain lattice orientation. A certain level of strain-induced misalignment between nanoislands of WSe$_2$ with respect to the graphene layer is evident from the azimuthal widths of the diffraction spots.

Our sample exhibits areas without WSe$_2$, since the top WSe$_2$ layer consists of spatially uniform small islands with some distance between the islands. However, the relevant processes discussed in our manuscript, \emph{i.e.}, both interlayer hot electron injection with 1.2 eV pump and Meitner-Auger type interlayer energy transfer with 1.55 eV pump, are based on the photoemission signals from the heterostructure, but will not occur at an isolated graphene or WSe$_2$ layer. For example, the observation of the excited state population at WSe$_2$ with below-bandgap excitation is not possible for an isolated monolayer WSe$_2$. With near-resonant pump, the deep-lying holes would not be excited in a pure graphene sample. Therefore, the signal discussed in our work is based on the heterostructure area.

Additionally, the doping level of epitaxial graphene is different from the pristine graphene without WSe$_2$. We discuss possible reasons for this observation: static charge (electron) transfer from graphene to WSe$_2$, or the decrease of electrons at the graphene/SiC interface with W or Se intercalation as the referee suggested. In the latter case that a substantial part of the Se or W elements is intercalated below the graphene layer, the intensity of the diffraction pattern of SiC, \emph{i.e.}, 6$\times$6 spots around the graphene (10) and diamond-shape distributed spots around $\sqrt{3}\times\sqrt{3}$ SiC, would be dramatically reduced. As we observed a clear SiC pattern in our LEED image (Supplementary Fig.\ref{figs1}\textbf{c}), this scenario seems unlikely. Second, we would expect graphene bilayer growth with Se or W intercalation, which is also not observed. Finally, there is no core level energy shift of SiC peaks after the growth of WSe$_2$\cite{nakamura2020spin}, which demonstrates the band alignment at the graphene/SiC interface (band bending of p-n junction) is not influenced by the top layer. These observations rule out a significant contribution of W or Se intercalation to the graphene's doping level. Therefore, we conclude on a static charge transfer from graphene to WSe$_2$ is the major contribution to the modification of the doping of graphene.

\begin{figure}
    \centering
    \includegraphics[width=1\linewidth,keepaspectratio=true]{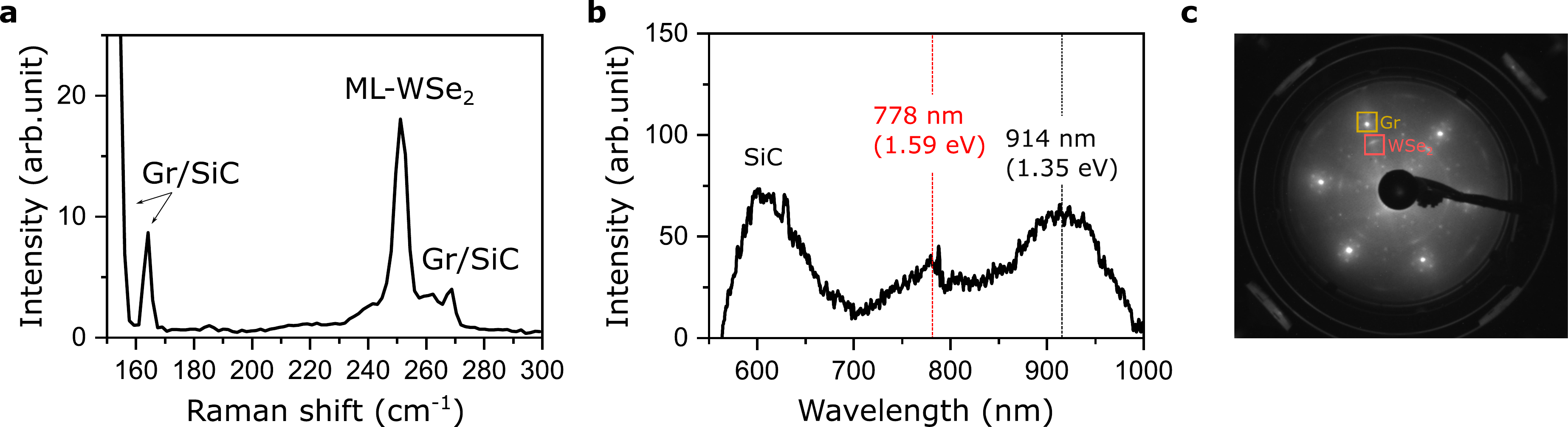}
    \caption{\textbf{Optical characteristics and surface analysis of the heterostructure sample.} \textbf{a}, Raman measurement of ML-WSe$_2$/graphene (Gr) at room temperature. An intense peak at 250 cm$^{-1}$ belongs to ML-WSe$_2$, whereas other peaks belong to the Gr/SiC substrate. \textbf{b}, Photoluminescence measurement of the ML-WSe$_2$/Gr heterostructure. The peak originating from the A-exciton of ML-WSe$_2$ is marked with the dashed red line. \textbf{c}, LEED pattern of the ML-WSe$_2$/Gr heterostructure at 95 eV after annealing.}
    \label{figs1}
\end{figure}

\noindent
\underline{\textbf{Delayed population rises at $\mathrm{K_{WSe_2}}$ and $\mathrm{Q_{WSe_2}}$ with the below-bandgap excitation}}

\noindent

Upon 1.2 eV excitation, the time trace at the $\mathrm{K_{Gr}}$ points shown in Fig.2\textbf{f} is by fitting with a single exponential decay function convolved with the instrument response function (IRF), $I(t)=H(t-t_0)\times(A\cdot exp(-(t-t_0)/\tau)+C)\otimes \mathrm{IRF}$. Here, $H(t)$ is the Heaviside step function, $\mathrm{IRF}$ is Gaussian envelope function, $A$ is the amplitude and $C$ is the offset. 
In contrast, as the $\mathrm{K_{WSe_2}}$ and $\mathrm{Q_{WSe_2}}$ valleys are populated by interlayer charge transfer (ICT) processes, the corresponding time traces are fitted with an exponential growth function to describe the ICT process adding with a single exponential decay for relaxation. 
The delayed population rise between graphene and WSe$_2$, $\Delta t=51\pm9~\mathrm{fs}$, is obtained by taking the time difference between the peak at $\mathrm{K_{Gr}}$ and the time delay when the population at the $\mathrm{K_{WSe_2}}$ valley reaches its maximum.

\noindent
\underline{\textbf{Exclusion of two/multiple-photon absorption in WSe$_2$}}

Upon 1.2 eV excitation, we could rule out the two/multiple-photon absorption of WSe$_2$ based on three experimental observations. First, we find no excited state population of WSe$_2$ in the energy range higher than the conduction band minimum from 1.0 to 2.5 eV as shown in Supplementary Fig.\ref{figS12}\textbf{a}. The valence band maximum of WSe$_2$ is identified at $E-E_F=-1.1$ eV by the EDC analysis in Fig.\textbf{3d} (main text). In the scenario of two-photon absorption, the photo-induced carriers would be directly populated at $E-E_F=1.3$ eV. This direct optical excitation is not observed in our measurement. The three-photon absorption, corresponding to the excitation at $E-E_F=2.5$ eV, is more difficult to happen and is neither observed.
Second, the delayed population rise in $\mathrm{K_{WSe_2}}$ and $\mathrm{Q_{WSe_2}}$ valleys compared to the rise of hot carriers in graphene, as shown in Fig.2\textbf{f} (main text), is strong evidence of the absence of the direct optical excitation in WSe$_2$. The hot carrier rise in graphene demonstrates the arrival of the pump pulse. However, the excited-state population of the WSe$_2$ layer appears $\sim$50 fs after the pump excitation. Finally, by integrating the photoemission intensity, we present the population dynamics of electron (ROI$_1$ in Supplementary Fig.\ref{figS12}\textbf{a}) and hole components (ROI$_2$ in Supplementary Fig.\ref{figS12}\textbf{a}) in WSe$_2$ upon 1.2 eV excitation. Different from the hot electron dynamics (blue curve in Supplementary Fig.\ref{figS12}\textbf{b}), the intensity of holes (red curve in Supplementary Fig.\ref{figS12}\textbf{b}) stays constant without any signature of optical excitation. The absence of hole dynamics could also be explained by an ultrafast interlayer hole transfer on a time scale much shorter than our temporal resolution. However, combined with the first two observations, we exclude the occurrence of two/multiple photon absorption.  With near-resonant excitation, the two-photon absorption is also excluded since no excited-state population is found at $E-E_F=2.0$ eV, which corresponds to the energy of excited-state electrons by absorbing two pump photons.

\begin{figure}
\centering
\includegraphics[width=16cm,keepaspectratio=true]{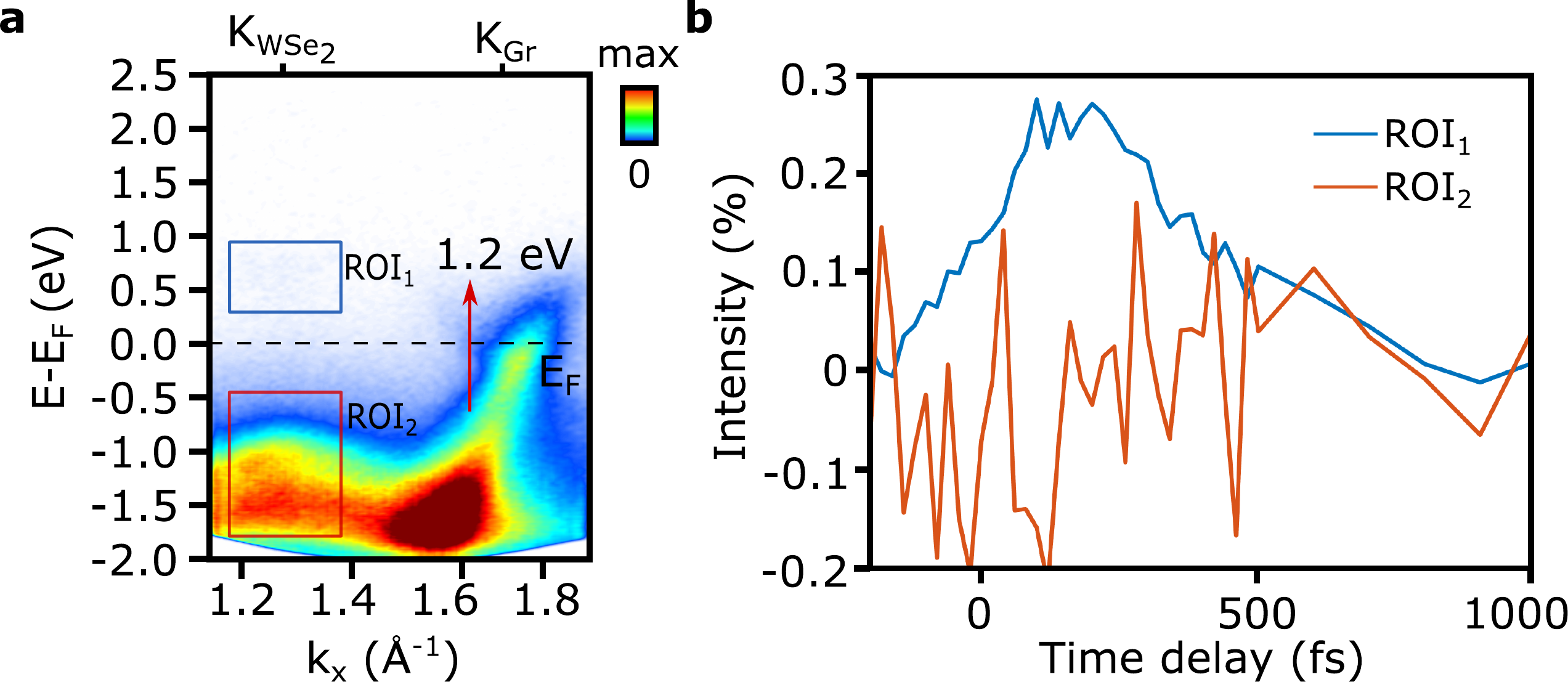}
\caption{\textbf{Carrier dynamics upon the below-bandgap excitation} 
\textbf{a}, Two dimensional energy-momentum cut with 1.2 eV pump. \textbf{b}, By integrating the photoemission intensity in ROI$_1$ and ROI$_2$ in \textbf{a}, the population dynamics of electrons (blue curve) and holes (red curve) are extracted, respectively. The conduction band of WSe$_2$ is filled via interlayer hot electron injection. The population of holes stay constant without the signature of optical excitation. The time traces are normalized to total photoemission intensity at negative time.
}
\label{figS12}
\end{figure}

\noindent
\underline{\textbf{Identification of the Fermi level and distribution of excited-state carriers at $\mathrm{K_{WSe_2}}$}}

\begin{figure}
\centering
\includegraphics[width=16cm,keepaspectratio=true]{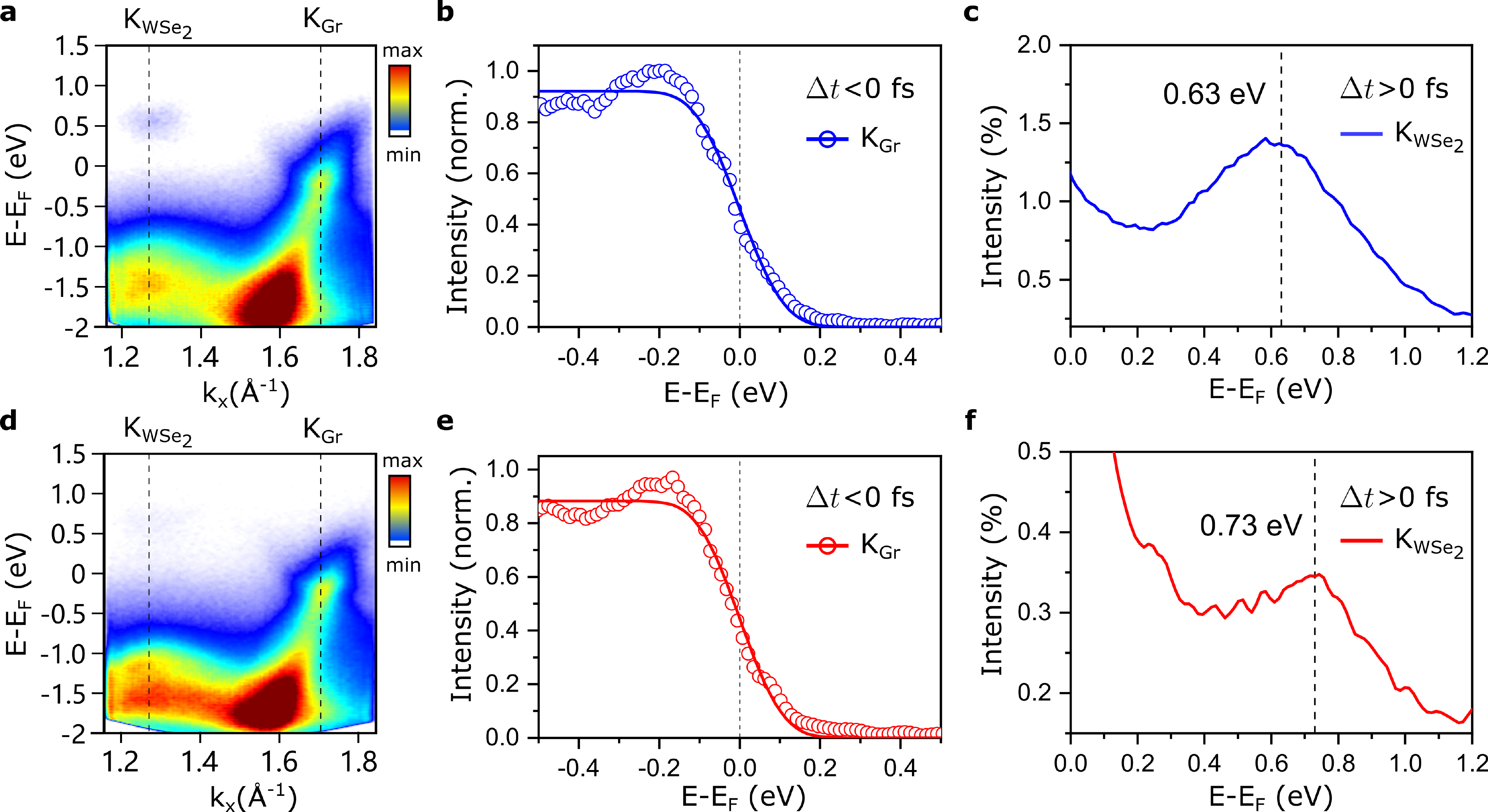}
\caption{\textbf{Fermi level calibration.} \textbf{a,d}, The 2D photoemission intensity spectra as a function of energy and momentum at time zero with 1.55 eV and 1.2 eV pump, respectively. \textbf{b,e}, EDCs of $\mathrm{K_{Gr}}$ at negative time delay fitted with a Fermi-Dirac distribution convolved with the energy IRF for 1.55 eV and 1.2 eV pump, respectively. The chemical potentials are aligned to zero by rigidly shifting the energy axis in both cases. \textbf{c, f} EDCs at $\mathrm{K_{WSe_2}}$ integrated within the first 100 $\mathrm{fs}$ obtained with 1.55 eV and 1.2 eV pump, respectively. The dashed lines represent the center of the excited-state carrier distributions extracted by a fitting procedure (see text).}
\label{figS3}
\end{figure}

\noindent
In both experiments, the energy axis calibration has been performed using the position of the Fermi level of graphene, which is obtained from the energy distribution curve (EDC) at the Dirac point of graphene (Supplementary Fig.~\ref{figS3}\textbf{a} and \textbf{d}). Before optical excitation, the EDCs at the $\mathrm{K_{Gr}}$ point (Supplementary Fig.~\ref{figS3}\textbf{b,e}) are fitted with a Fermi-Dirac distribution function at 300 K convolved with the IRF ($\sim$150 meV FWHM) determined by the energy resolution of the spectrometer and the bandwidth of the probe pulses\cite{maklar2020quantitative}. The chemical potentials are set to be zero for both experimental conditions to remove the XUV-probe-induced space charging effect in each measurement. The EDCs in Supplementary Fig.~\ref{figS3}\textbf{b,e} are integrated over a momentum window, $\Delta k=0.1 $ \si{\angstrom^{-1}} and selected at negative time delay. 
Based on the energy reference obtained from the Fermi level fits, the energy positions of the conduction band minima at the $\mathrm{K_{WSe_2}}$ point can be obtained from EDCs showing the excited state carrier distributions of WSe$_2$ upon resonant and off-resonant excitation as displayed in Supplementary Fig.~\ref{figS3}\textbf{c} and \textbf{f}, respectively. The photoemission intensity has been normalized by the total electron count of the spectrum. The energy difference of the carrier distribution, more specifically, the smaller kinetic energy of excited carriers with 1.55 eV pump, arises from the exciton formation upon the resonant excitation. By fitting with a single Gaussian lineshape on top of a empirical second-order polynomial background, the energies of excited-state carriers are extracted as $0.63~\textrm{eV}$ upon the resonant A-exciton excitation (Supplementary Fig.~\ref{figS3}\textbf{c}) and $0.73~\textrm{eV}$ upon the below-bandgap excitation (Supplementary Fig.~\ref{figS3}\textbf{f}). The background has been removed in the main text (side figures in Fig.3\textbf{a-b}).         

\noindent
\underline{\textbf{Identification of the Dirac point energy}}

\noindent

To identify the energy position of the Dirac point, we selected an energy-$k_y$ cut ($\Delta t<0 ~ \mathrm{fs}$) at Dirac point and along the green dashed line in Supplementary Fig.~\ref{figS4}\textbf{a}. The small titling angle allows us to see both valence bands clearly in Supplementary Fig.~\ref{figS4}\textbf{b}. We track the graphene valence band dispersion by fitting the momentum distribution curves of occupied bands with two Voigt lineshape functions. Then, each graphene valence band is fitted to a linear dispersion and the Dirac point is estimated at the intersection of two lines (red and black), $E-E_F=-0.10 \pm 0.05~\mathrm{eV}$, in a reasonable agreement with the previous characterization of similar heterostructures\cite{nakamura2020spin}. 
This energy/momentum cut is different than the one presented in the main text (Fig.3) along the $\mathrm{\Gamma}$-K direction, which is featured by the suppression of one side of the cone due to photoemission matrix element effects (sublattice interference)\cite{gierz2011illuminating}. In the main text, we choose this cut direction because it allows us to clearly resolve the excited state dynamics from both layers. 

\begin{figure}
\centering
\includegraphics[width=16.3cm,keepaspectratio=true]{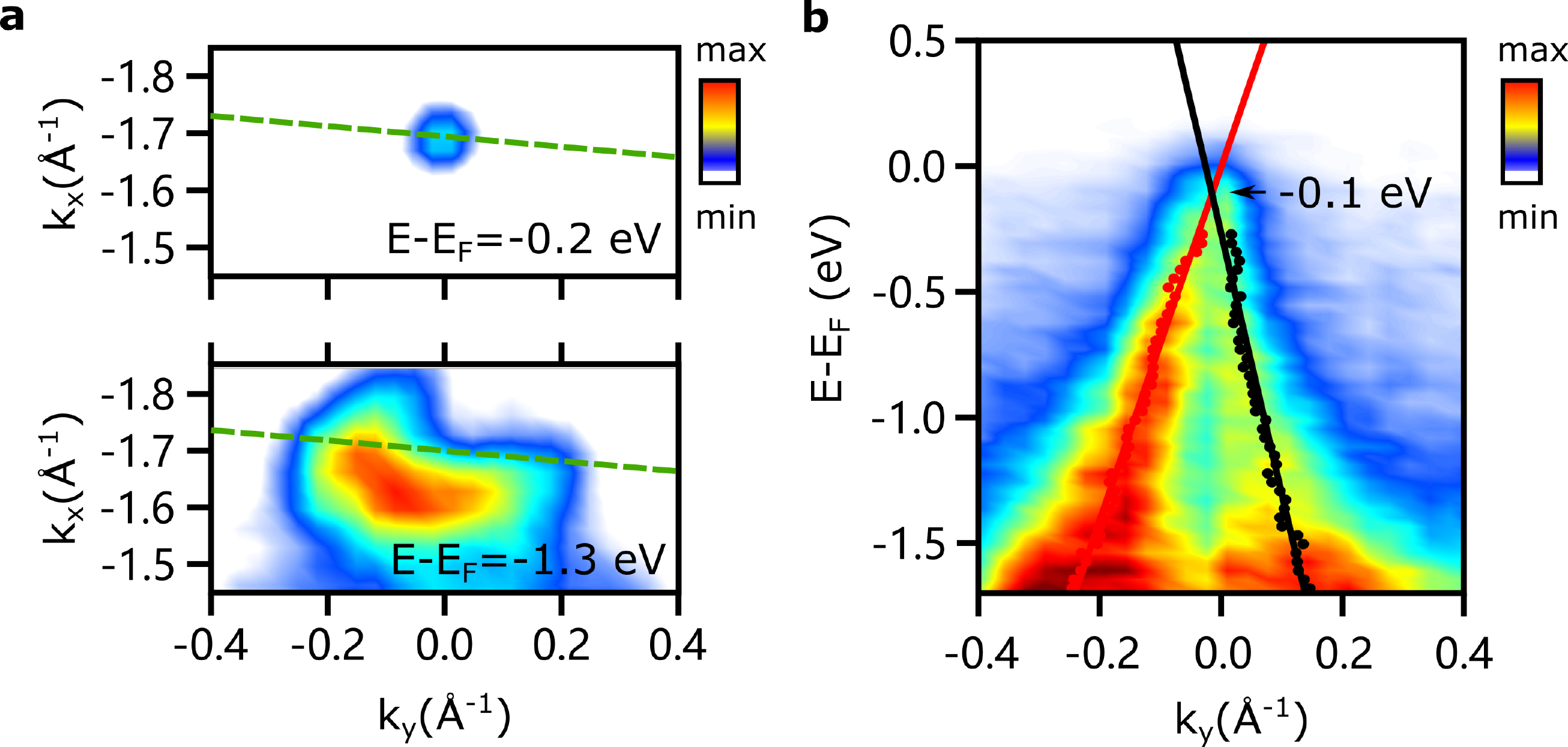}
\caption{\textbf{Experimental determination of the Dirac energy.} \textbf{a}, 2D momentum distribution map $I(k_x,k_y)$ at $E-E_F=-0.2$ eV and $E-E_F=-1.3$ eV. At the boundary of the Brillouin zone, it shows the quasi-triangular-shaped $\pi$ band of graphene. \textbf{b}, Energy/momentum cut $I(E,k)$ along the green dash line in \textbf{a} showing the conical band dispersion of graphene. Red and black markers indicate the band positions extracted from momentum distribution curve fits. Lines are linear fits of the band positions, yielding the energy position of the Dirac point of $E_D=-0.1\pm 0.05$~eV.}
\label{figS4}
\end{figure}

\noindent
\underline{\textbf{Identification of the Fermi velocity}}

\noindent

The Fermi velocity along $\Gamma$-K direction is extracted as $v_F=(1.8\pm0.1)\cdot 10^6~m/s$  as shown in Supplementary Fig.~\ref{figS11} using the same band dispersion tracking method in the above paragraph. The energy/momentum spectrum is selected at negative time delay ($\Delta t< 0~\mathrm{fs}$). The Fermi velocities in the direction perpendicular to $\Gamma$-K with a slightly tilting angle (Supplementary Fig.~\ref{figS4}\textbf{b}) are $v_F=(1.1\pm0.1)\cdot 10^6~m/s$ (red) and $v_F=(1.5\pm0.1)\cdot 10^6~m/s$ (black), respectively.
The Fermi velocity of graphene has been found to to be in the range of $1\cdot 10^6$ to $3\cdot 10^6$ $\mathrm{m/s}$, depending on the dielectric constant of the environment\cite{hwang2012fermi}. For epitaxially grown heterostructures, the dielectric constant of the embedding graphene layer between the bottom substrate and top TMDC layer could be modified by the coverage sizes of TMDC layer and substrate material, as the dielectric constant is determined by $\epsilon=(\epsilon_{top}$+$\epsilon_{substrate})/2$. At the same time, the Fermi velocity is also sensitive to the graphene's doping level\cite{siegel2011many}. We would like to note that because of the steep band dispersion of the graphene band, the momentum of transiently excited intraband electron-hole pairs is small. Therefore, it requires a relative small excitonic COM momentum $\mathbf{Q}$ to fulfill energy and momentum conservation, which favors the IET process. 

\begin{figure}
\centering
\includegraphics[width=8cm,keepaspectratio=true]{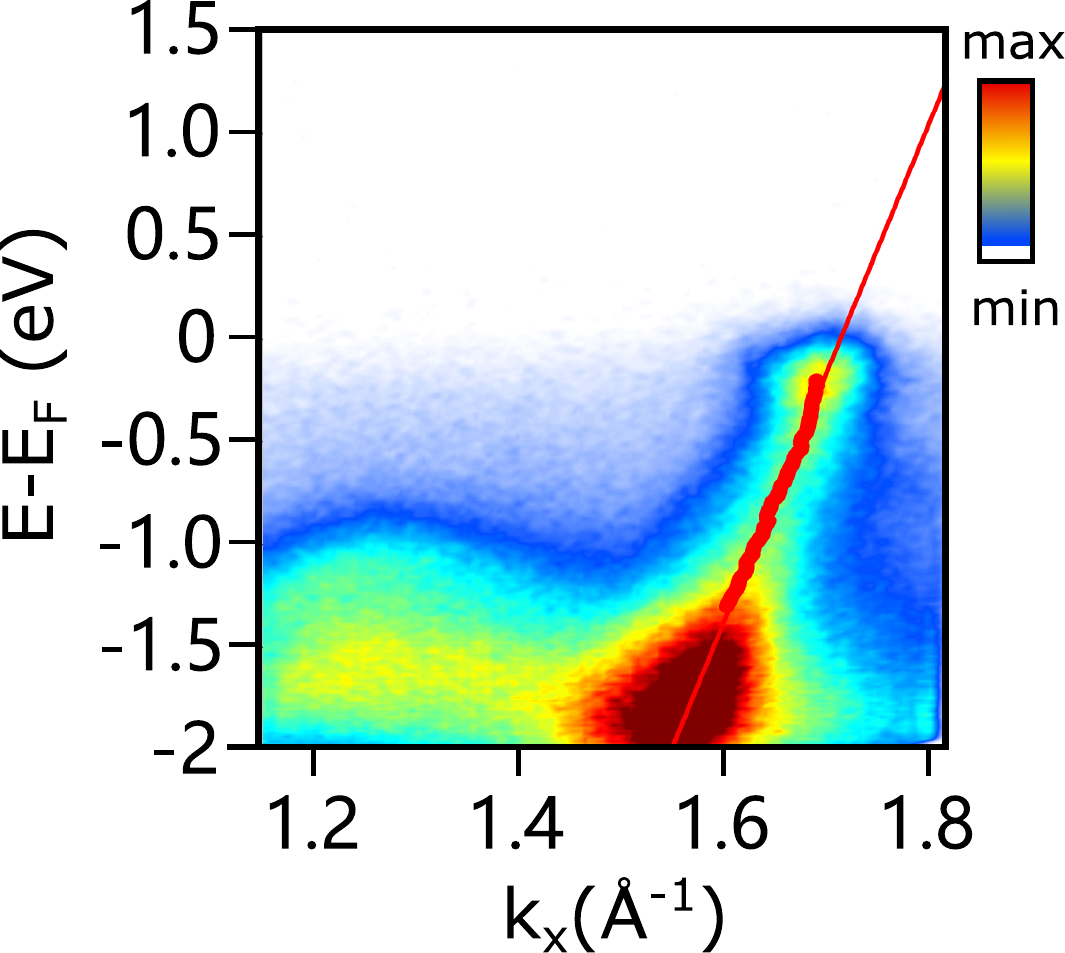}
\caption{\textbf{Experimental determination of the Fermi velocity.} The energy/momentum cut $I(E,k)$ along $\Gamma$-K direction with band positions (red markers) and the linear fit of the band dispersion (red line).}
\label{figS11}
\end{figure}

\noindent
\underline{\textbf{The near-unity efficiency of IET}}

\begin{figure}
\centering
\includegraphics[width=16cm,keepaspectratio=true]{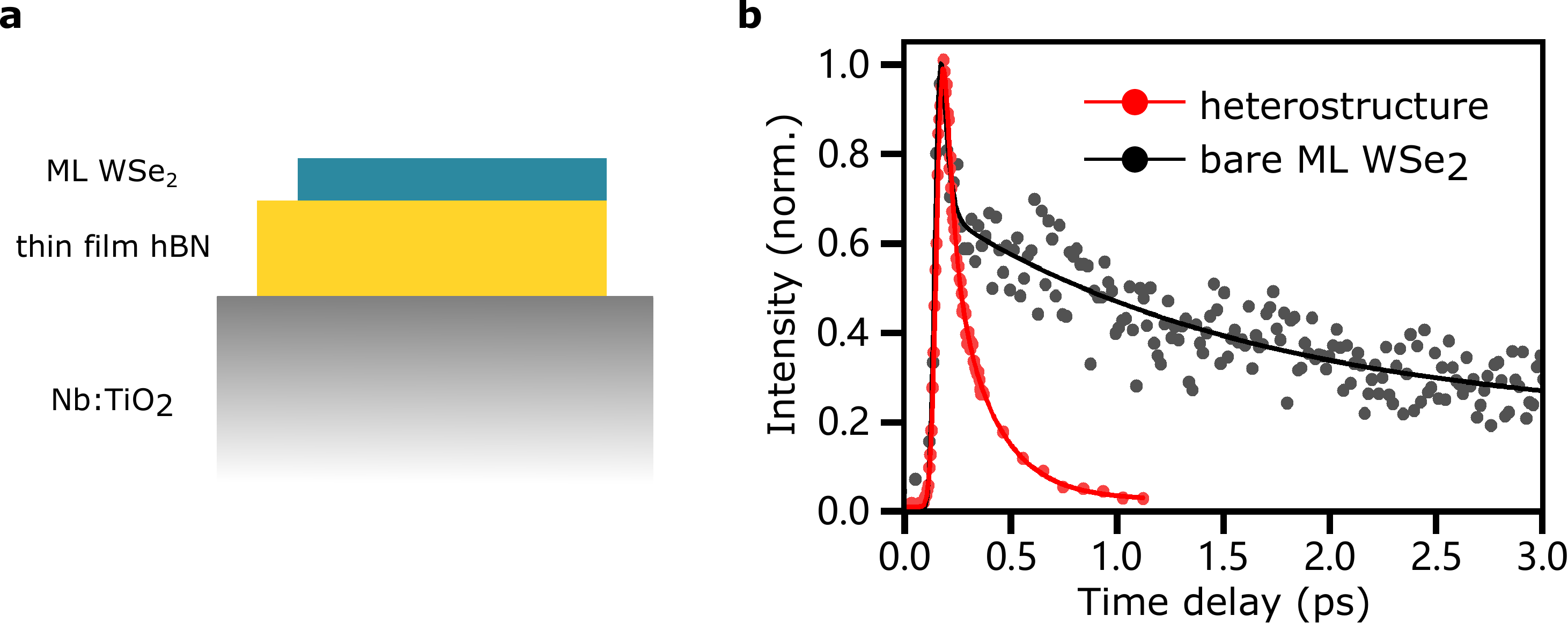}
\caption{\textbf{Estimation of the IET efficiency.} \textbf{a}, Schematic of the bare ML-WSe$_2$ sample (blue slab) with the bottom hBN layer (yellow) mounted on a Nb:TiO$_2$ substrate (grey). \textbf{b}, Time traces of the excited-state carriers at the $\mathrm{K_{WSe_2}}$ valleys of the bare ML sample (black) and heterostructure (red) sample, respectively.}
\label{figS5}
\end{figure}

\noindent
To estimate the efficiency of energy transfer, we performed the same measurement (1.55 eV excitation) on bare ML-WSe$_2$, which is prepared by scotch-tape exfoliation and transferred on top of thin hexagonal boron nitrid (hBN) with conductive TiO$_2$ substrate (Supplementary Fig.~\ref{figS5}\textbf{a}). 
The efficiency of the energy transfer process is commonly defined by the lifetime of the 'donor' material (here WSe$_2$) with and without the 'acceptor' material (here graphene) as: $\eta_{ET}=(\tau_{ML}-\tau_{hetero})/\tau_{ML}$, where $\tau_{ML}$ represents the exciton lifetime of the bare ML-WSe$_2$ and $\tau_{hetero}$ is the exciton lifetime in the WSe$_2$/graphene heterostructure. 
The excited-state population dynamics at the $\mathrm{K_{WSe_2}}$ valley within each system are presented in Supplementary Fig.~\ref{figS5}\textbf{b}. The lifetimes of the ML sample $\tau_{ML}=1616\pm345$ fs is extracted by fitting with an exponential decay function convolved with the IRF. The exciton lifetime of the heterostructure $\tau_{hetero}=67\pm7$ fs is obtained by solving the system of rate equations as described in the main text. Thus, we obtain for the interlayer energy transfer efficiency, $\eta_{ET}= 96\pm 1 \%$. This near-unity transfer efficiency is supported by the underlying conservation of energy and momentum. We note that the different sample fabrication methods of the bare ML and heterostructure may have influence on the exciton lifetime. However, picosecond to sub-nanosecond exciton lifetimes in ML samples, consistent with our observations, have been reported for samples fabricated with various methods\cite{yuan2017exciton,wang2015ultrafast}. Therefore, we believe that the comparison with the exciton lifetimes in the bare ML WSe$_2$ sample provides a reasonable estimate of the transfer efficiency.

\noindent
\underline{\textbf{Valence bands shifting and broadening effects}}

\begin{figure}
\centering
\includegraphics[width=16cm,keepaspectratio=true]{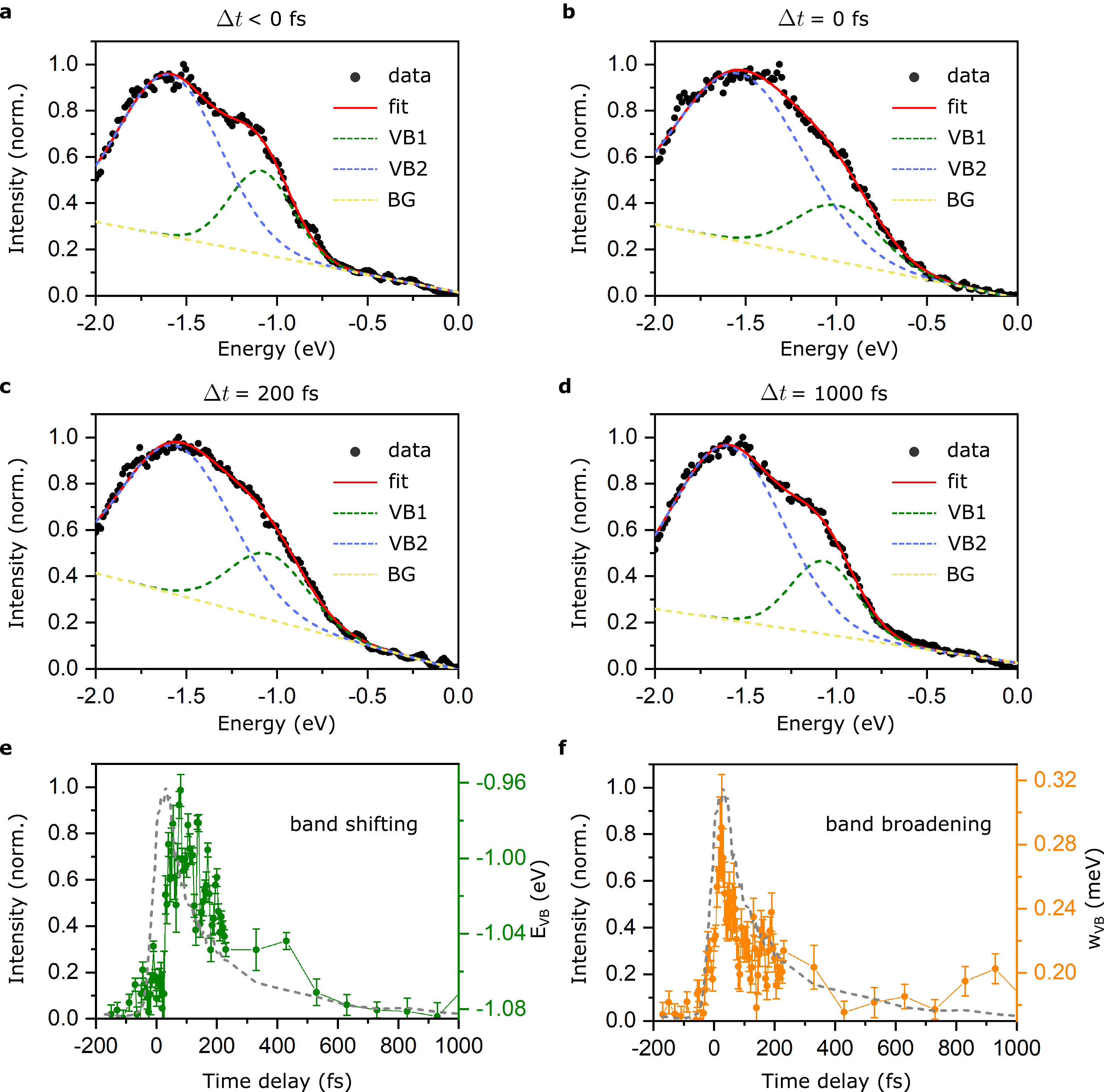}
\caption{ \textbf{Transient lineshape of the valence bands of WSe$_2$.} \textbf{a-d}, The EDCs at the $\mathrm{K_{WSe_2}}$ valley present the spectral features of the first two valence bands at selected time delays, $\Delta t<0$ fs, $\Delta t=0$ fs, $\Delta t=200$ fs and $\Delta t=1000$ fs. The EDCs are fitted with two Gaussian functions describing VB1 (green dashed curve) VB2 (blue dashed curve). The second-order polynomial background is shown as yellow dashed lines. \textbf{e}, Transient peak position of VB1 (green). \textbf{f} Peak linewidth both of VBs as function of time (yellow). The population dynamic of excited-states in the WSe$_2$ conduction band (grey) is shown in \textbf{e-f} as a reference. }
\label{figS6}
\end{figure}

\noindent
After photoexcitation, we observe shifting and broadening effects of the WSe$_2$ valence band as shown in the 2D difference spectrum (Fig.~3\textbf{c}) and EDCs at $\mathrm{K_{WSe_2}}$ (Fig.~3\textbf{d}). To extract the transient lineshape, we fit the EDC of the top two VBs (VB1 and VB2) with two Gaussian functions on top of an empirical second-order polynomial background (BG), $I(E)=A_1\cdot exp(-\frac{(E-E_1)^2}{2\omega_1^2})+A_2\cdot exp(-\frac{(E-E_2)^2}{2\omega_2^2})+\mathrm{BG}$, where $E_1$, $E_2$ are peak positions and $\omega_1$, $\omega_2$ are the peak width. Supplementary Fig.~\ref{figS6}\textbf{a-d} present representative fitting results at four time delays, $\Delta t<0$ fs, $\Delta t=0$ fs, $\Delta t=200$ fs and $\Delta t=1000$ fs. Because of the large spectral overlap between VB1 and VB2, the fitting is performed with the same shifting, $\Delta E_1(t)=\Delta E_2(t)$, and broadening parameter , $\Delta \omega_1(t)=\Delta \omega_2(t)$, for the two peak functions, assuming the VBs respond to the interfacial coupling and the excitation-induced modification in the same way. The extracted time dependent peak shift and linewidth parameter are shown in Supplementary Fig.~\ref{figS6}\textbf{e} and \textbf{f}, respectively. The band shifting reflects the electronic band gap renormalization due to the ICT-induced hot electrons\cite{dendzik2020observation,liu2019direct}. Thus, a relative time delay can be observed compared with the excited-state population dynamics in the conduction bands. In contrast, the linewidth is a measure of the photohole self-energy, which depends on the many-body interactions with photoexcited carriers and phonons\cite{ruppert2017role}. It follows more closely the transient of overall excited carriers in the system (grey dashed curve in Supplementary Fig.~\ref{figS6}\textbf{f}).     

\noindent
\underline{\textbf{Meitner-Auger type IET-induced hot electrons near the Fermi level}}

The Meitner-Auger-like interlayer energy transfer involves an intraband excitation of the deep-lying valence electrons to the photo-generated hot holes below the Fermi level. We observe the deep-lying hot holes, as shown in Fig.\textbf{3e}(main text). However, to separate the contribution of Meitner-Auger IET induced electrons from the photo-generated hot holes of graphene near the Fermi level requires an accurate description of the intrinsic energy-momentum dynamics in the graphene layer. The carrier dynamics in graphene after photoexcitation include multiple contributions: intraband carrier-carrier scattering, interband Auger heating, phonon-mediated cooling, and Auger recombination. These components contribute to the energy-dependent dynamics along the graphene bands\cite{tomadin2013nonequilibrium,liang2015carrier}. This is particularly important for states close to the Fermi level, which are subject to hot carrier accumulation from the highly excited states, carrier redistribution due to the thermalization, and carrier annihilation via Auger recombination. Together with the complication of interlayer charge and energy transfer, it is very challenging to quantitatively disentangle these dynamics with our current energy and time resolution.

Although a quantitative separation of the IET-induced electrons from the photo-generated hot holes remains difficult, some information can be gained from the energy distribution of hot carriers below and above the Fermi level in Fig.\textbf{3e}. In particular, comparing the near-resonant and below-bandgap excitation conditions proves insightful. The ratio of the relative intensity increase above the Fermi level ($E-E_F=0-1$ eV, red area) to the signal depletion below EF ($E-E_F=-1-0$ eV, blue area), $A_{el}/A_{h}$ can be regarded as a measure of the electron-hole imbalance in the system, and we find a significantly stronger imbalance $A_{el}/A_{h}=1.6$ for near-resonate excitation, compared to $A_{el}/A_{h}=1.1$ for below-bandgap excitation. This supports the occurrence of MA-type IET process. 
Compared with the almost identical spectral weights of electrons and holes using below-bandgap excitation, the unequal spectral areas with the 1.55 eV pump demonstrate the carrier redistribution due to the strong interlayer interactions.
While we would like to note that the spectral intensities are subject to photoemission matrix elements, therefore cannot be directly interpreted as electron and hole occupations. These probe-related effects can be reasonably assumed to be identical for the two excitation conditions, thus not affecting our conclusions.

\noindent
\underline{\textbf{Pump fluence dependent Meitner-Auger interlayer energy transfer}}

The pump fluence plays an important role on the MA-type IET process, because of the deep-lying valence electrons are excited to the photo-generated hot holes. The high pump fluence generates large density of hot holes, \emph{i.e.}, more vacancies for IET-induced intraband transition, which benefits the MA-type interlayer coupling. On the other hand, the weak pump fluence would suppress the MA-type IET process with the reduced photo-generated hole density. To check this effect, we perform the same measurement with the near-resonant excitation (1.55 eV) but weak pump fluence, $F=0.9~\mathrm{mJ/cm^2}$, around half of the pump fluence applied in the manuscript,  $F=1.7~\mathrm{mJ/cm^2}$. We find the decreased intensity of the deep-lying hole population, as shown in the Supplementary Fig.\ref{figs15}. From the energy distribution curve (EDC) of graphene band with the weak pump fluence (blue curve), a depletion signal around $E-E_F=-1.8$ eV could be observed but weaker than that with the high pump fluence.

\begin{figure}
\centering
\includegraphics[width=8cm,keepaspectratio=true]{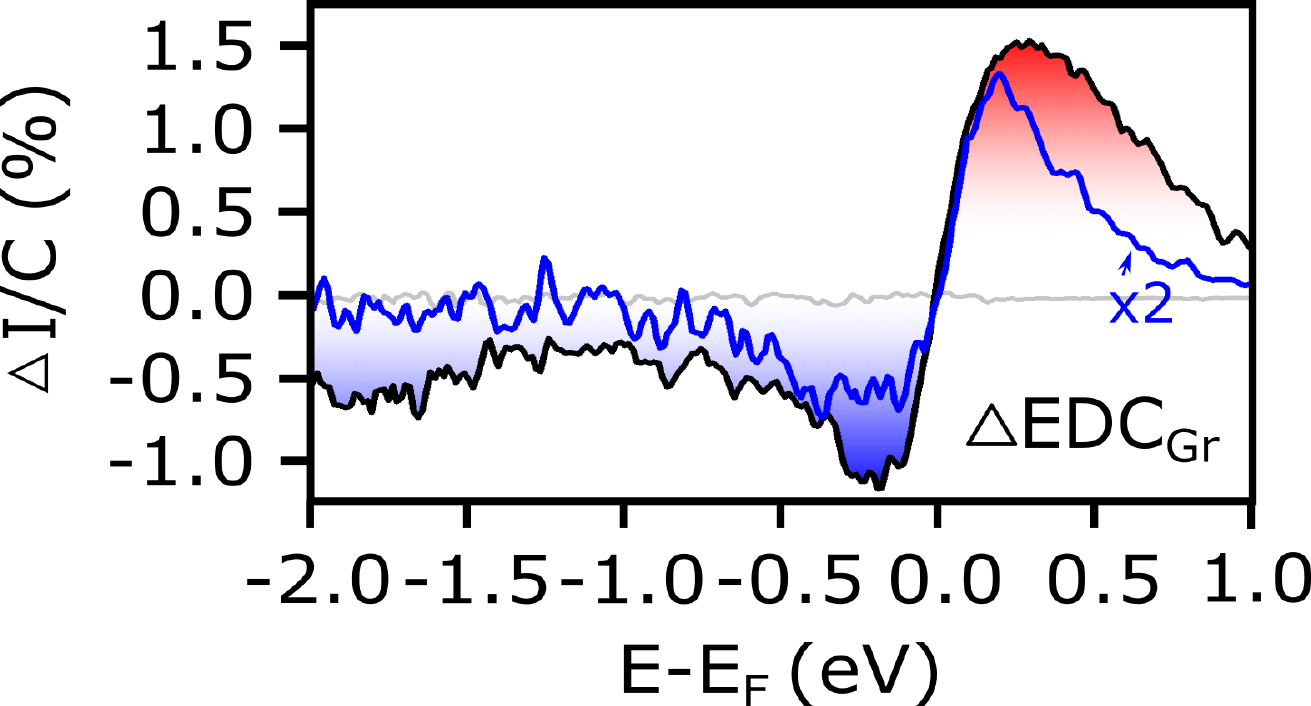}
\caption{\textbf{Pump fluence dependent Meitner-Auger interlayer energy transfer.} 
The momentum-integrated spectrum of the graphene band when pumping with a weak pump fluence (blue). The EDC of graphene with the high pump fluence (black; applied in the main text) is also shown as a comparison.
}
\label{figs15}
\end{figure}

\noindent
\underline{\textbf{Doping level dependence of interlayer energy transfer process}}

As we discussed in the main text, the optical excitation prepares the hot holes near the Fermi level, which enables the intraband transition under the mechanism of Meitner-Auger-type IET. The doping level of graphene plays an essential role in the energy position of the photo-generated hot holes, therefore, influencing the MA-type IET. To clarify the influence of the doping level, we calculate the Meitner-Auger type transfer rate by increasing the Fermi energy to $E_F=\unit[0.2]{eV}$. As shown in Supplementary Fig.\ref{figS14}\textbf{a}, a dramatically suppression of the Meitner-Auger IET is observed with the transfer rate on the order of $\unit[0.06\cdot 10^{-3}]{meV}$, much smaller than that of $\unit[3]{meV}$ based on our real sample system. Compared with the F\"orster-type coupling (Supplementary Fig.\ref{figS14}\textbf{c}), the maximum of the transfer rate is even larger than that of the MA-type coupling.

However, we would like to note that MA-type IET could also happen in the conduction bands with the n-doped graphene composition. The natural doping in the graphene layer provides electrons in the conduction band, below the Fermi level and above the Dirac point. After compensating with the photo-generated hot holes near the Fermi level, the net electron population could be excited to highly excited states via the MA-type IET process. By adding the equation of motion for the conduction band electron occupation, we find a rate of the MA transfer in the conduction band as follows.

\begin{align}
\Gamma_{\mathbf{Q}}=|W_{\mathbf{Q}}|^2\frac{4}{\hbar v_F}\left(Q-\frac{E_{\mathbf{Q}}}{\hbar v_F} \right)\left(f^c_{Q-\frac{E_{\mathbf{Q}}}{\hbar v_F}}-f^c_{\mathbf{Q}}\right)\theta\left(Q-\frac{E_{\mathbf{Q}}}{\hbar v_F} \right)
\end{align}
with $f^c_{\mathbf{k}}=(\exp(\hbar v_F|\mathbf{k}|-\mu)/k_BT)+1)^{-1}$ and chemical potential $\mu$. It is similar to the transfer rate in the valence band (details see section 'Meitner-Auger-like energy transfer'), but with the electron occupation $f^c_{\mathbf{k}}$ in the conduction band.

As shown in Supplementary Fig.\ref{figS14}\textbf{b}. the maximum transfer rate reaches $\unit[1.6]{meV}$. Therefore, although the F\"orster-type IET is faster than the intraband transition in the valence band, the MA-type IET in the conduction band is the dominant process with highly n-doped graphene. All the calculations are performed at room temperature.

\begin{figure}
\centering
\includegraphics[width=16cm,keepaspectratio=true]{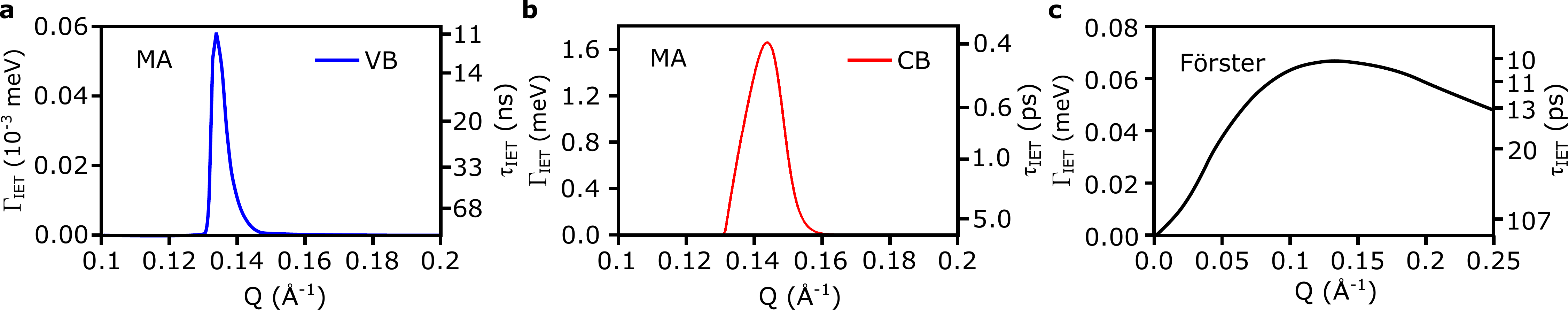}
\caption{\textbf{Energy transfer rate of a heretostructure with the highly n-doped graphene.} 
\textbf{a-b}, The calculated Meitner-Auger transfer rate with $E_F=\unit[0.2]{eV}$ in the valence band and conduction band, respectively. \textbf{c}, The calculated F\"orster-type transfer rate in the same condition. 
}
\label{figS14}
\end{figure}

\newpage
\noindent
\begin{Large}  
Supplementary Methods  
\end{Large} 

\noindent
\underline{\textbf{Characteristics of pump beams}}

\begin{figure}
\centering
\includegraphics[width=8cm,keepaspectratio=true]{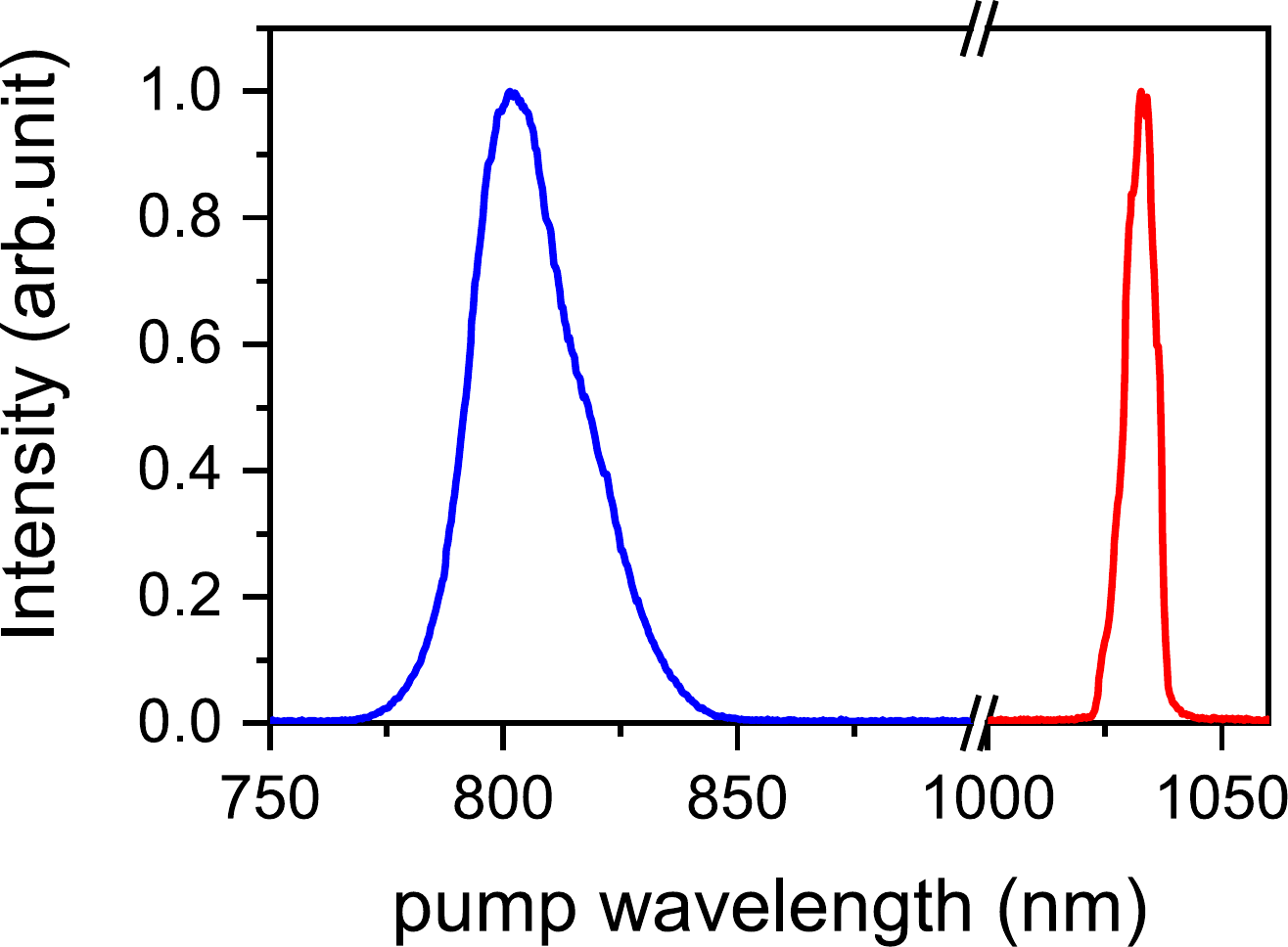}
\caption{\textbf{Pump spectra.} Excitation spectra of the two light sources used for pumping as a function of wavelength.}
\label{figS10}
\end{figure}

In this work, we use two different pump beams which wavelengths are centred at 800 nm and 1030 nm, respectively (Supplementary Fig.~\ref{figS10}).
The pulse duration of 1030nm pump line is $\sim$ 200 fs FWHM, while the transform-limited pulse duration of 800 nm pump is $\sim$ 35 fs FWHM. In the measurement, the pump fluence of 800 nm is $F_{800}=1.7 ~\mathrm{mJ/cm^2}$ and that of 1030 nm is $F_{1030}=5.3 ~\mathrm{mJ/cm^2}$, with the consideration of effective pump-probe overlap profile based on the formula, $a=\frac{1}{\pi(\omega_{pump}^2+\omega_{probe}^2)}$, in the work of Harb \emph{et~al}\cite{harb2006carrier}. The beam size of the pump is $\omega_{pump}=248\pm20$ $\mathrm{\mu m}$ and that of probe pulse is $\omega_{probe}=80\pm5$ $\mathrm{\mu m}$. Here, $\omega_{pump}$ and $\omega_{probe}$ are the respective beam widths of the pump and probe beams.  

\noindent
\underline{\textbf{Separating the interlayer charge and energy transfer by a rate equation model}}

\noindent
As discussed in the main text, we observed the photoemission signatures of both ICT and IET upon resonant A-exciton excitation. To extract the corresponding transfer rates, $\Gamma_{ICT}$ and $\Gamma_{IET}$, we develop a multi-state coupled rate equation model describing the interlayer charge and energy flow, as well as the hot carrier relaxations. 
In Fig.4\textbf{a}, the time trace of hot carriers in the CBM of WSe$_2$ (black curve) includes the dynamics of photo-generated excitons $N_T^{ex}$ and ICT-induced quasi-free electrons $N_T^{el}$. The VB1 shifting (green curve in Fig.4\textbf{a}) mainly reflects the dynamic of $N_T^{el}$. Therefore, it provides the possibility to disentangle the dynamics of these two kinds of quasiparticles. Here, subscript \emph{T} represents TMDC. Simultaneously, the deep valence band holes in graphene $N_{Gr}^h$ are populated by the IET process and recombine with the rate $\Gamma_h$, as shown in Fig.4\textbf{b}. Finally, the dynamics of hot electrons in graphene $N_{Gr}^{el}$ contains the ICT-induced charge flow (input and output towards WSe$_2$) and a decay process with the rate of $\Gamma_{el}$. 
With these considerations, the complete dynamics across the interface can be described with the following set of coupled rate equations:

\begin{gather}
        \dot N_T^{ex}=-\Gamma_{IET} N_T^{ex}+S(t) \\
    \dot N_T^{el}=-\Gamma_{ICT} N_T^{el}+\Gamma_{ICT}N_{Gr}^{el}\\
    \dot N_{Gr}^{el}=+\Gamma_{ICT}N_T^{el}-\Gamma_{ICT}N_{Gr}^{el}-\Gamma_{el}N_{Gr}^{el}+S(t)\\
    \dot N_{Gr}^h=-\Gamma_h N_{Gr}^h+\Gamma_{IET} N_T^{ex}
\end{gather}

Here, $S(t)$ represents the optical excitation as a Gaussian-shaped pump envelope function. By numerically solving the system of coupled differential equations, and a global fit of the solution to the data, we obtain the IET transfer time $\tau_{IET}=67\pm7$ fs, and the ICT transfer time $\tau_{ICT}=118\pm18$ fs ($\tau=\hbar/\Gamma$). At the same time, we get the relaxation times of electrons in graphene, $\tau_{Gr}^{el}=84\pm7$ fs, and that of the deep valence holes, $\tau_{Gr}^h=7\pm4$ fs. The fitting results are shown in Fig.4\textbf{a} and \textbf{b}.

\noindent
\underline{\textbf{Microscopic calculation of IET mechanisms}}

\noindent
We perform microscopic calculations of the IET process mediated by the Meitner-Auger, F\"orster and Dexter type mechanisms.

\subsection{Meitner-Auger-like energy transfer}

A schematic illustration of the Meitner-Auger type (MA) interlayer transfer is depicted in the main text in Fig.4\textbf{f}. Here, an exciton in the TMDC recombines non-radiatively; its energy excites an electron deep in the valence band of graphene to states close to the Dirac point but in the valence band. 

Starting point for the calculation of the MA-type interlayer coupling is the Hamiltonian
\begin{equation}
H_F = \sum_{\mathbf{k},\mathbf{q},\mathbf{k'},\mathbf{q'},\lambda,\lambda',\nu,\nu'} V^{\lambda \nu  \nu' \lambda'}_{\mathbf{k},\mathbf{q},\mathbf{q'},\mathbf{k'}} \lambda^{\dagger}_\mathbf{k} \nu^{\dagger}_\mathbf{q} \nu'_\mathbf{q'} \lambda'_\mathbf{k'}.
\end{equation}

As a convention, we use $\lambda^{(')}$ as band indices and $\mathbf{k}^{(')}$ as momenta in WSe$_2$ layer and $\nu^{(')}$ as band indices and $\mathbf{q}^{(')}$ as momenta in graphene. The appearing matrix element is formally given as
\begin{equation}
V^{\lambda \nu  \nu' \lambda'}_{\mathbf{k},\mathbf{q},\mathbf{q'},\mathbf{k'}} = \int_{\mathbb{R}^3} d^3 r \int_{\mathbb{R}^3} d^3 r' \Psi^{\lambda*}_\mathbf{k} (\mathbf{r}) \Psi^{\nu*}_\mathbf{q} (\mathbf{r'}) V (\mathbf{r} , \mathbf{r'})  \Psi^{\nu'}_\mathbf{q'} (\mathbf{r'}) \Psi^{\lambda'}_\mathbf{k'} (\mathbf{r}).
\end{equation}
The band indices in the TMD are restricted to interband transitions $\lambda\neq \lambda'$ but the band indices in graphene are taken as the valence band $\nu=\nu'=v$. The remaining integrals can be evaluated within a $\mathbf{k}\cdot\mathbf{p}$ expansion. Last we introduce exciton operators in WSe$_2$ $P_\mathbf{Q}^{\mu} = \sum_\mathbf{q} \varphi_\mathbf{q}^{\mu} c^\dagger_{\mathbf{q}+\frac{m_e}{m_h + m_e}\mathbf{Q}}v_{\mathbf{q}-\frac{m_h}{m_h + m_e}\mathbf{Q}}$ with quantum state $\mu$ and COM momentum $\mathbf{Q}$. The final Hamiltonian reads
\begin{equation}
H= \sum_{\mathbf{k},\mathbf{Q},\mu} W^{\mu}_\mathbf{Q} P_\mathbf{Q}^{\dagger \mu} v_\mathbf{k-Q}^\dagger v_\mathbf{k} + h.c.\, ,
\end{equation}
with the coupling element
\begin{equation}
W_\mathbf{Q} = \frac{1}{e} V_\mathbf{Q} \mathbf{d}^{cv}\cdot \mathbf{Q} \varphi^{* \mu} (\mathbf{r}=0).
\end{equation}
In the following we restrict ourselves to the lowest bound excitons $\mu = 1s$.
From this Hamiltonian we calculate the equation of motion for the exciton occupation in the TMD $N_\mathbf{Q} = \langle P^{\dagger}_\mathbf{Q} P_\mathbf{Q} \rangle$ and the electron occupation in the valence band of graphene $f_\mathbf{k} = \langle v^\dagger_\mathbf{k} v_\mathbf{k} \rangle$ by exploiting Heisenberg equation of motion.

The resulting equations of motion read
\begin{align}
\partial_t N_\mathbf{Q} = \frac{2\pi}{\hbar} \sum_\mathbf{k} |W_\mathbf{Q}|^2 &\left( f_\mathbf{k} (1 - f_\mathbf{k-Q}) - N_\mathbf{Q} (f_\mathbf{k-Q}-f_\mathbf{k})\right)\delta (\epsilon_\mathbf{k} - \epsilon_\mathbf{k-Q} - E_\mathbf{Q}) \\
\partial_t f_\mathbf{k} = \frac{2 \pi}{\hbar} \sum_\mathbf{Q} |W_\mathbf{Q}|^2&\left( f_\mathbf{k-Q} (1 - f_\mathbf{k}) - N_\mathbf{Q} (f_\mathbf{k}-f_\mathbf{k-Q})\right)\delta (\epsilon_\mathbf{k-Q} - \epsilon_\mathbf{k} - E_\mathbf{Q}) \\
+ \frac{2 \pi}{\hbar} \sum_\mathbf{Q}|W_\mathbf{Q}|^2 &\left( N_\mathbf{Q} (f_\mathbf{k-Q}-f_\mathbf{k}) - f_\mathbf{k}  (1 - f_\mathbf{k-Q}) \right)\delta (\epsilon_\mathbf{k} - \epsilon_\mathbf{k-Q} - E_\mathbf{Q}) 
\end{align}

\subsection{Estimation of the decay rate of WSe$_2$ excitons}

From the Boltzmann equation we can identify the decay rate of WSe$_2$ excitons as
\begin{equation}
\Gamma_\mathbf{Q} = 8\pi\sum_\mathbf{k} (f_\mathbf{k-Q}-f_\mathbf{k})\delta (\epsilon_\mathbf{k} - \epsilon_\mathbf{k-Q} - E_\mathbf{Q}),
\end{equation}
where we have added a factor of 4 to account for the valley and spin degree of freedom in graphene. Analyzing the Dirac distribution, we find that $\mathbf{k}$ accounts for electrons close the Dirac point, and $\mathbf{k}-\mathbf{Q}$ for electrons deep in the valence band. In order to get a simple expression for the decay rate, we assume that the electrons close to the Dirac point have much smaller momenta than the electrons deep in the valence band, i .e. $\mathbf{k}\ll\mathbf{Q}$ and 
$\mathbf{k+Q}\approx\mathbf{Q}$.

This way, the Dirac distribution and the $\mathbf{k}$ can be evaluated analytically yielding
\begin{equation}
\Gamma_\mathbf{Q} =  |W_\mathbf{Q}|^2 \frac{4}{\hbar v_F} \left( Q - \frac{E_\mathbf{Q}}{\hbar v_F} \right) \left( f_\mathbf{Q} - f_{Q-\frac{E_\mathbf{Q}}{\hbar v_F}} \right) \theta \left( Q - \frac{E_\mathbf{Q}}{\hbar v_F} \right)\label{meitner_auger}
\end{equation}
The rate depends on the matrix element of the MA transfer, the density of states in graphene and on the occupation difference of the involved states in graphene which accounts for the Pauli blocking. The heavyside function $\theta \left( Q - \frac{E_\mathbf{Q}}{\hbar v_F} \right)$ accounts for the fact, that a minimal momentum is required to fulfil the energy and momentum conservation during the intervalley transfer.

Fig.~4\textbf{c} in the main text illustrates the MA rate of WSe$_2$ excitons to graphene for the photo-induced hole vacancies at different energy of $\mu_{Gr}^{h*}$. We adjusted the graphene dispersion to the results from the ARPES measurement. From $\tau=\hbar/\Gamma$ we find scattering times of \unit[270]{fs} ($\mu_{Gr}^{h*}=\unit[-0.3]{eV}$), \unit[210]{fs} ($\mu_{Gr}^{h*}=\unit[-0.4]{eV}$), and \unit[175]{fs} ($\mu_{Gr}^{h*}=\unit[-0.5]{eV}$).

\subsection{Origin of the finite center-of-mass (COM) momentum}
The MA mediated IET requires nonzero COM momentum of the exciton. For example, the required COM momentum is $\sim$ \unit[1.3]{nm$^{-1}$} as shown in Fig.4\textbf{c} (main text) with $\mu_{Gr}^*=-0.3~\mathrm{eV}$, which corresponds to a kinetic energy of approximately \unit[100]{meV} based on the effective mass of the exciton $m_{ex}=0.65 m_e$.\cite{hong2020probing}
Then, where does the energy (momentum) come from? In the following, we discuss the possible origins of COM momentum of the exciton which is quantified by the kinetic energy assuming the parabolic excitonic band dispersion.  

At room temperature, the mean kinetic energy of the excitons is \unit[25.6]{meV} which is not enough to explain the required energy. Therefore, we calculate the temporal evolution of the exciton energy-momentum occupation during the optical pump for detuned excitation\cite{selig2019ultrafast}. The equation of motion for the excitonic coherence in the rotating frame reads
\begin{equation}
\dot{P}_\mathbf{0} (t) = \frac{1}{i\hbar} \left( E_\mathbf{0} - \hbar \omega_L - i \gamma \right) P_\mathbf{0} (t) + \mathbf{d}\cdot \mathbf{E} (t),\label{cohere}
\end{equation}
where the first term accounts for the detuning of the excitonic transition energy $E_\mathbf{0}$ from the light pulse energy $\hbar \omega_L$. $\gamma$ accounts for the dephasing of the excitonic coherence with contributions from radiative and exciton phonon coupling\cite{selig2016excitonic}. The last term accounts for the optical excitation with the dipole element $\mathbf{d}$ and the exciting electric field $\mathbf{E} (t)$.
The equation of motion for the incoherent exciton occupation reads
\begin{align}
\dot{N}_\mathbf{Q} = \Gamma^{Form}_\mathbf{Q} |P_\mathbf{0}|^2 + \sum_\mathbf{K} \Gamma^{in}_\mathbf{Q,K} N_\mathbf{K} - \sum_\mathbf{K} \Gamma_\mathbf{Q,K}^{out} N_\mathbf{Q}.\label{occu}
\end{align}
The first term accounts for the formation of incoherent exciton occupation from phonon induced dephasing from the excitonic coherence. The last two terms account for the thermalization of incoherent excitons\cite{selig2018dark}. The coupling element of the exciton formation reads
\begin{equation}
\Gamma_\mathbf{Q}^{Form} = \frac{2}{\hbar} \sum_{\pm, \alpha} |g_\mathbf{Q}|^2 \left(\frac{1}{2}\pm\frac{1}{2} + n_\mathbf{Q}^\alpha \right) \frac{\gamma}{(E_\mathbf{Q} - \hbar \omega_L \mp \hbar \Omega^\alpha)^2 + \gamma^2}.\label{gammaform}
\end{equation}
with the energy $\hbar \Omega_\mathbf{Q}^\alpha$ and the occupation $n_\mathbf{Q}^\alpha$ of phonons in the branch $\alpha$ with momentum $\mathbf{Q}$.
The $\pm$ summation accounts for phonon emission/absorption processes.

\begin{figure}
\centering
\includegraphics[width=16cm,keepaspectratio=true]{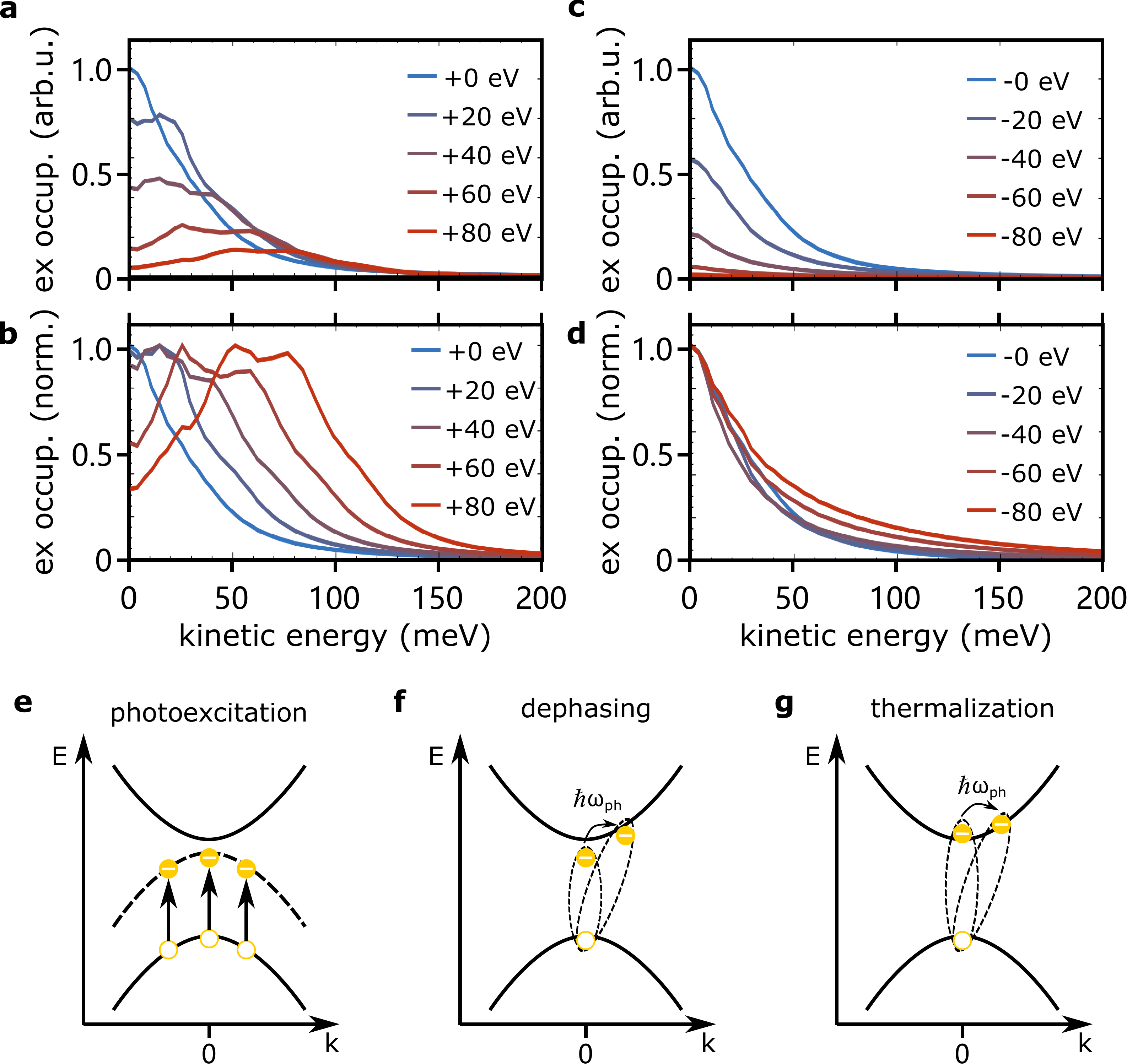}
\caption{ \textbf{Finite COM momentum of excitons at the $\mathrm{K_{WSe_2}}$ valley.} \textbf{a}, The calculated energy-dependent hot exciton occupation with detuned pump photon energy above the resonant excitonic transition energy. \textbf{b}, Normalized hot exciton distributions in \textbf{a}. \textbf{c} The hot exciton occupation with detuned pump photon energy below the resonant excitonic transition energy and \textbf{d} is the corresponding normalized hot exciton distributions. Schematic illustrations of involved ultrafast dynamics: \textbf{e}, photoexcitation creates the coherent exciton. The dash curve represents the coherent excitonic state. \textbf{f}, The phonon-assisted dephasing process transfers coherent excitons to incoherent excitons, at the same time, increases the COM momentum of excitons. \textbf{g}, The following thermalization also contributes to the finite COM momentum.}
\label{figS7}
\end{figure}

Supplementary Fig.~\ref{figS7}\textbf{a} illustrates the snapshots of the exciton occupation directly at the maximum of the pump pulse as a function of kinetic energy with selected detuning pump photon energy \textit{above} the excitonic transition energy. The temperatures are set as room temperature for all the calculation. 
With increasing pump photon energy, the amount of injected excitons decreases due to the non-resonant excitation, eq. \ref{cohere}. However, at larger detunings, excitons occupy larger energy states due to the excess energy of the pump pulse, which is provided by acoustic and optical phonon transitions. The Supplementary Fig.~\ref{figS7}\textbf{b} illustrates the exciton occupations but normalized to the maximum. Here it is even more obvious that the excitons obtain higher energies as the detuning increases. Interesting, for larger detunings two maxima can be observed, where the higher one originates from the formation of excitons via acoustic phonon scattering. The lower peak originates from the formation of excitons via optical phonon emission but also from relaxation of excitons from the higher peak via optical phonon emission. 

Supplementary Fig.~\ref{figS7}\textbf{c} illustrates snapshots of the exciton occupation directly at the maximum of the pump pulse as a function of kinetic energy for selected detunings \textit{below} the excitonic transition energy.
Similar to the previous scenario, the density of injected excitons decreases with increasing detuning due to the non-resonant excitation, eq. \ref{cohere}. In Supplementary Fig.~\ref{figS7}\textbf{d}, the energy-dependent exciton occupations are normalized to their maximum. Interestingly, for pumping with larger negative detunings, the exciton distribution broadens such that the relative exciton occupation at large kinetic energies increases. The reason is, that for larger detunings the Lorentzian in equation \ref{gammaform} flattens which results in higher occupation of hot exciton at large energy range. As a consequence, for the near-resonant excitation below the excitonic transition, a substantial amount of excitons is formed at energies above \unit[100]{meV} which contribute to the Meitner-Auger scattering. To conclude, non-resonant excitation of the exciton, both above and below the resonance, introduce hot excitons with high kinetic energy, which are subjected to the Meitner-Auger IET.

In our experiment, the pump photon energy is $\hbar \nu_{pump}=1.55~\mathrm{eV}$ and the A-exciton transition energy is $E_{ex}=1.63~\mathrm{eV}$ determined by the energy difference of excited-state particles at CBM and VBM. After the photoexcition which prepares the \textit{coherent} excitons with zero COM momentum (Supplementary Fig.~\ref{figS7}\textbf{e}), the phonon-assisted dephasing process transfers the coherent excitons to \textit{incoherent} exciton population which gain the finite COM momenta (Supplementary Fig.~\ref{figS7}\textbf{f}). This dephasing process has been observed by our previous study\cite{dong2020measurement}. The subsequent thermalization of excitons at the excitonic states (Supplementary Fig.~\ref{figS7}\textbf{g}) also contribute to the nonzero COM momenta which is already included in our calculation. 

\subsection{F\"orster coupling}

To calculate the F\"orster rate from WSe$_2$ to graphene, we start with the Hamiltonian
\begin{equation}
H_F = \sum_{\mathbf{k},\mathbf{q},\mathbf{k'},\mathbf{q'},\lambda,\lambda',\nu,\nu'} V^{\lambda \nu  \nu' \lambda'}_{\mathbf{k},\mathbf{q},\mathbf{q'},\mathbf{k'}} \lambda^{\dagger}_\mathbf{k} \nu^{\dagger}_\mathbf{q} \nu'_\mathbf{q'} \lambda'_\mathbf{k'}.
\end{equation}

As a convention, we use $\lambda^{(')}$ as band indices and $\mathbf{k}^{(')}$ as momenta in WSe$_2$ layer and $\nu^{(')}$ as band indices and $\mathbf{q}^{(')}$ as momenta in graphene. The appearing matrix element reads
\begin{equation}
V^{\lambda \nu  \nu' \lambda'}_{\mathbf{k},\mathbf{q},\mathbf{q'},\mathbf{k'}} = \int_{\mathbb{R}^3} d^3 r \int_{\mathbb{R}^3} d^3 r' \Psi^{\lambda*}_\mathbf{k} (\mathbf{r}) \Psi^{\nu*}_\mathbf{q} (\mathbf{r'}) V (\mathbf{r} , \mathbf{r'})  \Psi^{\nu'}_\mathbf{q'} (\mathbf{r'}) \Psi^{\lambda'}_\mathbf{k'} (\mathbf{r}).
\end{equation}
Here, $\Psi^{\lambda/\nu}_{\mathbf{k}/\mathbf{q}}$ account for the electronic Bloch waves in WSe$_2$ and graphene.
The appearing Coulomb potential shall take into account the dielectric environment of the heterostructure, including the WSe$_2$ and graphene layer which are distanced by a gap with dielectric constant $\epsilon_R$ and width $z$ (closely stacked structures have $z=0$)\cite{ovesen2019interlayer}. Additionally we take substrates below and above the structure into account.

We can evaluate the matrix element by Fourier transforming the Coulomb potential and calculating the real space integrals within a $\mathbf{k\cdot p}$ expansion\cite{katsch2019theory}. We introduce exciton operators in WSe$_2$ $P_\mathbf{Q}^{\mu} = \sum_\mathbf{q} \varphi_\mathbf{q}^{\mu} c^\dagger_{\mathbf{q}+\frac{m_e}{m_h + m_e}\mathbf{Q}}v_{\mathbf{q}-\frac{m_h}{m_h + m_e}\mathbf{Q}}$ with quantum state $\mu$ and COM momentum $\mathbf{Q}$ as well as pair operators in graphene $R_\mathbf{Q}^{\mathbf{q}} = c^\dagger_{\mathbf{q} + \frac{1}{2}\mathbf{Q}}v_{\mathbf{q}-\frac{1}{2}\mathbf{Q}}$. 
The Hamiltonian then reads
\begin{equation}
H_F = \sum_{\mathbf{Q},\mathbf{q},\mu} F_\mathbf{Q}^{\mu} (z) P^{\dagger \mu}_\mathbf{Q} R_\mathbf{Q}^{\mathbf{q}} + h.c.
\end{equation}
The appearing coupling element reads
\begin{equation}
 F_\mathbf{Q}^{\mu} (z) = \frac{1}{e^2 \sqrt{A}} V_\mathbf{Q} (z) \varphi^\mu (\mathbf{r} = 0) \mathbf{Q}\cdot \mathbf{d_T}^{cv}  \mathbf{Q}\cdot \mathbf{d}_G^{vc}\label{exc_coupling_element}
\end{equation}

with $\mathbf{d}_T^{vc}$ the dipole element in WSe$_2$, $\mathbf{d}_G^{vc}$ the dipole element in graphene, $\varphi^\mu (\mathbf{r})$ the excitonic wave function in real space with quantum number $\mu$ in WSe$_2$. We restrict our analysis to the lowest lying excitons $\mu = 1s$.

The F\"orster induced transition rate is given as \cite{selig2019theory}
\begin{equation}
\Gamma_\mathbf{Q} (z) = 4 \pi \sum_\mathbf{q} |F_\mathbf{Q} (z)|^2 \delta \left( E^\mathbf{q}_\mathbf{Q} - E_\mathbf{Q}^{1s}  \right),\label{def_rate}
\end{equation}
where we have already added a factor of 2 to account for the electron spin in graphene. We analytically treat the summation over the delta function, where the area which appears in equation (\ref{exc_coupling_element}) cancels. We arrive at
\begin{equation}
\Gamma_\mathbf{Q} (z) = \frac{|F_\mathbf{Q} (z)|^2 A E_\mathbf{Q}^{1s}}{2 \hbar^2 v_F^2}.
\end{equation} 
$v_F$ is the Fermi velocity in graphene. The area $A$ cancels with the area in $|F_\mathbf{Q} (z)|$. In a last step we average over the angle dependence of $|F_\mathbf{Q} (z)|^2$, und sum the result over the $K$ and $K'$ point in graphene, which is already included in the $\mathbf{q}$ summation in eq. \ref{def_rate}. This way we arive at the final expression
\begin{equation}
\Gamma_\mathbf{Q} (z) = \frac{|V_\mathbf{\mathbf{Q}}(z)|^2 |\varphi^{1s}(\mathbf{r}=0)|d_T^2 d_G^2 E_\mathbf{Q}^{1s}Q^4}{8 \hbar^2 v_F^2 e^2}.
\end{equation}

The Coulomb potential $V_\mathbf{\mathbf{Q}} (z)$ is given as
\begin{equation}
V_\mathbf{Q} (z) = \frac{e^2}{\epsilon_0 |\mathbf{Q}| \epsilon_\mathbf{Q} (z)},
\end{equation}
where the momentum dependent dielectric function $\epsilon_\mathbf{Q} (z)$ accounts for the dielectric screening from the surrounding\cite{ovesen2019interlayer}. As input parameters, we require the thickness of graphene and WSe$_2$ layers and their respective dielectric constants. Note, that in the limit of infinitely thin films and a uniform background, our results coincides with our previous one\cite{selig2018dark}.
The required parameters are listed in table \ref{parameters}.

Fig.~4\textbf{d} in the main text illustrates the F\"orster transfer rate as a function of COM momentum and for different WSe$_2$ - graphene distances. For $\mathbf{Q}=0$ we find a vanishing F\"orster rate followed by a monotonous increase. The large $\mathbf{Q}$ behavior is dictated by the interplay of the momentum dependence of the Coulomb potential and the factor $Q^4$. For the closest stacking, i.e. \unit[0.0]{nm} we find a peak transition rate of about \unit[0.08]{meV}.



Now, we would like to emphasize the difference between MA and F\"orster transfer theoretically. The MA and F\"orster transfer are related and can be derived from the same Hamiltonian since the physical origin of both mechanisms is the Coulomb interaction between the involved two-dimensional materials. However, the MA-type IET can be described as monopole-dipole interaction which is often neglected in the Coulomb potential and different from the dipole-dipole coupling term corresponding to the F\"orster-type IET. To illustrate this, we present the detailed derivations of Dexter, F\"orster, and MA transfer using a multipole expansion of the Coulomb interaction. We start from the Coulomb Hamiltonian involving the typical Coulomb matrix element. Here, we use a simple Coulomb potential $V(\mathbf{r-r'})=\frac{e_0^2}{4\pi\epsilon_0\epsilon}\frac{1}{|\mathbf{r-r'}|}$ with a constant uniform background. However, in the manuscript, we use a more complex nonlinear dielectric function valid for the TMDC-graphene interface and treat it in the momentum space. The Hamiltonian involving all transfer processes (F\"orster, Dexter, MA) reads
\begin{align}
H&=\sum_{\substack{\lambda,\lambda',\nu,\nu' \\ \mathbf{k,k',q,q'}}} V^{\lambda\nu\nu'\lambda'}_{\mathbf{k},\mathbf{q},\mathbf{q}',\mathbf{k}'} \lambda_{\mathbf{k}}^{\dagger}\nu^{\dagger}_{\mathbf{q}}\nu'_{\mathbf{q}'}\lambda'_{\mathbf{k}'} \label{eq:CoulombHamilton} \\
V^{\lambda\nu\nu'\lambda'}_{\mathbf{k},\mathbf{q},\mathbf{q}',\mathbf{k}'}&=\int d^3r\int d^3r' ~ \Psi^{*}_{\lambda,\mathbf{k}}(\mathbf{r})\Psi^*_{\nu,\mathbf{q}}(\mathbf{r'}) V(\mathbf{r-r'}) \Psi_{\nu',\mathbf{q}'}(\mathbf{r}')\Psi_{\lambda',\mathbf{k}'}(\mathbf{r}) \;, \label{eq:CoulombMatrix}
\end{align}
where $(\lambda^{(')},\mathbf{k}^{(')})$ stand for graphene quantum numbers and $(\nu^{(')},\mathbf{q}^{(')})$ for WSe$_2$ quantum numbers. We provide now a detailed derivation of all these processes:

The spatial vector $\mathbf{r}$ can be decomposed into a lattice vector $\mathbf{R}_n$ pointing to the $n$th unit cell and a vector $\mathbf{r}_n$ defined locally in the $n$th unit cell: $\mathbf{r}\rightarrow\mathbf{r}_n+\mathbf{R}_n$. Then the spatial integral is changed into a sum over all unit cells and an integration over one unit cell. For electronic Bloch functions of the form $\Psi_{\lambda,\mathbf{k}}=\xi_{\mathbf{k}}(\mathbf{r})u_{\lambda,\mathbf{k}}(\mathbf{r})$, with lattice periodic function $u_{\lambda,\mathbf{k}}(\mathbf{r})$ and envelope $\xi_{\mathbf{k}}(\mathbf{r})$, we obtain
\begin{align}
V^{\lambda\nu\nu'\lambda'}_{\mathbf{k},\mathbf{q},\mathbf{q}',\mathbf{k}'}&=\sum_{\mathbf{R}_n,\mathbf{R}_n'}\xi^*_{\mathbf{k}}(\mathbf{R}_n)\xi^*_{\mathbf{q}}(\mathbf{R}_n')\xi_{\mathbf{q}'}(\mathbf{R}_n')\xi_{\mathbf{k}'}(\mathbf{R}_n) \nonumber \\
&\times\int_{UC} d^3r_n\int_{UC} d^3r_n' ~ u^{*}_{\lambda,\mathbf{k}}(\mathbf{r})u^*_{\nu,\mathbf{q}}(\mathbf{r'}) V(\mathbf{r}_n-\mathbf{r}'_n+\mathbf{R}_n-\mathbf{R}'_n) u_{\nu',\mathbf{q}'}(\mathbf{r}')u_{\lambda',\mathbf{k}'}(\mathbf{r}) \;. \label{eq:CoulombMatrix2}
\end{align}
We assumed that the envelope is spatially constant over one unit cell and exploited the periodicity of the functions $u_{\lambda,\mathbf{k}}(\mathbf{r}).$ Next, we Taylor expand the Coulomb potential at two points $(\mathbf{R}_n-\mathbf{R}_n')$:
\begin{align}
\frac{1}{|\mathbf{r}_n-\mathbf{r}_n'+\mathbf{R}_n-\mathbf{R}_n'|}&=\frac{1}{|\mathbf{R}_n-\mathbf{R}_n'|} \nonumber \\
&+ \frac{(\mathbf{R}_n-\mathbf{R}_n')}{|\mathbf{R}_n-\mathbf{R}_n'|^3}\cdot(\mathbf{r}_n'-\mathbf{r}_n) \nonumber \\
&+\frac{\mathbf{r}_n\cdot\mathbf{r}_n'}{|\mathbf{R}_n-\mathbf{R}_n'|^3} - 3 \frac{(\mathbf{r}_n'\cdot(\mathbf{R}_n-\mathbf{R}_n'))((\mathbf{R}_n-\mathbf{R}_n')\cdot\mathbf{r}_n)}{|\mathbf{R}_n-\mathbf{R}_n'|^5} \;. \label{eq:Taylor}
\end{align}
The first line is the zeroth-order and corresponds to monopole-monopole interaction between the two parts of the heterostructure. The second line is the monopole-dipole interaction which corresponds to an intraband-interband coupling. This term is often neglected in a rotating wave approximation when applied to gaped structures, such as a semiconductor heterostructure. However, in the case of graphene, this term can not be ignored since the vanishing bandgap allows energetically favourite intraband transitions (if not Pauli-blocked). Here, a coupling of intraband excitation in graphene and excitons in TMDCs can occur. A typical effect is carrier multiplication\cite{rana2007electron,wendler2014carrier}. In our study, we include this term which gives rise to the MA coupling, as a monopole (graphene)-dipole (TMDC) interaction. 
The last line, appearing in the first order of both arguments $\mathbf{r}_n$, $\mathbf{r}_n'$, constitutes the typical dipole-dipole interaction coupling (F\"orster-type transfer). 

Coming back to the Hamiltonian and inserting the last line of Taylor expansion (Eq. \eqref{eq:Taylor}) into Eq. \eqref{eq:CoulombMatrix2}, we obtain the F\"orster-type coupling,
\begin{align}
V^{\lambda\nu\bar{\nu}\bar{\lambda}}_{\mathbf{k},\mathbf{q},\mathbf{q}',\mathbf{k}'} &= \frac{1}{4\pi\epsilon_0\epsilon}\sum_{\mathbf{R}_n,\mathbf{R}'_n} \xi^*_{\mathbf{k}}(\mathbf{R}_n)\xi^*_{\mathbf{q}}(\mathbf{R}_n')\xi_{\mathbf{q}'}(\mathbf{R}_n')\xi_{\mathbf{k}'}(\mathbf{R}_n) \nonumber \\
&\times\left( \frac{\mathbf{d}^{\lambda\bar{\lambda}}_{\mathbf{k,k'}}\cdot \mathbf{d}^{\nu\bar{\nu}}_{\mathbf{q,q'}}}{|\mathbf{R}_n-\mathbf{R}'_n|^3} - 3 \frac{\mathbf{d}^{\lambda\bar{\lambda}}_{\mathbf{k,k'}}\cdot (\mathbf{R}_n-\mathbf{R}'_n) \mathbf{d}^{\nu\bar{\nu}}_{\mathbf{q,q'}} (\mathbf{R}_n-\mathbf{R}'_n)}{|\mathbf{R}_n-\mathbf{R}'_n|^5}\right)
\end{align}
with the dipole matrix elements of both materials defined as $\mathbf{d}^{\lambda\bar{\lambda}}_{\mathbf{k,k'}}=e_0\int_{UC} d^3r~u^*_{\lambda,\mathbf{k}}(\mathbf{r})\mathbf{r}u_{\bar{\lambda},\mathbf{k}'}(\mathbf{r})$ where $\bar{\lambda}\neq\lambda$. As explicitly expressed, the F\"orster-type transfer is referred to as a dipole-dipole coupling.

Next, we insert the second line of Eq. \eqref{eq:Taylor} into Eq. \eqref{eq:CoulombMatrix2} to investigate the MA-type coupling:
\begin{align}
V^{\lambda\nu\nu'\lambda'}_{\mathbf{k},\mathbf{q},\mathbf{q}',\mathbf{k}'} &=\frac{e_0^2}{4\pi\epsilon_0\epsilon}\sum_{\mathbf{R}_n,\mathbf{R}_n'}\xi^*_{\mathbf{k}}(\mathbf{R}_n)\xi^*_{\mathbf{q}}(\mathbf{R}_n')\xi_{\mathbf{q}'}(\mathbf{R}_n')\xi_{\mathbf{k}'}(\mathbf{R}_n) \frac{(\mathbf{R}_n-\mathbf{R}'_n)}{|\mathbf{R}_n-\mathbf{R}'_n|^3} \nonumber \\
&\times\left(\int_{UC}d^3r_n ~ u^*_{\lambda,\mathbf{k}}(\mathbf{r}_n)u_{\lambda',\mathbf{k}'}(\mathbf{r}_n)\int_{UC}d^3r'_n ~ u^*_{\nu,\mathbf{q}}(\mathbf{r}_n')\mathbf{r}_n'u_{\nu',\mathbf{q}}(\mathbf{r}_n') \right.\nonumber \\
&\left. - \int_{UC}d^3r_n ~ u^*_{\lambda,\mathbf{k}}(\mathbf{r}_n)\mathbf{r}_n u_{\lambda',\mathbf{k}'}(\mathbf{r}_n)\int_{UC}d^3r'_n ~ u^*_{\nu,\mathbf{q}}(\mathbf{r}_n')u_{\nu',\mathbf{q}}(\mathbf{r}_n') \right) \;. \label{eq:MA}
\end{align}
Here, the second line of Eq. \eqref{eq:MA} describes the interaction of a TMDC interband transition ($r_n'$-integral, $(\nu,\mathbf{q})$ WSe$_2$ quantum numbers) and an intraband graphene transition ($r_n$-integral, $(\lambda,\mathbf{k})$ graphene quantum numbers). Note, the third line of Eq. \eqref{eq:MA} describes an interband transition in the graphene layer and an intraband transition in the WSe$_2$ layer. It is not feasible in our experiment and also unlikely from theoretical considerations. Thus, we neglect this term. 
Finally, we obtain
\begin{align}
V^{\lambda\nu\bar{\nu}\lambda}_{\mathbf{k},\mathbf{q},\mathbf{q}',\mathbf{k}'} &=\frac{e_0}{4\pi\epsilon_0\epsilon}\sum_{\mathbf{R}_n,\mathbf{R}_n'}\xi^*_{\mathbf{k}}(\mathbf{R}_n)\xi^*_{\mathbf{q}}(\mathbf{R}_n')\xi_{\mathbf{q}'}(\mathbf{R}_n')\xi_{\mathbf{k}'}(\mathbf{R}_n) \frac{(\mathbf{R}_n-\mathbf{R}'_n)}{|\mathbf{R}_n-\mathbf{R}'_n|^3} \mathbf{d}^{\nu\bar{\nu}}_{\mathbf{q,q'}} \delta_{\mathbf{k,k'}} \;, \label{eq:MAHamilton}
\end{align}
where we used the orthogonality of the lattice periodic functions to solve the $r_n$-integral. As explicitly shown in Eq. \eqref{eq:MAHamilton}, only one dipole element is involved in MA-type transfer, distinguishable from the F\"orster-type dipole-dipole coupling.

\subsection{Distance dependence of Meitner-Auger-like and F\"orster-type energy transfer}

To study the distance dependence of the Meitner-Auger transfer and compare it to the known dependence of the F\"orster process, we calculate MA- and F\"orster-type transfer rates as a function of interlayer distance $z$. In Supplementary Fig.~\ref{figS13} we show the $z$-dependence of the Meitner-Auger transfer and also the F\"orster result, from which we observe a very different behaviour at long distances. For the F\"orster transfer, we obtain an exponential decay ($\exp(-z)$) in the near-field and the well-known $z^{-4}$ dependence in the far-field. In contrast to the F\"orster transfer, the Meitner-Auger transfer decays solely exponentially. In principle, these two processes could be distinguished by the different behaviours at long distances. However, both processes decay exponentially at short distances, where they can not be distinguished. To formally study the $z$-dependence of both mechanisms, we assume a thermal distribution of the initial states. The corresponding $Q$-dependent rate is then averaged over this thermal distribution (cp. for instance, the approach of reference 14 in the main text). To study the behaviour analytically, we assume a uniform dielectric environment. \\

The detailed computations are shown in the following:
The Meitner-Auger rate reads $\Gamma_{\mathbf{Q}}=|W_{\mathbf{Q}}|^2\frac{4}{\hbar v_F}(Q-E_{\mathbf{Q}}/\hbar v_F)(f_{\mathbf{Q}}-f_{E_{\mathbf{Q}}-Q/\hbar v_F})\theta (Q-E_{\mathbf{Q}}/\hbar v_F)$, where $W_{\mathbf{Q}}$ carries a general Coulomb potential depending on the dielectric environment. In the manuscript we consider a complex non-linear dielectric function to accurately describe the two-dimensional materials on the substrate and evaluate the matrix element in momentum space. However, to have analytical insights, we assume now a Coulomb potential of the form $V(\mathbf{r-r'})=\frac{1}{4\pi\epsilon\epsilon_0}\frac{1}{|\mathbf{r-r'}|}$ with uniform background. We can then start from the Coulomb matrix element Eq. \eqref{eq:MAHamilton} forming $W_{\mathbf{Q}}$ in the Meitner-Auger rate. We perform a coordinate transformation with the new coordinates $\mathbf{s}=(\mathbf{R}_n+\mathbf{R}_n'+\mathbf{z})/2$ and $\mathbf{S}=\mathbf{R}_n-\mathbf{R_n'-\mathbf{z}}$.  Equation \eqref{eq:MAHamilton} becomes
\begin{align}
V^{\lambda\nu\bar{\nu}\lambda}_{\mathbf{k,q,q',k'}}=\frac{1}{4\pi\epsilon_0\epsilon }\sum_{\mathbf{s,Q}}e^{-i\mathbf{Q}\cdot(\mathbf{s+z})} \frac{\mathbf{s+z}}{|\mathbf{s+z}|^3}\delta_{\mathbf{Q,k-k'}}\delta_{\mathbf{Q,q-q'}} \mathbf{d}^{\nu\bar{\nu}}_{\mathbf{q,q-Q}} \;.
\end{align}
The dipole element can be written as $\mathbf{d}^{\nu\bar{\nu}}_{\mathbf{q,q-Q}}=d^{\nu\bar{\nu}}_{\mathbf{q,q-Q}}\mathbf{e}$ with polarization vector $\mathbf{e}$. The Hamiltonian Eq. \eqref{eq:CoulombHamilton} reads 
\begin{align}
H&=\sum_{\substack{\lambda,\nu \\ \mathbf{q,Q,k}}} \frac{1}{4\pi\epsilon_0\epsilon}d^{\nu\bar{\nu}}_{\mathbf{q-Q}/2,\mathbf{q+Q}/2,} a(\mathbf{Q,z}) ~ \lambda^{\dagger}_{\mathbf{k+Q}/2}\nu^{\dagger}_{\mathbf{q-Q}/2}\bar{\nu}^{\dagger}_{\mathbf{q+Q}/2}\lambda^{\dagger}_{\mathbf{k-Q}/2} \\
\text{with}\quad a(\mathbf{Q,z})&=\int d^2s ~ \frac{e^{-i\mathbf{Q}\cdot(\mathbf{s+z})}}{|\mathbf{s+z}|^3} (\mathbf{s+z})\cdot\mathbf{e} \;.
\end{align}
The integral can be solved and reads
\begin{align}
a(\mathbf{Q},z)=-2\pi i\frac{e^{-Qz}}{Q} \mathbf{Q}\cdot\mathbf{e} \;.
\end{align}
To evaluate the $z$-dependence of the Meitner-Auger rate, we perform a thermal average according to our previous work\cite{selig2019theory}:
\begin{align}
\frac{1}{\tau_T}=\langle\frac{1}{\tau}\rangle_T=\frac{1}{\Omega} \int d^2Q ~ e^{-\beta E_{\mathbf{Q}}} \frac{1}{\tau_{\mathbf{Q}}}=\frac{2}{\hbar\Omega} \int d^2Q ~ e^{-\beta E_{\mathbf{Q}}} \Gamma_{\mathbf{Q}}
\end{align}
with $\Omega=\int d^2Q~\exp(-\beta E_{\mathbf{Q}})$ and $\beta=1/k_BT$. To calculate the integral, we set the out-scattering and in-scattering occupation $f_{\mathbf{Q}}$ and $f_{E_{\mathbf{Q}}-Q/\hbar v_F}$ to one and zero, which correspond to a fairly rough approximation. The matrix element $|W_{\mathbf{Q}}|^2$ is determined by the function $|a(\mathbf{Q},z)|^2=2\pi^2\exp(-2Qz)$ after an angle average. The solution of the integral reads
\begin{align}
\frac{1}{\tau_T}&=\frac{\pi^2}{2\hbar\lambda(T)}e^{-\frac{E_{1s}(E_{1s}\lambda^2(T)+2\hbar v_F z)}{\hbar^2v_F^2}} \left[ -2\hbar\lambda(T)v_F z \right. \nonumber \\
&\left. + e^{\frac{(E_{1s}\lambda^2(T)+\hbar v_F z)^2}{\hbar^2v_F^2\lambda^2(T)}}\sqrt{\pi} \left(2E_{1s}\lambda^2(T)z + \hbar v_F (\lambda^2(T)+2z^2)\right) \text{erfc}\left(\frac{E_{1s}\lambda(T)}{\hbar v_F}+\frac{z}{\lambda(T)}\right) \right] \;, \label{eq:ThermAver}
\end{align}
where we introduced the thermal wavelength $\lambda(T)=\hbar/\sqrt{2Mk_BT}$. In Supplementary Fig.~\ref{figS13} we show the $z$-dependence of the Meitner-Auger transfer and also the F\"orster result. In contrast to the F\"orster tansfer, the Meitner-Auger transfer decays exponentially. From the function $a(Q,z)\propto\exp(-Qz)$ we see that the main contribution stems from momenta $Q$ at $Q\approx 0$. But to account for energy and momentum conservation, the Meitner-Auger transfer requires large $Q$, ensured by the Heaviside function in $\Gamma_{\mathbf{Q}}$. We see from the exponential function in front of the parenthesis, that the strength of the exponential decay is determined by the slope of the linear graphene band structure. For the F\"orster transfer the function $a(Q,z)$ is proportional to $Q\exp(-Qz)$ resulting in the different far-field behavior. Interestingly, the different far-field behavior can be traced back to the linear band structure of graphene, reflected by the Heaviside function in $\Gamma_{\mathbf{Q}}$, and not to the difference in the interaction. When we artificially set the TMDC band gap to zero, the Meitner-Auger transfer shows the same $z$-dependence as the F\"orster transfer in the near- and far-field. 
At last, we want to stress, that to calculate Eq. \eqref{eq:ThermAver} we assumed thermalized electron occupations in graphene, which is obviously not the case for a transient experiment.

Compared with a heterostructure made by the exfoliation and stacking technique, the epitaxially grown heterostructure provides a shorter interlayer distance, which benefits the interfacial energy transfer. We have identified the closely stacked WSe$_2$ and graphene layer without any significant spatial gap with atomic force microscopy in a similar heterostructure\cite{nakamura2020spin}. The $z$-dependence of the energy transfer rates is shown in Supplementary Fig.\ref{figS13}. It demonstrates a larger transfer energy rate at a shorter z-distance for both MA and F\"orster-type IET processes. A different IET mechanism could be dominant with increased interlayer distance. However, not only the interlayer distance $z$ determines the interlayer coupling, but also the rotational orientation between two layers. The band structure alignment of an epitaxially grown heterostructure is favoured in a few discrete orientations, for example, twisting angle of $0\deg$ and $60\deg$ between the WSe$_2$ and graphene layer. The heterostructures prepared by the exfoliation method provide the flexibility for engineering the band structure alignment by turning the twisting angel. Therefore, the heterostructure samples prepared by these methods provide their specific way for us to understand the interlayer interaction.  

\begin{figure}
\centering
\includegraphics[width=8cm,keepaspectratio=true]{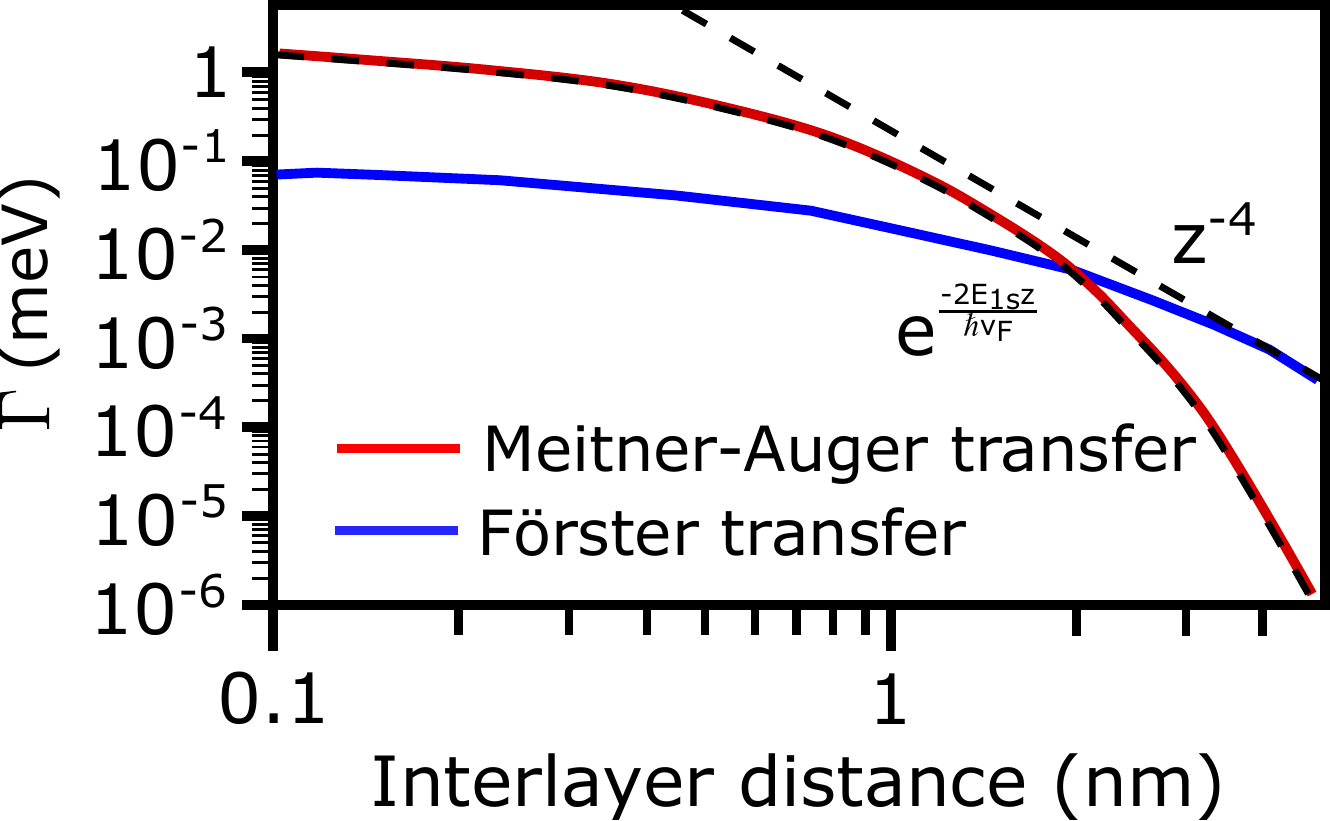}
\caption{\textbf{Meitner-Auger and F\"orster transfer rate as function of layer distance.} 
 The F\"orster transfer rate decays exponentially at short distance and goes over into a $z^{-4}$ dependence (blue). The Meitner-Auger transfer rate continues its exponential decay (red). The strength of the exponential decay of Meitner-Auger rate depends on the slope of the linear graphene band structure
}
\label{figS13}
\end{figure}

\subsection{Dexter Coupling}   
The IET process could also mediated by Dexter-type two-particle exchange, whose transfer rate is determined by the wave function overlap\cite{dexter1953theory}.   
Starting point for the calculation is the Hamiltonian
\begin{equation}
H_D = \sum_{\mathbf{k},\mathbf{q},\mathbf{k'},\mathbf{q'}} V^{c v  v c}_{\mathbf{k},\mathbf{q},\mathbf{k'},\mathbf{q'}} c^{\dagger}_\mathbf{k} v^{\dagger}_\mathbf{q} v_\mathbf{k'} c_\mathbf{q'} + h.c.,
\end{equation}   
with the same conventions for the notation as for the calculation of the F\"orster transfer. The appearing coupling element is defined as
\begin{equation}
V^{c v v c }_{\mathbf{k},\mathbf{q},\mathbf{k'},\mathbf{q'}} = \int_{\mathbb{R}^3} d^3 r \int_{\mathbb{R}^3} d^3 r' \Psi^{ c*}_\mathbf{k} (\mathbf{r}) \Psi^{v*}_\mathbf{q} (\mathbf{r'}) V (\mathbf{r} , \mathbf{r'}) \Psi^{v}_\mathbf{k'}  (\mathbf{r'})  \Psi^{c}_\mathbf{q'}(\mathbf{r}).
\end{equation}

The Coulomb potential is translational invariant in the in-plane direction, i.e. $V (\mathbf{r} , \mathbf{r'}) = V (\mathbf{r}_\parallel - \mathbf{r'}_\parallel,z,z')$. Fourier transforming the Coulomb potential w.r.t. the in-plane components, writing the electronic wave functions as Bloch waves and decomposing the spatial coordinates into one component inside the unit cell and one which addresses the unit cells $\mathbf{r} \rightarrow \mathbf{r}  + \mathbf{R} $, yields for the coupling element

\begin{equation}
V^{cvvc }_{\mathbf{k},\mathbf{q},\mathbf{k'},\mathbf{q'}} =\frac{1}{A}\sum_\mathbf{K} \int_{uc} dz \int_{uc} dz' \chi^c (z) \chi^v (z') V_\mathbf{K}(z,z') \delta_\mathbf{k,q'+K}\delta_\mathbf{q,k'-K},
\end{equation}
with
\begin{equation}
\chi^\lambda (z)= \frac{1}{V_{uc}} \int_{uc} d^2 r_\parallel u^{*\lambda}_{TMD} (\mathbf{r}_\parallel,z)u^{\lambda}_{Graphene} (\mathbf{r}_\parallel,z) 
\end{equation}

To evaluate the coupling element further, we restrict ourselves to the case with vanishing interlayer spacing. We decompose the $z$ and $z'$ integration into two integrals over WSe$_2$ and graphene

\begin{align}
V^{cvvc}_{\mathbf{k},\mathbf{q},\mathbf{k'},\mathbf{q'}} =\frac{1}{A}\sum_\mathbf{K} \delta_\mathbf{k,q'+K}\delta_\mathbf{q,k'-K} \times \nonumber \\ 
\left( \int_{WSe_2} dz \int_{WSe_2} dz' \chi^c (z) \chi^v (z') V_\mathbf{K}(z,z') \right. \nonumber \\
 + \int_{WSe_2} dz \int_{Graphene} dz' \chi^c (z) \chi^v (z') V_\mathbf{K}(z,z')  \nonumber \\
 + \int_{Graphene} dz \int_{WSe_2} dz' \chi^c (z) \chi^v (z') V_\mathbf{K}(z,z')  \nonumber \\
\left. + \int_{Graphene} dz \int_{Graphene} dz' \chi^c (z) \chi^v (z') V_\mathbf{K}(z,z') \right) 
\end{align}

Given that the Coulomb potential varies only weakly with each layer, we can replace the $z/z'$ dependence by the position of the layer $z=z_{WSe_2},z_{Graphene}$ in the Coulomb potential. This way we arrive at
\begin{align}
V^{cvvc}_{\mathbf{k},\mathbf{q},\mathbf{k'},\mathbf{q'}} =\frac{1}{A}\sum_\mathbf{K} \delta_\mathbf{k,q'+K}\delta_\mathbf{q,k'-K} \times \nonumber \\ 
\left( \chi^c_{WSe_2} \chi^v_{WSe_2} V_\mathbf{K}(z=z_{WSe_2},z'=z_{WSe_2}) \right. \nonumber \\
 + \chi^c_{WSe_2} \chi^v_{Graphene} V_\mathbf{K}(z=z_{WSe_2},z'=z_{Graphene})  \nonumber \\
 + \chi^c_{Graphene} \chi^v_{WSe_2} V_\mathbf{K}(z=z_{Graphene},z'=z_{WSe_2})  \nonumber \\
\left. + \chi^c_{Graphene} \chi^v_{Graphene} V_\mathbf{K}(z=z_{Graphene},z'=z_{Graphene}) \right) ,
\end{align}
with 
\begin{equation}
\chi^\lambda_{WSe_2 / Graphene} = \int_{WSe_2 / Graphene} dz \chi^\lambda (z),
\end{equation}
i.e. the contribution of the wave function overlap of the band $\lambda$ in the individual layers. Assuming, that the integration in both layers contributes equally to the wave function overlap between WSe$_2$ and graphene in conduction and valence band, i.e. $\chi^\lambda_{WSe_2}=\chi^\lambda_{ Graphene}  = \frac{1}{2} \chi^\lambda$, we obtain the final expression for the matrix element
\begin{align}
V^{ cvvc}_{\mathbf{k},\mathbf{q},\mathbf{k'},\mathbf{q'}} = \frac{1}{4 A} \chi^c \chi^v \sum_\mathbf{K} \delta_\mathbf{k,q'+K}\delta_\mathbf{q,k'-K} V^{Dex}_\mathbf{K} \end{align}
with
\begin{align}
V^{Dex}_\mathbf{K} = \left( V_\mathbf{K}(z=z_{WSe_2},z'=z_{WSe_2}) \right.  + V_\mathbf{K}(z=z_{WSe_2},z'=z_{Graphene})  \nonumber \\
 +  V_\mathbf{K}(z=z_{Graphene},z'=z_{WSe_2})  \left. + V_\mathbf{K}(z=z_{Graphene},z'=z_{Graphene}) \right) .
\end{align}

Reinserting the result back into the Hamiltonian (and indexing the operators according to the layer, since due to the momentum conservation our convention breaks down) yields
\begin{equation}
H_D = \sum_{\mathbf{K},\mathbf{k},\mathbf{q}} \frac{1}{4 A} \chi^c \chi^v  V^{Dex}_\mathbf{K+k-q}  c^{\dagger WSe_2}_\mathbf{k+K} v^{\dagger Gr}_\mathbf{q-K} v^{WSe_2}_\mathbf{k} c^{Gr}_\mathbf{q} + h.c..
\end{equation}   

So far, the momenta are defined w.r.t. the $\Gamma$ point in graphene and WSe$_2$. Redefining the coordinates $\mathbf{k} \rightarrow K^W + \mathbf{k}$ and $\mathbf{q} \rightarrow K^G + \mathbf{q}$ expresses them w.r.t. the $K$ point in WSe$_2$/graphene.
A projection on excitonic wave functions in WSe$_2$ yields 

\begin{equation}
H_D = - \sum_{\mathbf{K},\mathbf{q},\nu} \left(  \frac{1}{4 \sqrt{A}} \chi^c \chi^v  \sum_\mathbf{k} \varphi^{* \nu}_{K^W+\mathbf{k}} V^{Dex}_{K^W-K^G+\mathbf{K+k-q}} \right)   P^{\dagger \lambda}_\mathbf{K}R^{K^G+\mathbf{q}}_\mathbf{K} + h.c..
\end{equation}

In the Dexter coupling element, the momentum distance between the $K$ points in graphene and the TMD directly enters. We have $|K^G-K^W| \approx$ \unit[3.6]{nm$^{-1}$}. As a first approximation we can ignore the COM and relative momenta inside the Coulomb potential $ V^{Dex}_{K^W-K^G+\mathbf{K+k-q}} \approx  V^{Dex}_{K^W-K^G}$ and get a first estimate for the Dexter coupling

\begin{equation}
H_D = - \sum_{\mathbf{K},\mathbf{q},\nu} D_\mathbf{K} P^{\dagger \nu}_\mathbf{K}R^{K^G+\mathbf{q}}_\mathbf{K} + h.c.,
\end{equation}

with
\begin{equation}
D_\mathbf{K} =  \frac{1}{4 \sqrt{A}} \chi^c \chi^v  \varphi^{*\nu}(\mathbf{r = 0}) V^{Dex}_{K^W-K^G}.
\end{equation}

Similar to the F\"orster transfer, we can evaluate the Dexter induced scattering rate for WSe$_2$ excitons to graphene
\begin{equation}
\Gamma_\mathbf{Q} = 2 \pi \sum_\mathbf{q} |D_\mathbf{Q}|^2 \delta (E_\mathbf{Q}^\mathbf{q}-E^{1s}_\mathbf{Q})
\end{equation}
with yields
\begin{equation}
\Gamma_\mathbf{Q} = \frac{ |D_\mathbf{Q}|^2 A E_\mathbf{Q}}{4 \hbar^2 v_F^2}.
\end{equation}
 
\begin{figure}
 \begin{center}
\includegraphics[width=1\linewidth]{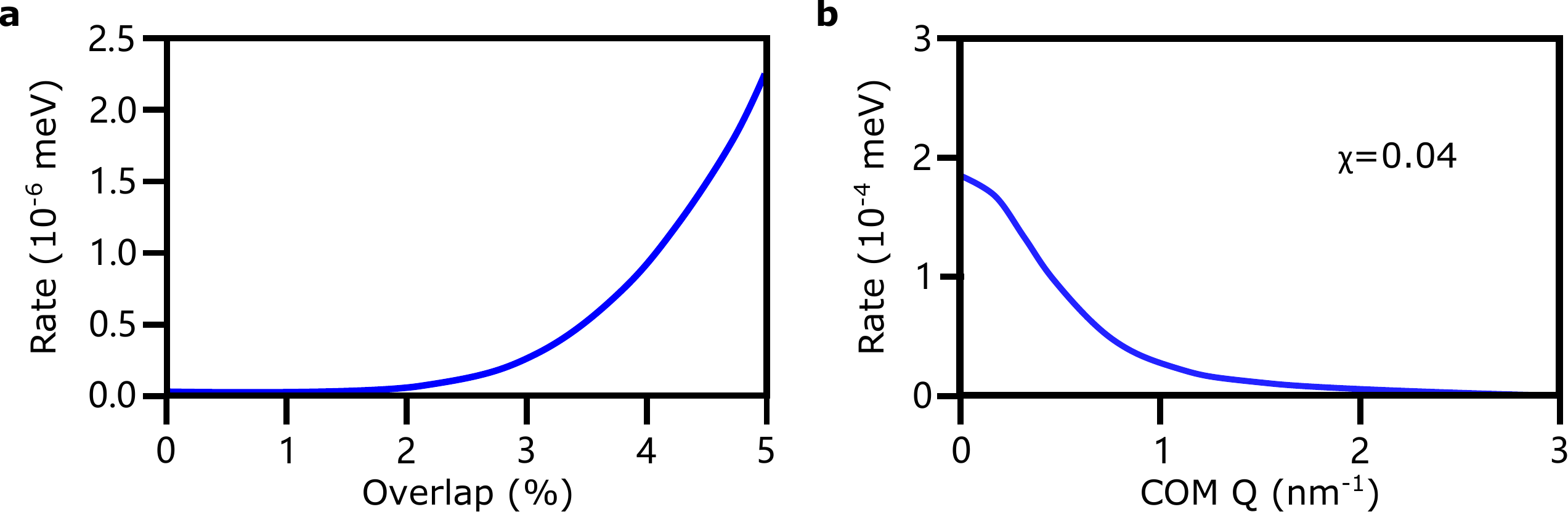}
 \end{center}
 \caption{\textbf{Estimation for the Dexter transfer rate.} \textbf{a}, Dexter transfer rate from WSe$_2$ to graphene as a function of the overlap of the electronic wave functions of graphene and WSe$_2$. \textbf{b}, Dexter transfer rate from WSe$_2$ to graphene as a function of the excitonic COM momentum $\mathbf{Q}$, with the $\mathrm{K_{WSe_2}}$ valley and graphene being shifted on top of each others. }
 \label{figS8}
\end{figure}

Supplementary Fig.~\ref{figS8}\textbf{a} illustrates the Dexter transfer rate from WSe$_2$ to graphene as a function of the wave function overlap for the same structure as considered for the F\"orster transfer. Assuming an overlap between the TMD and graphene wave functions of $\chi =$0.039 (see estimation below, Supplementary Fig.~\ref{Tunnel}\textbf{a}), we arrive at a Dexter rate of \unit[1.0]{$\cdot$10$^{-6}$meV}. This number is small due to the mismatch of the $K$ points of WSe$_2$ and graphene and due to the small overlap of the wave functions which enters with the fourth power.
As long as the COM momentum $\mathbf{Q}$ is much smaller compared to the distance between the $K$ points, the Dexter rate is independent of the COM momentum. 

\subsection{Maximum estimation for the Dexter process}

The reason for the very weak Dexter process is the large momentum mismatch between the $K$ points in WSe$_2$ and graphene. This generates a momentum-bottleneck. To get an estimation of the maximally possible Dexter rate (and Dexter-like processes, such as Dexter-two-phonon processes), we remove this bottleneck artificially and move the $K$ points of both layers on top of each other. The Hamiltonian reads
\begin{equation}
H_D = - \sum_{\mathbf{K},\mathbf{q},\nu} D_\mathbf{q,K}   P^{\dagger \lambda}_\mathbf{K}R^{\mathbf{q}}_\mathbf{K} + h.c..
\end{equation}
with the coupling element
\begin{equation}
D_\mathbf{q,K} = \frac{1}{4 \sqrt{A}} \chi^c \chi^v  \sum_\mathbf{k} \varphi^{* \nu}_{\mathbf{k}} V^{Dex}_{\mathbf{K+k-q}}.
\end{equation}

The relaxation rate of excitons to graphene is then given by
\begin{equation}
\Gamma_\mathbf{K} = \frac{A}{2\pi\hbar} \int_0^{2\pi} d\phi \frac{E_\mathbf{K}}{\hbar^2 v_F^2} |D_{\frac{E_\mathbf{K}}{\hbar v_F} \left(cos \phi , sin \phi \right)^T,\mathbf{K}}|^2.
\end{equation}

Supplementary Fig.~\ref{figS8}\textbf{b} illustrates the maximum estimation of the Dexter rate as the function of the COM momentum. We find a relatively weak rate in the order of $10^{-3}~\mathrm{meV}$ due to the poor wave function overlap. If we would set $\chi=1$, we would obtain \unit[70]{meV} at the maximum. Last, we calculate the thermal average of the Dexter rate, i.e. integrate the momentum dependent Dexter rate together with a normalized Boltzmann distribution at \unit[300]{K}. We obtain \unit[1.5]{$\cdot 10^{-4}~\mathrm{meV}$}.

We would like to note the computation of orbital overlaps is very difficult. Practically, we do not calculate the overlap explicitly but compute the Dexter rate as function of wavefunction overlap. Note, the overlap is not a "totally" unknown parameter in our sample, since we can use the trARPES result as an input. With the below-bandgap excitation (outside the range of the MA transition and therefore constituting as an independent result), we observe an interlayer electron scattering from graphene to WSe$_2$. By comparing with our calculation of the phonon-assisted charge tunneling rate as function of wavefunction overlap, we could extract an overlap value of around 4\%. It can be used to calculate the Dexter-type energy transfer rate in a reasonable range. This wavefunction overlap value is comparable to the \textit{ab-initio} calculated value of the neighboured atoms overlap in graphene, 7\%\cite{reich2002tight}, and the interlayer overlap in a MoSe$_2$/WSe$_2$ heterostructure, 1\%\cite{ovesen2019interlayer}.

Besides the wavefunction overlap, we present the details of the Dexter transfer calculation and the applied approximations in the following. We start from the general Coulomb Hamiltonian involving all transfer processes (F\"orster, Dexter, Meiner-Auger)
\begin{align}
H&=\sum_{\substack{\lambda_1,\lambda_2,\lambda_3,\lambda_4 \\ \mathbf{k}_1,\mathbf{k}_2,\mathbf{k}_3,\mathbf{k}_4 \\ l_1,l_2,l_3,l_4}} V^{\lambda_1\nu_2\lambda_3\lambda_4 l_1l_2l_3l_4}_{\mathbf{k}_1,\mathbf{k}_2,\mathbf{k}_3,\mathbf{k}_4} \lambda_{1,\mathbf{k}_1}^{\dagger l_1}\lambda^{\dagger l_2}_{2,\mathbf{k}_2}\lambda^{l_3}_{3,\mathbf{k}_3}\lambda^{l_4}_{4,\mathbf{k}_4} \label{eq:D} \\
V^{\lambda_1\nu_2\lambda_3\lambda_4 l_1l_2l_3l_4}_{\mathbf{k}_1,\mathbf{k}_2,\mathbf{k}_3,\mathbf{k}_4}&=\int d^3r\int d^3r' ~ \Psi^{*l_1}_{\lambda_1,\mathbf{k}_1}(\mathbf{r})\Psi^{*l_2}_{\lambda_2,\mathbf{k}_2}(\mathbf{r'}) V(\mathbf{r-r'}) \Psi^{l_3}_{\lambda_3,\mathbf{k}_3}(\mathbf{r}')\Psi^{l_4}_{\lambda_4,\mathbf{k}_4}(\mathbf{r}) 
\end{align}
with band $\lambda_i$, wave vector $\mathbf{k}_i$ and layer $l_i$. In Eq. \eqref{eq:D} we can perform the sum over the band and layer indices. The combination $l_1=l_3$ and $l_2=l_4$ and $\lambda_1=\lambda_4=c$ and $\lambda_2=\lambda_3=v$ corresponds to the Dexter Hamiltonian:
\begin{align}
H=\sum_{\mathbf{k}_1,\mathbf{k}_2,\mathbf{k}_3,\mathbf{k}_4} V^{c v  v c GWGW}_{\mathbf{k}_1,\mathbf{k}_2,\mathbf{k}_3,\mathbf{k}_4} c^{\dagger G}_{\mathbf{k}_1} v^{\dagger W}_{\mathbf{k}_2} v^G_{\mathbf{k}_3} c^W_{\mathbf{k}_4} + \text{H.c.}=\sum_{\mathbf{k},\mathbf{q},\mathbf{k'},\mathbf{q'}} V^{c v  v c}_{\mathbf{k},\mathbf{q},\mathbf{k'},\mathbf{q'}} c^{\dagger}_\mathbf{k} v^{\dagger}_\mathbf{q} v_\mathbf{k'} c_\mathbf{q'} + \text{H.c.} \label{eq:D2}
\end{align}
where in the second step, we define $\mathbf{k}^{(')}$ as wave vectors in graphene and $\mathbf{q}^{(')}$ as wave vectors in WSe$_2$, such that we can drop the layer index for graphene ($G$) and WSe$_2$ ($W$). The Hamiltonian Eq. \eqref{eq:D2} describes the electron transfer from the conduction band of WSe$_2$ to graphene, together with the valence electron transfer from graphene to WSe$_2$. The matrix element reads
\begin{align}
V^{cvvc }_{\mathbf{k},\mathbf{q},\mathbf{k'},\mathbf{q'}} &= \frac{1}{A}\sum_\mathbf{K}\int d^3r\int d^3r' ~ \Psi^{*}_{c,\mathbf{k}}(\mathbf{r})\Psi^*_{v,\mathbf{q}}(\mathbf{r'}) V_{\mathbf{K}}(z,z') \Psi_{v,\mathbf{q}'}(\mathbf{r}')\Psi_{c,\mathbf{k}'}(\mathbf{r}) \;, \label{eq:D3}
\end{align}
where we denote the in-plane Fourier transformed Coulomb potential by $V_{\mathbf{K}}(z,z')$. \\

Now, we calculate the Dexter transfer rate. First, we set the interlayer distance to zero. This yields an upper limit of the Dexter rate in the heterostructure using the proper Coulomb potential. From Eq. \eqref{eq:D3} we perform the standard steps of shifting the integrals into the first unit cell $\mathbf{r}_{\parallel}^{(')}\rightarrow \mathbf{r}_{\parallel,n}^{(')}+\mathbf{R}_n^{(')}$ and summing over all unit cells $\sum_{n,n'}$. After inserting Bloch functions we obtain
\begin{align}
V^{cvvc }_{\mathbf{k},\mathbf{q},\mathbf{k'},\mathbf{q'}} &=\frac{1}{A}\sum_\mathbf{K} \int_{UC} dz \int_{UC} dz' \chi^c (z) \chi^v (z') V_\mathbf{K}(z,z') \delta_\mathbf{k,q'+K}\delta_\mathbf{q,k'-K} \label{eq:D4}\;,
\end{align}
where the Kronecker delta is obtained from the Bloch wave envelopes. In Eq. \eqref{eq:D4} we define the orbital overlap $\chi^\lambda (z)= \frac{1}{V_{uc}} \int_{UC} d^2 r_\parallel u^{*\lambda}_{W} (\mathbf{r}_\parallel,z)u^{\lambda}_{G} (\mathbf{r}_\parallel,z) $ of WSe$_2$ ($W$) and graphene ($G$) wavefunction. The $z^{(')}$-integration runs over the two materials, such that Eq. \eqref{eq:D4} can also be written as
\begin{align}
V^{cvvc}_{\mathbf{k},\mathbf{q},\mathbf{k'},\mathbf{q'}} &=\frac{1}{A}\sum_\mathbf{K} \delta_\mathbf{k,q'+K}\delta_\mathbf{q,k'-K} \left( \int_{W} dz \int_{W} dz' \chi^c (z) \chi^v (z') V_\mathbf{K}(z,z') + \int_{W} dz \int_{G} dz' \chi^c (z) \chi^v (z') V_\mathbf{K}(z,z')  \right. \nonumber \\
&\hspace{-5mm}\left.  + \int_{G} dz \int_{W} dz' \chi^c (z) \chi^v (z') V_\mathbf{K}(z,z') + \int_{G} dz \int_{G} dz' \chi^c (z) \chi^v (z') V_\mathbf{K}(z,z') \right) \;. \label{eq:Dexter1}
\end{align}
To treat the $z$-integrations in Eq. \eqref{eq:Dexter1}, we assume that the Coulomb potential varies only weakly in the out-of-plane direction (due to the atomic thickness of the two materials). The approach presented here is also performed for the Rytova-Keldysh potential in 2D materials\cite{chernikov2014exciton,wu2015exciton,robert2018optical} and serves as a standard assumption\cite{rytova2018screened,keldysh1979coulomb,PhysRevB.84.085406}. Then, we can use the Coulomb potential directly at the material position and take it out of the integral. We obtain Eq. \eqref{eq:Dexter1}
\begin{align}
V^{cvvc}_{\mathbf{k},\mathbf{q},\mathbf{k'},\mathbf{q'}} &=\frac{1}{A}\sum_\mathbf{K} \delta_\mathbf{k,q'+K}\delta_\mathbf{q,k'-K} \left( \chi^c_{W} \chi^v_{W} V_\mathbf{K}(z=z_{W},z'=z_{W}) \right.\nonumber \\
&\left. + \chi^c_{W} \chi^v_{G} V_\mathbf{K}(z=z_{W},z'=z_{G})  + \chi^c_{G} \chi^v_{W} V_\mathbf{K}(z=z_{G},z'=z_{W})   + \chi^c_{G} \chi^v_{G} V_\mathbf{K}(z=z_{G},z'=z_{G}) \right) 
\end{align}
with $\chi^\lambda_{W / G} = \int_{W / G} dz \chi^\lambda (z)$ describing the wavefunction overlap of the bands $\lambda$ of the individual layers (since $\chi^{\lambda}=\int_{UC}d^2r_{\parallel}u_W^*(\mathbf{r}_{\parallel},z)u_G(\mathbf{r}_{\parallel},z)$). Next, we assume that both layers contribute equally to the wavefunction overlap. This has the consequence that the interlayer and intralayer Coulomb potential contribute equally to the Dexter rate. In an extreme scenario, we calculate the Dexter-type transfer rate with the wavefunction overlap occurring only in one material. It means two times of the intralayer potential will appear instead of the interlayer potential. The Dexter-type transfer rate only increases by a factor of $\sim$1.5, indicating that the transfer process depends mainly on the wavefunction overlap value (which enters to the power of 4 in the Dexter rate) but not where the overlap occurs. By using $\chi^\lambda_{W}=\chi^\lambda_{ G}  = \frac{1}{2} \chi^\lambda$ we obtain the final expression
\begin{align}
V^{ cvvc}_{\mathbf{k},\mathbf{q},\mathbf{k'},\mathbf{q'}} &= \frac{1}{4 A} \chi^c \chi^v \sum_\mathbf{K} \delta_\mathbf{k,q'+K}\delta_\mathbf{q,k'-K} V^{Dex}_\mathbf{K} \\
V^{Dex}_\mathbf{K} &= V_\mathbf{K}(z=z_{W},z'=z_{W})   + V_\mathbf{K}(z=z_{W},z'=z_{G})   +  V_\mathbf{K}(z=z_{G},z'=z_{W})   + V_\mathbf{K}(z=z_{G},z'=z_{G}) \;.
\end{align}
When we reinsert the matrix element into the Hamiltonian Eq. \eqref{eq:D2} and shift the origin of the wave vector on the K points of the respective materials, i.e. $\mathbf{k} \rightarrow \mathbf{K}^W + \mathbf{k}$ and $\mathbf{q} \rightarrow \mathbf{K}^G + \mathbf{q}$ we obtain
\begin{align}
H_D = \sum_{\mathbf{K},\mathbf{k},\mathbf{q}} \frac{1}{4 A} \chi^c \chi^v  V^{Dex}_{\mathbf{K}^W-\mathbf{K}^G+\mathbf{K+k-q}}  c^{\dagger W}_{\mathbf{K}^W+\mathbf{k+K}} v^{\dagger G}_{\mathbf{K}^G+\mathbf{q-K}} v^{W}_{\mathbf{K}^W+\mathbf{k}} c^{G}_{\mathbf{K}^G+\mathbf{q}} + \text{H.c.} \label{eq:Dexter2}
\end{align}
Finally, we find that the Fourier transformed Coulomb potential $V^{Dex}_{\mathbf{K}^W-\mathbf{K}^G+\mathbf{K+k-q}}$ depends on the momentum difference $\mathbf{K}_W-\mathbf{K}_G+\mathbf{K}+\mathbf{k}-\mathbf{q}$, where $\mathbf{K}_W$ and $\mathbf{K}_G$ are the K points of WSe$_2$ and graphene respectively, $\mathbf{k}$ and $\mathbf{q}$ are the electronic momenta in WSe$_2$ and graphene, respectively and $\mathbf{K}$ is the Coulomb induced momentum transfer. From $V^{Dex}_{\mathbf{K}^W-\mathbf{K}^G+\mathbf{K+k-q}}$ we see that if a large momentum mismatch between the valleys occurs, the Dexter contribution is reduced. To obtain an upper limit estimate, we approximate $\mathbf{K}_W-\mathbf{K}_G+\mathbf{K}+\mathbf{k}-\mathbf{q}\approx \mathbf{K}_W-\mathbf{K}_G$, to be the dominant distance. \\
The computation of the Dexter rate is based on the above approximations and the trARPES result of ICT.

\noindent
\underline{\textbf{Microscopic calculation of interlayer phonon-assisted tunneling process}}

\noindent
In this section we will derive an expression for the phonon-assisted tunneling of carriers between the layers. The Hamiltonian of phonon scattering and tunneling can be generally written as
\begin{align}
H= H_0 + H_1 ,
\end{align}
with $H_0$ accounting for the dispersion of electrons and phonons.
\begin{align}
H_0=\sum_a \epsilon^a a^\dagger_a a_a + \sum_b \hbar \omega^b b_b^\dagger b_b.
\end{align}

The first term accounts for the dispersion of carriers with operators $a^{(\dagger)}_a$ and the second term accounts for the dispersion of phonons with operators $b^{(\dagger)}_b$. The quantum numbers $a,b$ account for layer and momentum of the carriers.
The interaction Hamiltonian $H_1$ reads
\begin{equation}
H_1 = \sum_{ab} t^{ab} a^\dagger_a a_b + \sum_{abc} g^{abc} a^\dagger_a a_b (b_c + b^\dagger_{-c}),
\end{equation}
where the first term represents the tunneling and the second term the scattering of carriers with phonons. Here the notation $-c$ implies, that the momentum has to be inverted, but all other quantum numbers stay the same. 

While we are interested in the second order processes of phonon-assisted tunneling, we apply a canocical transformation to the Hamiltonian
\begin{equation}
H' = e^{-S} H e^S = H_0 + \underbrace{\left( H_1 + [H_0,S] \right)}_{\text{first order}} + \underbrace{\frac{1}{2} [H_1,S]}_{\text{second order}},
\end{equation}
and claim that the first order in the interaction vanishes. This holds true for the choice
\begin{equation}
S = \sum_{ab} \alpha_{ab} t^{ab} a^\dagger a_b + \sum_{abc}  g^{abc} a^\dagger_a a_b (\beta_{abc}b_c + \gamma_{abc} b^\dagger_{-c}),
\end{equation}
with coefficients
\begin{align}
\alpha_{ab} &= \frac{1}{\epsilon^b - \epsilon^a}, \\
\beta_{abc} &= \frac{1}{\epsilon^b-\epsilon^a + \hbar\omega^c},\\
\gamma_{abc} &= \frac{1}{\epsilon^b-\epsilon^a - \hbar\omega^c}.
\end{align}

The second order Hamiltonian is now given as
\begin{equation}
H_2 = \frac{1}{2} [H_1 , S].
\end{equation}
Restricting ourselves only to the tunneling-phonon contribution (besides this, also higher order tunneling terms, two-phonon processes as well as attractive electron-electron interaction through phonon interaction are contained in this Hamiltonian) we obtain

\begin{align}
H&=\frac{1}{2}\sum_{abcd}t^{db}g^{adc}a^{\dagger}_a a_{b}((\frac{1}{\epsilon^{b}-\epsilon^{d}}-\frac{1}{\epsilon^{d}-\epsilon^{a}+\hbar\omega^{c}})b_{c}+(\frac{1}{\epsilon^{b}-\epsilon^{d}}-\frac{1}{\epsilon^{d}-\epsilon^{a}-\hbar\omega^{-c}})b_{-c}^{\dagger})\nonumber\\
&-\frac{1}{2}\sum_{abcd}t^{ad}g^{dbc}a^{\dagger}_a a_b((\frac{1}{\epsilon^{d}-\epsilon^{a}}-\frac{1}{\epsilon^{b}-\epsilon^{d}+\hbar\omega^{c}})b_{c}+(\frac{1}{\epsilon^{d}-\epsilon^{a}}-\frac{1}{\epsilon^{b}-\epsilon^{d}-\hbar\omega^{-c}})b_{-c}^{\dagger})
\end{align}

Now we insert the compounds: $a=(\mathbf{k}_{a},\lambda_a,l_a)$ for electrons, where $\mathbf{k}_{a}$ accounts for the momentum, $\lambda_a$ accounts for the band and $l_a$ accounts for the layer quantum number. For phonons we insert the compounds $c=(\mathbf{k}_{c},l_c,\xi_c)$, with momentum $\mathbf{k}_c$, layer $l_c$ and branch $\xi_c$ and apply the selection rules from the matrix elements:

\begin{align}
t_{\mathbf{k}_{b}\mathbf{k}_{d}}^{\lambda_b \lambda_d l_b l_d}=t^{\lambda_b \lambda_d l_b l_d}\delta_{\mathbf{k}_{b}\mathbf{k}_{d}}^{l_b\bar{l_d}}\delta^{\lambda_b \lambda_d},
\end{align}

i.e. the tunneling conserves momentum, and band but changes the layer index,
and

\begin{align}
g_{\mathbf{k}_{a}\mathbf{k}_{d}\mathbf{k}_{c}}^{\lambda_a \lambda_d l_a l_d l_c}= g_{\mathbf{k}_{c}}^{\lambda_a \lambda_d l_a l_d l_c} \delta^{l_a l_d}\delta_{\mathbf{k}_{c},\mathbf{k}_{a}-\mathbf{k}_{d}}^{l_d l_c}\delta^{\lambda_a \lambda_d}.
\end{align}

i.e. phonon scattering conserves the layer index, but changes the
momentum of the carriers.

As a result we obtain

\begin{align}
H&=\frac{1}{2}\sum_{\mathbf{k}\mathbf{K} \lambda l\xi}\lambda_{\mathbf{k}+\mathbf{K}}^{\dagger\bar{l}} \lambda_{\mathbf{k}}^{l}\left( \underbrace{t^{\lambda l\bar{l}}g_{\mathbf{K}}^{\lambda \bar{l}\xi}(\alpha^{\lambda l\bar{l}}_{\mathbf{k}}+\gamma^{\lambda \bar{l}\xi}_{\mathbf{k},\mathbf{K}})}_{s_\mathbf{k,K}^{\lambda l \bar{l}\xi}}b_{\mathbf{K}}^{\bar{l}\xi}+\underbrace{t^{\lambda l\bar{l}}g_{\mathbf{K}}^{\lambda \bar{l}\xi}(\alpha^{\lambda l\bar{l}}_{\mathbf{k}}+\beta^{\lambda \bar{l}\xi}_{\mathbf{k},\mathbf{K}})}_{\tilde{s}_\mathbf{k,K}^{\lambda l \bar{l}\xi}}b_{-\mathbf{K}}^{\dagger\bar{l}\xi}\right)\nonumber \\
&-\frac{1}{2}\sum_{\mathbf{k}\mathbf{K} \lambda l\xi}\lambda_{\mathbf{k}+\mathbf{K}}^{\dagger\bar{l}}\lambda_{\mathbf{k}}^{l}\left( \underbrace{t^{\lambda \bar{l}l}g_{\mathbf{K}}^{\lambda l\xi}(\alpha^{\lambda l\bar{l}}_{\mathbf{k+K}}+\gamma^{\lambda l\xi}_{\mathbf{k},\mathbf{K}})}_{u_\mathbf{k,K}^{\lambda l \bar{l}\xi}}b_{\mathbf{K}}^{l\xi}+\underbrace{t^{\lambda \bar{l}l}g_{\mathbf{K}}^{\lambda l\xi}(\alpha^{\lambda l\bar{l}}_{\mathbf{k+K}}+\beta_{\mathbf{k},\mathbf{K}}^{\lambda l\xi})}_{\tilde{u}_\mathbf{k,K}^{\lambda l \bar{l}\xi}}b_{-\mathbf{K}}^{\dagger l\xi}\right),
\end{align}
with 
\begin{align}
\alpha^{\lambda ij}_{\mathbf{k}} &= \frac{1}{\epsilon_{\mathbf{k}}^{\lambda i}-\epsilon_{\mathbf{k}}^{\lambda j}}\\
\beta^{\lambda i\xi}_{\mathbf{k},\mathbf{K}} &= \frac{1}{\epsilon_{\mathbf{k}+\mathbf{K}}^{\lambda i}-\epsilon_{\mathbf{k}}^{\lambda i}+\hbar\omega_{-\mathbf{K}}^{i\xi}}\\
\gamma^{\lambda i\xi}_{\mathbf{k},\mathbf{K}}  &= \frac{1}{\epsilon_{\mathbf{k}+\mathbf{K}}^{\lambda i}-\epsilon_{\mathbf{k}}^{\lambda i}-\hbar\omega_{\mathbf{K}}^{i\xi}}
\end{align}
Both lines describe phonon assisted tunneling from $\mathbf{k},b$ to $\mathbf{k}+\mathbf{K},\bar{b}$. However, in the first line first the tunneling and next the phonon scattering takes place, whereas in the second line, first the phonon scattering and next the tunneling takes place. Considering the intermediate states$\mathbf{k},\bar{b}$ and $\mathbf{k}+\mathbf{K},b$ being much larger in energy, in both lines, all contributions have the same sign and add up. The relative sign between both lines ($-$)
is compensated by the opposite signs of the appearing $u$ and $s$ functions.
 To further evaluate the phonon assisted tunnel Hamiltonian, we carry out the summation over the layer
\begin{align}
H&=\frac{1}{2}\sum_{\mathbf{k}\mathbf{K} \lambda G\xi}\lambda_{\mathbf{k}+\mathbf{K}}^{\dagger W} \lambda_{\mathbf{k}}^{G}\left( \underbrace{t^{\lambda GW}g_{\mathbf{K}}^{\lambda W\xi}(\alpha^{\lambda GW}_{\mathbf{k}}+\gamma^{\lambda W\xi}_{\mathbf{k},\mathbf{K}})}_{s_\mathbf{k,K}^{\lambda G W\xi}}b_{\mathbf{K}}^{W\xi}+\underbrace{t^{\lambda GW}g_{\mathbf{K}}^{\lambda W\xi}(\alpha^{\lambda GW}_{\mathbf{k}}+\beta^{\lambda W\xi}_{\mathbf{k},\mathbf{K}})}_{\tilde{s}_\mathbf{k,K}^{\lambda G W\xi}}b_{-\mathbf{K}}^{\dagger W\xi}\right)\nonumber \\
&-\frac{1}{2}\sum_{\mathbf{k}\mathbf{K} \lambda W\xi}\lambda_{\mathbf{k}+\mathbf{K}}^{\dagger G}\lambda_{\mathbf{k}}^{W}\left( \underbrace{t^{\lambda GW}g_{\mathbf{K}}^{\lambda W\xi}(\alpha^{\lambda WG}_{\mathbf{k+K}}+\gamma^{\lambda W\xi}_{\mathbf{k},\mathbf{K}})}_{u_\mathbf{k,K}^{\lambda W G\xi}}b_{\mathbf{K}}^{W\xi}+\underbrace{t^{\lambda GW}g_{\mathbf{K}}^{\lambda W\xi}(\alpha^{\lambda WG}_{\mathbf{k+K}}+\beta_{\mathbf{k},\mathbf{K}\xi}^{\lambda W})}_{\tilde{u}_\mathbf{k,K}^{\lambda W G\xi}}b_{-\mathbf{K}}^{\dagger W\xi}\right)\nonumber \\
&+\frac{1}{2}\sum_{\mathbf{k}\mathbf{K} \lambda \xi}\lambda_{\mathbf{k}+\mathbf{K}}^{\dagger G} \lambda_{\mathbf{k}}^{W}\left( \underbrace{t^{\lambda WG}g_{\mathbf{K}}^{\lambda G\xi}(\alpha^{\lambda WG}_{\mathbf{k}}+\gamma^{\lambda G\xi}_{\mathbf{k},\mathbf{K}})}_{s_\mathbf{k,K}^{\lambda W G\xi}}b_{\mathbf{K}}^{G\xi}+\underbrace{t^{\lambda WG}g_{\mathbf{K}}^{\lambda G\xi}(\alpha^{\lambda WG}_{\mathbf{k}}+\beta^{\lambda G\xi}_{\mathbf{k},\mathbf{K}})}_{\tilde{s}_\mathbf{k,K}^{\lambda W G\xi}}b_{-\mathbf{K}}^{\dagger G\xi}\right)\nonumber \\
&-\frac{1}{2}\sum_{\mathbf{k}\mathbf{K} \lambda\xi}\lambda_{\mathbf{k}+\mathbf{K}}^{\dagger W}\lambda_{\mathbf{k}}^{G}\left( \underbrace{t^{\lambda WG}g_{\mathbf{K}}^{\lambda G\xi}(\alpha^{\lambda GW}_{\mathbf{k+K}}+\gamma^{\lambda G\xi}_{\mathbf{k},\mathbf{K}})}_{u_\mathbf{k,K}^{\lambda G W\xi}}b_{\mathbf{K}}^{G\xi}+\underbrace{t^{\lambda WG}g_{\mathbf{K}}^{\lambda G\xi}(\alpha^{\lambda GW}_{\mathbf{k+K}}+\beta_{\mathbf{k},\mathbf{K}}^{\lambda G\xi})}_{\tilde{u}_\mathbf{k,K}^{\lambda G W\xi}}b_{-\mathbf{K}}^{\dagger G\xi}\right)
\end{align}
Here, the first two lines involve phonons in WSe$_2$ whereas the last two lines involve phonons in graphene.  The $-$ signs in front of line two and line four are compensated by the opposite signs of the $s$ and $u$ functions. To investigate the different contributions to the tunnel-phonon coupling in more detail, we evaluate the functions $s$ and $u$ in the different terms: (a) we fix the momenta in WSe$_2$ to the vicinity of the $K$ point. (b) While all scattering processes conserve the energy, the approximation (a) settles the energy and momentum range of involved carriers. (c) In the next step, we approximate the prefactors $\alpha^{\lambda i j}_\mathbf{k}$ by their values in the region of interest, where we find $\alpha^{c G W}_{\mathbf{k}\approx K_W} = \frac{1}{1 \text{eV}} $ and $\alpha^{c W G}_{\mathbf{k}\approx K_G} = \frac{1}{250 \text{meV}} $, which are read out from the DFT calculation in the main manuscript. To evaluate the prefactors $\beta^{\lambda i\xi}_{\mathbf{k},\mathbf{K}}$ and $\gamma^{\lambda i\xi}_{\mathbf{k},\mathbf{K}}$ we first realize, that that $\Delta E_{K_G}$ appears whenever WSe$_2$ phonons are involved, and  $\Delta E_{K_W}$ appears whenever graphene phonons are involved. $\Delta E_{K_G}\approx$\unit[250]{meV} is large in comparison to typical phonon energies of \unit[30]{meV} in WSe$_2$ and $\Delta E_{K_W}\approx$\unit[1]{eV} is large in comparison to typical phonon energies of \unit[200]{meV} in graphene. Consequently we ignore the appearing phonon energies in $\beta^{\lambda i\xi}_{\mathbf{k},\mathbf{K}}$ and $\gamma^{\lambda i\xi}_{\mathbf{k},\mathbf{K}}$. As a result, the Hamiltonian simplifies to
\begin{align}
H&=\sum_{\mathbf{k}\mathbf{K} \lambda ,\xi} \underbrace{\frac{t^{\lambda GW}g_{\mathbf{K}}^{\lambda W \xi}}{\epsilon^W_{K_{gr}}-\epsilon^G_{K_{gr}}}}_{h^{W\xi}_\mathbf{K}\approx const.} \lambda_{\mathbf{k}+\mathbf{K}}^{\dagger W} \lambda_{\mathbf{k}}^{G}\left(b_{\mathbf{K}}^{W\xi}+b_{-\mathbf{K}}^{\dagger W\xi}\right) + h.c.\,\nonumber \\
&+\sum_{\mathbf{k}\mathbf{K} \lambda \xi} \underbrace{\frac{t^{\lambda WG}g_{\mathbf{K}}^{\lambda G \xi}}{\epsilon^G_{K_{W}}-\epsilon^W_{K_{W}}}}_{h^G_\mathbf{K}\approx const.} \lambda_{\mathbf{k}+\mathbf{K}}^{\dagger W} \lambda_{\mathbf{k}}^{G}\left(b_{\mathbf{K}}^{G\xi}+b_{-\mathbf{K}}^{\dagger G\xi}\right) + h.c.\,.
\end{align}

For a carrier in graphene, the relaxation rate to WSe$_2$ via phonon-assisted tunneling is given as
\begin{align}
\Gamma^G_\mathbf{k} &= 2\pi\sum_{\pm,\mathbf{K},\xi,i\in \{W,G\}} |h^{i\xi}_\mathbf{K}|^2 \left(\frac{1}{2} \pm \frac{1}{2} + n^{i\xi}_\mathbf{K} \right) \delta (\epsilon_\mathbf{k}^G - \epsilon_\mathbf{k+K}^W \mp \hbar \omega^{i\xi}_\mathbf{K} )\nonumber \\
\end{align}

Assuming $h^{i\xi}_\mathbf{K}\approx h^{i\xi} $ and $\hbar \omega^{i\xi}_\mathbf{K} \approx \hbar \omega^{i\xi}$, we obtain
\begin{equation}
\Gamma^G_\mathbf{k} = A \sum_{i\in \{W,G \},\xi,\pm} \frac{m^W}{\hbar^2} |h^i|^2 \left(\frac{1}{2} \pm \frac{1}{2} + n^{i\xi} \right) 1_{\epsilon^G_\mathbf{k} \mp \hbar \omega^{i\xi} - \epsilon^W_\mathbf{0} >0},
\end{equation}
which is constant for graphene electrons which have at least the energy of the conduction band plus the phonon energy in graphene. This reflects the constant density of states in WSe$_2$. The area $A$ cancels with the area in the phonon coupling element which is contained in $h^i$.

In contrast for carriers initially located in WSe$_2$, we get
\begin{align}
\Gamma^W_\mathbf{k} &= 2\pi\sum_{\pm,\mathbf{K},\xi,i\in \{W,G\}} |h^{i\xi}_\mathbf{K}|^2 \left(\frac{1}{2} \pm \frac{1}{2} + n^{i\xi}_\mathbf{K} \right) \delta (\epsilon_\mathbf{k}^W - \epsilon_\mathbf{k+K}^G \mp \hbar \omega^{i\xi}_\mathbf{K} ).
\end{align}
With similar approximation as above, we end up at
\begin{equation}
\Gamma^W_\mathbf{k} = A \sum_{i\in \{W,G \},\xi,\pm} \frac{\epsilon_\mathbf{k} \mp \hbar \omega^{i\xi}}{\hbar^2 v^2_F} |h^i|^2  \left(\frac{1}{2} \pm \frac{1}{2} + n^{i\xi} \right) ,
\end{equation}

\begin{figure}
 \begin{center}
\includegraphics[width=1\linewidth]{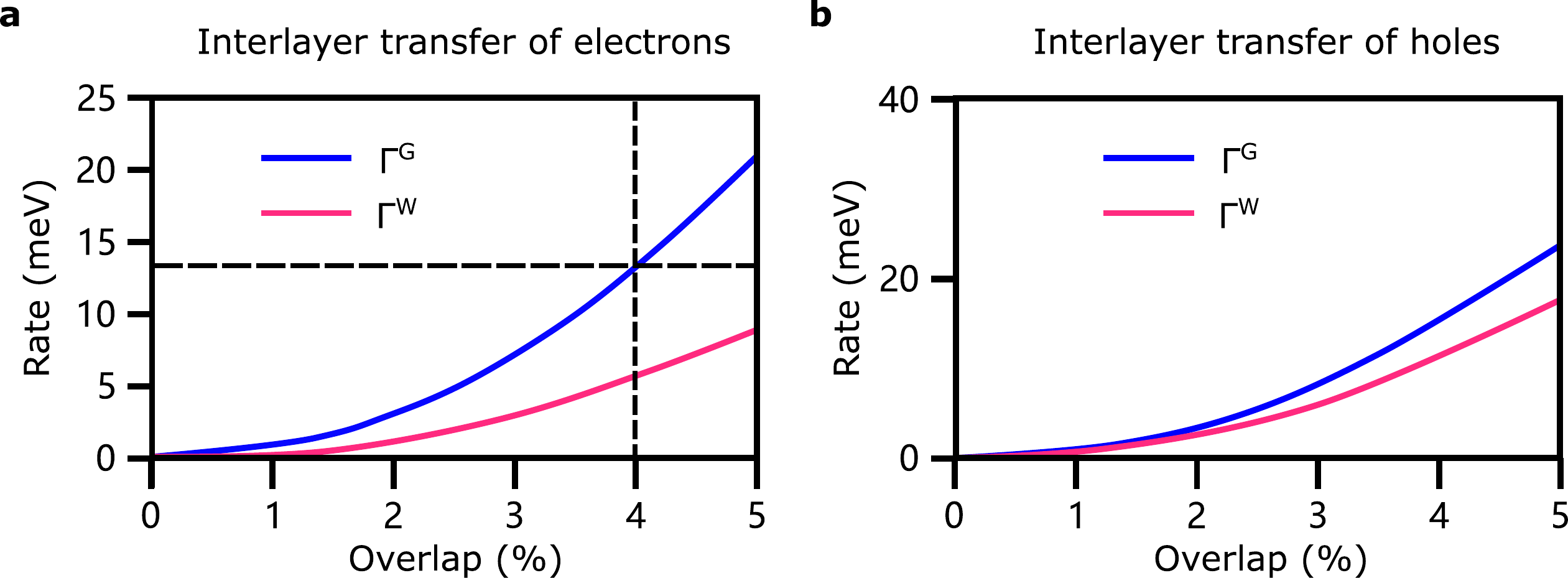}
 \end{center}
 \caption{\textbf{Tunnel transfer for electrons and holes.} \textbf{a}, The ICT-induced relaxation rate of electrons in graphene towards WSe$_2$ (blue) as a function of the electronic wavefunction overlap between the involved conduction bands. The relaxation rate of electrons in WSe$_2$ towards graphene (pink). The dashed lines indicate the overlapping values which can be expected from the experiment results. \textbf{b}, The relaxation rate of holes in graphene, which have larger energies compared to the valence band maximum in WSe$_2$ (blue) and the relaxation rate of holes in WSe$_2$ (pink).}
 \label{Tunnel}
\end{figure}

To calculate the relaxation rates, we assume a potential barrier of $E_B=$\unit[5]{eV} corresponding to the energy of the WSe$_2$ conduction band w.r.t. to the vacuum level. Then we approximate the tunneling element as $t = \chi E_B$, with $\chi$ being the wavefunction overlap\cite{ovesen2019interlayer}. For graphene, we include two optical phonon branches with energies of $\hbar \omega=$\unit[200]{meV} and the coupling strength of $g=$\unit[200]{meV}\cite{piscanec2004kohn}. For WSe$_2$, we include two optical phonon branches with energies of $\hbar \omega=$\unit[30]{meV} and the coupling strength of $g=$\unit[10]{meV}\cite{jin2014intrinsic}. 
Supplementary Fig.~\ref{Tunnel}\textbf{a} illustrates the approximate relaxation rates of electrons from graphene to WSe$_2$ ($\Gamma^G$) and WSe$_2$ to graphene ($\Gamma^W$) as a function of the overlap of the wavefunctions between WSe$_2$ and graphene. We find for both a quadratic increase as a function of the overlap, since the latter enters quadratic in both relaxation rates. The difference between both relaxation rates arise from different final densities of states of the carrier relaxation. In the experiment, a delayed rise of the WSe$_2$ signal w.r.t. to the graphene signal of about \unit[50]{fs} (\unit[13]{meV}) was found, which indicates an overlap of 4.0\%. The order of magnitude of this value appears reasonable, since the overlap between two neighboring graphene atoms is 7\% as an example\cite{reich2002tight}.

To evaluate the phonon-assisted tunneling rates for holes, we assume similar phonon coupling elements in conduction and valence band in WSe$_2$\cite{jin2014intrinsic} and graphene but account for the different dispersion of the valence band\cite{kormanyos2014spin}. Supplementary Fig.~\ref{Tunnel}\textbf{b} illustrates the tunneling rates of holes. We find qualitatively similar tunneling rates as for electrons. However, the tunneling of holes from WSe$_2$ to graphene is stronger compared to the electrons. This arises from the larger density of states of graphene for the involved final states. The reason for this, is that the Fermi energy of the system is closer to the conduction band minimum compared to the valence band maximum, cp. Fig.~2\textbf{a} in the manuscript.

\begin{table}
\centering 
\caption{Parameters used in the computation. $^*$ exemplary value at $z=0$ determined numerically by evaluating the Wannier equation for WSe$_2$ on a SiC substrate. \cite{berghauser2014analytical,selig2016excitonic}. $^{**}$ taken as double distance between the chalcogen atoms. The Fermi velocity taken the experimental result.} 
\begin{tabular}{l l | l l l}\hline 
	Param. &  & Param. & & Ref. \\ 
   $\hbar$ & \unit[0.658]{eV fs}  & $d_G$ & \unit[0.25]{e nm}  & \cite{malic2011microscopic} \\ 
   $e$ & \unit[1]{e} &  $v_F$ & \unit[1.8]{nm fs$^{-1}$} & exp\\
   $\epsilon_0$ & \unit[5.5$\cdot$10$^{-2}$]{e$^2$eV$^{-1}$nm$^{-1}$}  & $d_{WSe_2}$ & \unit[0.32]{e nm}  &\cite{selig2016excitonic}\\ 
   $k_B$ & \unit[8.6$\cdot$10$^{-5}$]{eV K$^{-1}$}   & $|\varphi_{WSe_2}(\mathbf{r}=0)|$ & \unit[0.36]{nm$^{-1}$} & $^*$\\
   $\epsilon_{SiC}$ & \unit[9.6]{}  & $E^{1s}_{WSe_2}$ & \unit[1.7]{eV} & \cite{christiansen2017phonon} \\
    & & $M_{WSe_2}$ & \unit[3.7]{eVfs$^2$nm$^{-2}$} & \cite{kormanyos2015k} \\ 
    $\epsilon_{G}$ & 6\cite{bessler2019dielectric}  & $\epsilon_{WSe_2}$ &  13.36  & \\
    $a_{G}$ & \unit[0.33]{nm} \cite{ye2014thickness} & $a_{WSe_2}$ & \unit[0.67]{nm}  & $^{**}$ \cite{berkelbach2013theory} \\
   \hline

\end{tabular} \label{parameters}
\end{table}

\newpage
\section*{References}
\bibliographystyle{plain}